\documentclass[12pt]{article}  
\usepackage{cite}
\usepackage{epsfig}
\usepackage{graphicx}
\usepackage{amsmath}
\usepackage{amssymb}
\usepackage{ulem}
\usepackage{mdwlist}
\usepackage{color}

\usepackage{a41}
\usepackage{color}
\usepackage[rflt]{floatflt}
\usepackage{float}
\usepackage{slashed}
\newcommand{\lsim}{\raisebox{-0.07cm   }
{$\, \stackrel{<}{{\scriptstyle\sim}}\, $}}
\newcommand{\gsim}{\raisebox{-0.07cm   }
{$\, \stackrel{>}{{\scriptstyle\sim}}\, $}}

\usepackage{array}

\usepackage[english]{babel}
\usepackage[latin1]{inputenc}
\usepackage[T1]{fontenc}
\usepackage{ae}

\usepackage{url}

\usepackage{amsmath, amsthm, amssymb}

\newcommand{\HA}{{\rm H}}
\newcommand{\GA}{{\rm G}}
\newcommand{\GeV}{\rm GeV}
\newcommand{\MS}{\overline{{\sf MS}}}

\newcommand{\Ahathat}{\hat{\hspace*{0mm}\hat{A}}}

\newcommand{\liFhalf}{{\rm Li}_4\left(\frac{1}{2}\right)}
\newcommand{\zF}{\zeta_4}

\newcommand{\ds}{\displaystyle}
\newcommand{\Li}{{\rm Li}}
\newcommand{\Mvec}{{\rm {\bf M}}}

\newcommand{\ep}{\varepsilon}

\usepackage{rotating}

\usepackage{graphicx}

\newcounter{mmacnt}
\def\restartmma{\setcounter{mmacnt}{0}}
\restartmma \catcode`|=\active
\def|#1|{\mathrm{#1}}
\catcode`|=12

\usepackage{color}

\usepackage{tikz}
\usetikzlibrary{matrix}

\allowdisplaybreaks[4]

\begin{document}
\setlength{\baselineskip}{0.515cm}
\sloppy
\thispagestyle{empty}
\begin{flushleft}
DESY 15--112
\\
DO--TH 22/26\\
CERN-TH-2022-179 \\
ZU-TH 53/22\\
RISC Report Series 22--25\\
MSUHEP-22-036\\
November 2022\\
\end{flushleft}

\mbox{}
\vspace*{\fill}
\begin{center}

{\Large\bf The Unpolarized and Polarized Single-Mass Three-Loop} 

\vspace*{2mm} {\Large\bf 
\boldmath{Heavy Flavor Operator Matrix Elements $A_{gg,Q}$ and $\Delta A_{gg,Q}$}}

\vspace{4cm}
{\large
J.~Ablinger$^a$,
A.~Behring$^{b,c}$,
J.~Bl\"umlein$^b$,
A.~De Freitas$^{a,b}$,
A.~Goedicke$^{b}$,\\

\vspace*{2mm}
A.~von~Manteuffel$^d$, 
C.~Schneider$^a$,
and K.~Sch\"onwald$^{b,e}$}

\vspace*{5mm}
{\it $^a$~Johannes Kepler University Linz, 
Research Institute for Symbolic Computation (RISC),
Altenbergerstra\ss{}e 69,
                          A--4040, Linz, Austria}\\

\vspace*{3mm}
{\it  $^b$ Deutsches Elektronen--Synchrotron DESY, Platanenallee 6, 15738 Zeuthen, Germany}

\vspace*{3mm}
{\it $^c$ 
Theoretical Physics Department, CERN, 1211 Geneva 23, Switzerland}

\vspace*{3mm}
{\it $^d$~Department of Physics and Astronomy, Michigan State University, East Lansing, MI 48824, USA}

\vspace*{3mm}
{\it $^e$~
Physik-Institut, Universit\"at Z\"urich, Winterthurerstrasse 190, CH-8057 Z\"urich, Switzerland}
\\

\end{center}
\normalsize
\vspace{\fill}
\begin{abstract}
\noindent 
We calculate the gluonic massive operator matrix elements in the unpolarized and polarized 
cases, $A_{gg,Q}(x,\mu^2)$ and $\Delta A_{gg,Q}(x,\mu^2)$, at three--loop order for a single 
mass. These quantities contribute to the matching of the gluon distribution in the variable 
flavor number scheme. The polarized operator matrix element is calculated in the Larin scheme. 
These operator matrix elements contain finite binomial and inverse binomial sums in Mellin 
$N$--space and iterated integrals over square root--valued alphabets in momentum fraction 
$x$--space. We derive the necessary analytic relations for the analytic continuation of these 
quantities from the even or odd Mellin moments into the complex plane, present analytic expressions 
in momentum fraction $x$--space and derive numerical results. The present results complete the 
gluon transition matrix elements both of the single-- and double--mass variable flavor number 
scheme to three--loop order.
\end{abstract}

\vspace*{\fill}
\noindent
\numberwithin{equation}{section}
\newpage
\section{Introduction}
\label{sec:1}

\vspace{1mm}
\noindent
The heavy flavor corrections to deeply inelastic scattering exhibit different scaling violations
if compared to the light flavor contributions. At the present experimental accuracy of the
deep--inelastic data, precision determinations of the strong coupling constant $\alpha_s(M_Z^2)$ 
\cite{ALPHA} and the determination of the parton distribution functions \cite{PDF} require their 
calculation to next-to-next-to leading order (NNLO). If the virtuality, $Q^2$, is significantly larger 
than the heavy quark mass squared, $m_Q^2$, i.e. $Q^2/m_Q^2\gsim 10$, one may calculate the heavy 
flavor corrections analytically \cite{Buza:1995ie}. 
Here $m_Q$ denotes the heavy quark mass. This corresponds 
to $Q^2$ values above about $25~\GeV^2$ in the case of charm. In Refs.~\cite{Bierenbaum:2009mv,
Ablinger:2017err,Behring:2014eya} it has been shown to NNLO in the single-- and double--mass cases, 
how to express the massive coefficient functions for the structure functions in terms of 
massive operator matrix elements (OMEs) $A_{ij}$ and the massless coefficient functions 
\cite{WILS3} 
at large scales, as well as the corresponding relations for the parton distribution functions 
in the variable flavor number scheme (VFNS)\footnote{The two--loop corrections can be found in 
Refs.~\cite{Buza:1995ie,TWOL,DELG2,Buza:1996wv}.}. From NNLO onward, five of the OMEs 
\cite{THREE,Ablinger:2014nga} are needed to express the deep--inelastic structure functions \cite{SF}. 
The additional OMEs, $A_{gg,Q}$ and $A_{gq,Q}$, at NNLO contribute only to the relations in the 
VFNS. 
The three--loop corrections to $A_{gq,Q}$ have been computed in Ref.~\cite{Ablinger:2014lka}.

In this  paper we calculate the massive OMEs $A_{gg,Q}$ and $\Delta A_{gg,Q}$ to three--loop order. 
In Refs.~\cite{Blumlein:2012vq,Ablinger:2014uka} the $O(T_F^2 N_F)$ and $O(T_F^2)$ contributions to 
the operator matrix element  have been calculated in the unpolarized case already. All logarithmic 
contributions have 
been derived in Refs.~\cite{Behring:2014eya,Blumlein:2021xlc}. A series of moments and scalar 
integrals in the double--mass case were obtained in Refs.~\cite{Ablinger:2017err,TWOMASS}. 
The $O(T_F)$ contribution to the three--loop anomalous dimensions 
$(\Delta)\gamma_{gg,Q}^{(2)}$, resulting 
from the single pole term of the unrenormalized OME, has been 
calculated in 
Refs.~\cite{ Ablinger:2017tan,Ablinger:2019etw,Blumlein:2021ryt}. The double--mass 
three--loop contributions were computed in Refs.~\cite{Ablinger:2018brx,Ablinger:2020snj}. Here both charm and 
bottom quark lines are contained in the corresponding Feynman diagrams. Contributions of this kind emerge
first at two--loop order due to reducible diagrams \cite{Blumlein:2018jfm,DELG2}.

We calculate the massive OME using the techniques described in Ref.~\cite{Ablinger:2015tua}.
These include the method of (generalized) hypergeometric functions \cite{HYP}, the Mellin--Barnes method 
\cite{MB}, the method of ordinary differential equations \cite{DIFF,Ablinger:2018zwz}, and the 
(multivartiate) Almkvist-Zeilberger algorithm \cite{AZ,Ablinger:21}. All these methods finally map the 
problem to multiply nested sums, which are solved using the packages {\tt Sigma} \cite{SIG1,SIG2}, 
{\tt EvaluateMultisums} and {\tt SumProduction} \cite{EMSSU}, based on difference--ring theory 
\cite{Karr:81,Schneider:01,Schneider:05a,Schneider:07d,Schneider:10b,Schneider:10c,Schneider:15a,
Schneider:08c}, as well as the packages {\tt OreSys} \cite{ORESYS}, and {\tt MultiIntegrate} \cite{MI,
Ablinger:21}.\footnote{For surveys on the computation methods and the related 
function and number spaces, see Refs.~\cite{Blumlein:2018cms,Blumlein:2022zkr}.} For an efficient 
treatment of the 
different sum- and function algebras, such as the harmonic sums \cite{Vermaseren:1998uu,Blumlein:1998if} 
and harmonic polylogarithms \cite{Remiddi:1999ew}, generalized harmonic sums and polylogarithms 
\cite{Moch:2001zr,Ablinger:2013cf}, cyclotomic harmonic sums and polylogarithms 
\cite{Ablinger:2011te}, iterated integrals induced by quadratic forms \cite{Ablinger:2021fnc}, 
and finite 
binomial sums and square root--valued iterated integrals \cite{Ablinger:2014bra}, 
we use the package {\tt HarmonicSums} \cite{HARMSU,Blumlein:2009fz,
Blumlein:2003gb,Ablinger:2011te,Ablinger:2013cf,Ablinger:2014bra,
Vermaseren:1998uu,Blumlein:1998if,Remiddi:1999ew,Blumlein:2009ta,Ablinger:2021fnc}.
We mention, in particular, that our solution methods do not require to refer to special bases of 
master integrals.

One obtains a general expression for the constant part of the unrenormalized 
massive OMEs $(\Delta) A_{gg,Q}^{(3)}$, $(\Delta) a_{gg,Q}^{(3)}$, at even (odd) integer values of the 
Mellin variable $N$, according to the light--cone expansion in the unpolarized and polarized cases, 
cf.~\cite{Blumlein:2012bf}. The representation is given in terms of synchronized sum--product 
representations. In the present case it turned out that all but two diagrams could be calculated in 
this way and for the latter cases the problem had to be solved in $x$--space, deriving the $N$--space 
representation afterwards.

We consider the case of a single heavy quark. With the present results, both the unpolarized 
and polarized gluon distributions can be mapped in the single--mass VFNS \cite{Buza:1996wv} 
and the double--mass VFNS \cite{Ablinger:2017err} to three--loop order.

The paper is organized as follows. In Section~\ref{sec:2} we present the basic formalism and 
describe the new functional aspects in the present case, both in Mellin $N$ and $x$--space.
In Section~\ref{sec:3} the results in Mellin $N$-space are presented for 
$a_{gg,Q}^{(3)}$ and $\Delta a_{gg,Q}^{(3)}$. The characteristic aspects of the $x$--space 
results are discussed in Section~\ref{sec:4}. Because the full $x$--space results 
are rather lengthy, we
describe its principal structure and  
present it in full form in a  computer--readable ancillary file to this paper.
Numerical results  are discussed in Section~\ref{sec:5} deriving a fast and precise numerical 
representation,
and Section~\ref{sec:6} contains 
the conclusions. A series of technical aspects are presented in the Appendices \ref{sec:A}--\ref{sec:D}.
\section{Basic Formalism and the computation method}
\label{sec:2}

\vspace{1mm}
\noindent
The unrenormalized massive OMEs $A_{gg,Q}^{(k)}$ from one-- to three--loop order for a 
single heavy quark 
and 
$N_F$ massless quarks, cf.~Ref.~\cite{Bierenbaum:2009mv}, is given by
\begin{eqnarray}
    \Ahathat_{gg,Q}^{(1)}&=& 
             \Bigl(\frac{\hat{m}^2}{\mu^2}\Bigr)^{\ep/2}\Biggl[
                          \frac{\hat{\gamma}_{gg}^{(0)}}{\ep}
                         +a_{gg,Q}^{(1)}
                         +\ep\overline{a}_{gg,Q}^{(1)}
                         +\ep^2\overline{\overline{a}}_{gg,Q}^{(1)}
                        \Biggr] + O(\ep^3),
\\
    \Ahathat_{gg,Q}^{(2)}&=&
             \Bigl(\frac{\hat{m}^2}{\mu^2}\Bigr)^{\ep}
                  \Biggl[
                          \frac{1}{\ep^2} c_{gg,Q,(2)}^{(-2)}
                         + \frac{1}{\ep} c_{gg,Q,(2)}^{(-1)}
                         +  c_{gg,Q,(2)}^{(0)}
                         + \ep c_{gg,Q,(2)}^{(1)} \Biggr]+ O(\ep^2),
\\
    \Ahathat_{gg,Q}^{(3)}&=&
             \Bigl(\frac{\hat{m}^2}{\mu^2}\Bigr)^{3\ep/2}
                  \Biggl[
                          \frac{1}{\ep^3} c_{gg,Q,(3)}^{(-3)}
                         + \frac{1}{\ep^2} c_{gg,Q,(3)}^{(-2)}
                         + \frac{1}{\ep} c_{gg,Q,(3)}^{(-1)}
                         + a_{gg,Q}^{(3)}\Biggr] + O(\ep)~.
 \end{eqnarray}
Here and in the following $\ep = D - 4$ denotes the dimensional parameter, 
$c_{gg,Q,(k)}^{(-l)}$ are the expansion coefficients of the unrenormalized OME $\Ahathat_{gg,Q}$,
$\hat{m}$ denotes the unrenormalized heavy quark 
mass, $\mu$ is both the factorization and
renormalization scale, $\zeta_i$ is Riemann's $\zeta$-function at integer values $i \geq 2$, 
$\beta_k$ and $\beta_{k,Q}$ are expansion coefficients of the  $\beta$-function in Quantum Chromodynamics 
(QCD) in different 
schemes, $\gamma_{ij}^{(k)}$ are anomalous 
dimensions ~\cite{ANOM,Ablinger:2014nga,Ablinger:2017tan,Blumlein:2021ryt,WILS3}
and $m_k^{(l)}$ are the expansion coefficients
of the heavy mass, while $a_{ij}^{(k)}$ and $\bar{a}_{ij}^{(k)}$  
are the expansion coefficients in $\ep$ to $O(\ep^0)$ and $O(\ep)$ of the different 
OMEs. All explanations have been given in Ref.~\cite{Bierenbaum:2009mv}, to which we refer.
Analogous expressions hold for $\Delta A_{gg,Q}^{(k)}$. Here the anomalous dimensions have to be replaced
by the polarized ones, and the coefficients $a_{ij}^{(k)} (\bar{a}_{ij}^{(k)})$ by $\Delta a_{ij}^{(k)}
(\Delta \bar{a}_{ij}^{(k)})$.

The renormalization of $\Ahathat_{gg,Q}^{(3)}$ includes mass and coupling renormalization, as well as 
the renormalization of the local operator, and the subtraction of the collinear
singularities, see~\cite{Bierenbaum:2009mv}. It is essential to work in a MOM-scheme
for charge renormalization first and then transform to the $\overline{\rm MS}$ scheme.
The renormalized OME from first to third order are then obtained by
\begin{eqnarray}
    A_{gg,Q}^{(1), \MS}&=& - \beta_{0,Q} 
\ln\left(\frac{m^2}{\mu^2}\right)~,\\
    A_{gg,Q}^{(2), \MS}&=&
      \frac{1}{8}\Biggl\{
                          2\beta_{0,Q}
                             \Bigl(
                                    \gamma_{gg}^{(0)}
                                   +2\beta_0 
                             \Bigr)
                         +\gamma_{gq}^{(0)}\hat{\gamma}_{qg}^{(0)}
                         +8\beta_{0,Q}^2
                 \Biggr\}
                    \ln^2 \Bigl(\frac{m^2}{\mu^2}\Bigr)
                +\frac{\hat{\gamma}_{gg}^{(1)}}{2}
                     \ln \Bigl(\frac{m^2}{\mu^2}\Bigr)
\nonumber\\ &&
                -\frac{\zeta_2}{8}\Bigl[
                                         2\beta_{0,Q}
                                            \Bigl(
                                                   \gamma_{gg}^{(0)}
                                                  +2\beta_0
                                            \Bigr)
                                        +\gamma_{gq}^{(0)}
                                            \hat{\gamma}_{qg}^{(0)}
                                  \Bigr]
                +a_{gg,Q}^{(2)}~,   \label{AggQ2MSren} 
\\
  A_{gg,Q}^{(3), \MS}&=&
                    \frac{1}{48}\Biggl\{
                            \gamma_{gq}^{(0)}\hat{\gamma}_{qg}^{(0)}
                                \Bigl(
                                        \gamma_{qq}^{(0)}
                                       -\gamma_{gg}^{(0)}
                                       -6\beta_0
                                       -4n_f\beta_{0,Q}
                                       -10\beta_{0,Q}
                                \Bigr)
                           -4
                                \Bigl(
                                        \gamma_{gg}^{(0)}\Bigl[
                                            2\beta_0
                                           +7\beta_{0,Q}
                                                         \Bigr]
\nonumber\\ &&
                                       +4\beta_0^2
                                       +14\beta_{0,Q}\beta_0
                                       +12\beta_{0,Q}^2
                                \Bigr)\beta_{0,Q}
                     \Biggr\}
                     \ln^3 \Bigl(\frac{m^2}{\mu^2}\Bigr)
                    +\frac{1}{8}\Biggl\{
                            \hat{\gamma}_{qg}^{(0)}
                                \Bigl(
                                        \gamma_{gq}^{(1)}
                                       +(1-n_f)\hat{\gamma}_{gq}^{(1)}
                                \Bigr)
\nonumber\\ &&
                           +\gamma_{gq}^{(0)}\hat{\gamma}_{qg}^{(1)}
                           +4\gamma_{gg}^{(1)}\beta_{0,Q}
                           -4\hat{\gamma}_{gg}^{(1)}[\beta_0+2\beta_{0,Q}]
                           +2\gamma_{gg}^{(0)}\beta_{1,Q}
                           +4[\beta_1+\beta_{1,Q}]\beta_{0,Q}
                     \Biggr\}
                     \ln^2 \Bigl(\frac{m^2}{\mu^2}\Bigr)
\nonumber\\ &&
                    +\frac{1}{16}\Biggl\{
                            8\hat{\gamma}_{gg}^{(2)}
                           -8n_fa_{gq,Q}^{(2)}\hat{\gamma}_{qg}^{(0)}
                           -16a_{gg,Q}^{(2)}(2\beta_0+3\beta_{0,Q})
                           +8\gamma_{gq}^{(0)}a_{Qg}^{(2)}
                           +8\gamma_{gg}^{(0)}\beta_{1,Q}^{(1)}
\nonumber\\ &&
                   +\gamma_{gq}^{(0)}\hat{\gamma}_{qg}^{(0)}\zeta_2
                                \Bigl(
                                        \gamma_{gg}^{(0)}
                                       -\gamma_{qq}^{(0)}
                                       +6\beta_0
                                       +4n_f\beta_{0,Q}
                                       +6\beta_{0,Q}
                                \Bigr)
\nonumber\\ &&
                   +4\beta_{0,Q}\zeta_2
                                \Bigl( 
                                       \gamma_{gg}^{(0)}
                                      +2\beta_0
                                \Bigr)
                                \Bigl(
                                       2\beta_0
                                      +3\beta_{0,Q}
                                \Bigr)
                     \Biggr\}
                     \ln \Bigl(\frac{m^2}{\mu^2}\Bigr)
                   +2(2\beta_0+3\beta_{0,Q})\overline{a}_{gg,Q}^{(2)}
\nonumber\\ &&
                   +n_f\hat{\gamma}_{qg}^{(0)}\overline{a}_{gq,Q}^{(2)}
                   -\gamma_{gq}^{(0)}\overline{a}_{Qg}^{(2)}
                   -\beta_{1,Q}^{(2)} \gamma_{gg}^{(0)}
                   +\frac{\gamma_{gq}^{(0)}\hat{\gamma}_{qg}^{(0)}\zeta_3}{48}
                                \Bigl(
                                        \gamma_{qq}^{(0)}
                                       -\gamma_{gg}^{(0)}
                                       -2[2n_f+1]\beta_{0,Q}
\nonumber\\ &&
                                       -6\beta_0
                                \Bigr)
                   +\frac{\beta_{0,Q}\zeta_3}{12}
                                \Bigl(
                                        [\beta_{0,Q}-2\beta_0]\gamma_{gg}^{(0)}
                                       +2[\beta_0+6\beta_{0,Q}]\beta_{0,Q}
                                       -4\beta_0^2
                                \Bigr)
\nonumber\\ &&
                   -\frac{\hat{\gamma}_{qg}^{(0)}\zeta_2}{16}
                                \Bigl(
                                        \gamma_{gq}^{(1)}
                                       +\hat{\gamma}_{gq}^{(1)}
                                \Bigr)
                   +\frac{\beta_{0,Q}\zeta_2}{8}
                                \Bigl(
                                        \hat{\gamma}_{gg}^{(1)}
                                      -2\gamma_{gg}^{(1)}
                                      -2\beta_1
                                      -2\beta_{1,Q}
                                \Bigr)
                           +\frac{\delta m_1^{(-1)}}{4}
                                \Bigl(
                                     8 a_{gg,Q}^{(2)}
\nonumber\\ &&
                                    +24 \delta m_1^{(0)} \beta_{0,Q}
                                    +8 \delta m_1^{(1)} \beta_{0,Q} 
                                    +\zeta_2 \beta_{0,Q} \beta_0
                                    +9 \zeta_2 \beta_{0,Q}^2
                                \Bigr)
                           +\delta m_1^{(0)}
                                \Bigl(
                                     \beta_{0,Q} \delta m_1^{(0)}
                                    +\hat{\gamma}_{gg}^{(1)}
                                \Bigr)
\nonumber\\ &&
                           +\delta m_1^{(1)}
                                \Bigl(
                                     \hat{\gamma}_{qg}^{(0)} \gamma_{gq}^{(0)}
                                    +2 \beta_{0,Q} \gamma_{gg}^{(0)}
                                    +4 \beta_{0,Q} \beta_0
                                    +8 \beta_{0,Q}^2
                                \Bigr)
                           -2 \delta m_2^{(0)} \beta_{0,Q}
                 +a_{gg,Q}^{(3)}~. \label{Agg3QMSren}
   \end{eqnarray}
Here the heavy quark mass is renormalized in the on--shell scheme (OMS) and the strong coupling constant in 
the $\overline{\sf MS}$ scheme. The transformation from the OMS to the $\overline{\sf MS}$ scheme for the 
quark mass is described e.g. in Eq.~(5.13) of Ref.~\cite{Klein:2009ig}, see also 
Refs.~\cite{MASS}.

The Feynman diagrams are generated by using {\tt QGRAF} \cite{Nogueira:1991ex,Bierenbaum:2009mv}, the 
spinor and 
Lorentz--algebra 
is performed using {\tt FORM} \cite{FORM}, and the color algebra by using {\tt Color} 
\cite{vanRitbergen:1998pn}. The 
operator insertions are resummed into linear propagators as has been 
discussed in Ref.~\cite{Ablinger:2014yaa}, e.g. by 
\begin{eqnarray}
\sum_{k = 0}^\infty (\Delta. p)^k t^k = \frac{1}{1 - t \Delta.p},
\label{eq:rest}
\end{eqnarray}
where $t$ denotes an auxiliary parameter. Similar relations hold for the more complicated operator insertions. 
Taking the $k$th moment of (\ref{eq:rest}),
i.e. the coefficient of $t^k$, one obtains the contribution due to $(\Delta. p)^k$. Here $\Delta$ denotes 
a light--like vector.

In total, 642 irreducible Feynman diagrams contribute to the OMEs. The reduction to 
master 
integrals using the integration-by-parts relations
\cite{IBP} has been performed using {\tt Reduze~2} \cite{Studerus:2009ye,vonManteuffel:2012np}.

Unlike the case in later computations starting in 2017, we did not use the method of arbitrary high moments
\cite{Blumlein:2017dxp}, establishing the corresponding difference equations by the method of guessing \cite{GUESS} 
and solving them using the package {\tt Sigma} \cite{SIG1,SIG2}. Instead we calculated the master integrals directly in 
Mellin $N$--space using the different 
methods mentioned in Section~\ref{sec:1}, from which the individual Feynman diagrams were calculated. 
Their first few 
moments were compared to Mellin moments computed using {\tt Matad} 
\cite{Steinhauser:2000ry}, 
cf.~Ref.~\cite{Bierenbaum:2009mv} in the unpolarized case. In the case of two Feynman 
diagrams,
which are related to each other by reversal of the internal fermion line and which give
the same result, we had
to use a different computation method. Here we applied the method of differential equations 
\cite{Ablinger:2018zwz} to 
the master integrals working in the auxiliary variable $t$ used for the operator 
resummation. 
The transition to 
momentum fraction $x$--space by an analytic continuation is described in detail in Ref.~\cite{BS22}. This method avoids
going to Mellin $N$--space. The result can be Mellin transformed analytically at the end of the calculation, 
cf.~Section~\ref{sec:4}. Working in Mellin $N$--space implies 
that one decides for expressions at either even or odd integer values of $N$ starting with a value $N_0$ implied by 
the crossing relations \cite{Politzer:1974fr,Blumlein:1996vs}. If one would like to extract the small $x$ behavior 
of the OME, one has to do this in $x$--space, since the corresponding poles are situated at $N=1$ in the unpolarized 
case and at $N=0$ in the polarized case, for which no $N$--space representation is available.

In Mellin $N$--space, the present OMEs exhibit nested finite binomial sums \cite{Ablinger:2014bra}. 
Transforming to $x$--space, 
these structures lead to $\GA$--functions \cite{Ablinger:2014bra} containing also 
square root--valued 
letters. Along with these very many new constants $\GA(\{a_i\},1)$ occur in intermediary steps. They can be 
rationalized by procedures contained in the package {\tt 
HarmonicSums}, leading 
to tabulated cyclotomic constants \cite{Ablinger:2011te}. The latter turn out to 
further reduce to multiple zeta values \cite{Blumlein:2009cf} in all cases. We show a series of examples in
Appendix~\ref{sec:B}. Also the iterated integrals $\GA(\{a_i\},x)$ containing root--valued letters can be 
rationalized, leading to cyclotomic harmonic polylogarithms \cite{Ablinger:2011te}. The latter 
representation can in principle be used in the numerical representation, cf.~Section~\ref{sec:5}.

Let us now discuss the new structures appearing in the present OMEs, which are nested finite binomial sums in Mellin 
$N$--space.
Their Mellin inversion can be expressed by iterative $\GA$--functions and 
usual harmonic polylogarithms over the alphabet $\mathfrak{A}$
\begin{eqnarray}
\mathfrak{A} = \left. \left\{f_{k}(x)\right\}\right|_{k=1..6} =
\Biggl\{\frac{1}{x}, \frac{1}{1-x}, \frac{1}{1+x}, 
\frac{\sqrt{1-x}}{x}, \sqrt{x(1-x)}, \frac{1}{\sqrt{1-x}}\Biggr\}.
\label{eq:alph}
\end{eqnarray}
The $\GA$--functions are defined by \cite{Ablinger:2014bra}
\begin{eqnarray}
\GA(\{b, \vec{a}\},x) = \int_0^x dy f_b(y) \GA(\{\vec{a}\},y),~~~\GA(\{\emptyset\},y) = 1,~~~ f_b(x), 
f_{a_i}(x) \in 
\mathfrak{A}.
\end{eqnarray}
In $N$--space we consider the objects
\begin{eqnarray} 
{\sf BS}_0(N) &=& \frac{1}{2N - (2l+1)},~~~l \in \mathbb{N},
\\
{\sf BS}_1(N) &=& 4^N \frac{(N!)^2}{(2N)!},
\\
{\sf BS}_2(N) &=& \frac{1}{4^N} \frac{(2N)!}{(N!)^2},
\\
{\sf BS}_3(N) &=& 
\sum_{\tau_1=1}^N \frac{\ds 4^{-\tau_1} \big(
        2 \tau_1\big)!}{\big(
        \tau_1!\big)^2 \tau_1},
\\
{\sf BS}_4(N) &=& 
\sum_{\tau_1=1}^N \frac{\ds 4^{\tau_1} \big(
        \tau_1!\big)^2}{\big(
        2 \tau_1\big)! \tau_1^2},
\\
{\sf BS}_5(N) &=& 
\sum_{\tau_1=1}^N \frac{4^{\tau_1} \big(
        \tau_1!\big)^2}{\big(
        2 \tau_1\big)! \tau_1^3},
\\
{\sf BS}_6(N) &=& 
\sum_{\tau_1=1}^N \frac{\ds 4^{-\tau_1} \big(
        2 \tau_1\big)! 
\sum_{\tau_2=1}^{\tau_1} \frac{\ds 4^{\tau_2} \big(
        \tau_2!\big)^2}{\big(
        2 \tau_2\big)! \tau_2^2}}{\big(
        \tau_1!\big)^2 \tau_1},
\\
{\sf BS}_7(N) &=& 
\sum_{\tau_1=1}^N \frac{\ds 4^{-\tau_1} \big(
        2 \tau_1\big)! 
\sum_{\tau_2=1}^{\tau_1} \frac{\ds 4^{\tau_2} \big(
        \tau_2!\big)^2}{\big(
        2 \tau_2\big)! \tau_2^3}}{\big(
        \tau_1!\big)^2 \tau_1},
\\
{\sf BS}_8(N) &=& 
\sum_{\tau_1=1}^N \frac{\ds
\sum_{\tau_2=1}^{\tau_1} \frac{\ds 4^{\tau_2} \big(
        \tau_2!\big)^2}{\big(
        2 \tau_2\big)! \tau_2^2}}{\tau_1},
\\
{\sf BS}_9(N) &=& 
\sum_{\tau_1=1}^N \frac{\ds 4^{-\tau_1} \big(
        2 \tau_1\big)! 
\sum_{\tau_2=1}^{\tau_1} \frac{\ds 4^{\tau_2} \big(
        \tau_2!\big)^2 
\sum_{\tau_3=1}^{\tau_2} \ds \frac{1}{\tau_3}}{\big(
        2 \tau_2\big)! \tau_2^2}}{\big(
        \tau_1!\big)^2 \tau_1},
\\
{\sf BS}_{10}(N) &=& \sum_{\tau_1=1}^N \frac{4^{\tau_1}}{\ds \binom{2 \tau_1}{\tau_1}} 
\frac{1}{\tau_1^2} S_1(\tau_1),
\end{eqnarray}
where $S_{\vec{a}} \equiv S_{\vec{a}}(N)$ denote the nested harmonic 
sums \cite{Vermaseren:1998uu,Blumlein:1998if}.

The above finite binomial sums obey the following recursion relations
\begin{eqnarray}
{\sf BS}_3(N) - {\sf BS}_3(N-1)  &=& \frac{1}{N} {\sf BS}_2(N),
\\
{\sf BS}_4(N) - {\sf BS}_4(N-1)  &=& \frac{1}{N^2} {\sf BS}_1(N),
\\
{\sf BS}_5(N) - {\sf BS}_5(N-1)  &=& \frac{1}{N^3} {\sf BS}_1(N),
\\
{\sf BS}_6(N) - {\sf BS}_6(N-1)  &=& \frac{1}{N} {\sf BS}_2(N) {\sf BS}_4(N),
\\
{\sf BS}_7(N) - {\sf BS}_7(N-1)  &=& \frac{1}{N} {\sf BS}_2(N) {\sf BS}_5(N),
\\
{\sf BS}_8(N) - {\sf BS}_8(N-1)  &=& \frac{1}{N} {\sf BS}_4(N),
\\
{\sf BS}_9(N) - {\sf BS}_9(N-1)  &=& \frac{1}{N} {\sf BS}_2(N) {\sf BS}_{10}(N),
\\
{\sf BS}_{10}(N) - {\sf BS}_{10}(N-1)  &=& \frac{1}{N^2} {\sf BS}_1(N) S_1.
\end{eqnarray}
In Appendix~\ref{sec:D} we will also calculate representations of $a_{gg,Q}^{(3)}(N)$ and $\Delta 
a_{gg,Q}^{(3)}(N)$ in the asymptotic
region $|N| \gg 1$  up to $O(1/N^{10})$. In the analyticity region this and 
the recurrences allow one to compute $(\Delta)a_{gg,Q}^{(3)}(N)$
for $N \in \mathbb{C}$, see Ref.~\cite{Blumlein:2009ta}.
The asymptotic expansion of $(\Delta) a_{gg,Q}^{(3)}$ can be obtained from the asymptotic expansions of its building 
blocks. The asymptotic representations of the finite binomial sums ${\sf BS}_k(N),~k = 
0 \dots 10$ are also given in Appendix~\ref{sec:D}.
The constants contributing to the above sums can be calculated using infinite binomial and inverse binomial sums
\cite{Davydychev:2003mv,Weinzierl:2004bn,Ablinger:2014bra}.
In the calculation, first partly different binomial sums appear after the sum reduction performed by {\tt 
Sigma}, which have the following relations
\begin{eqnarray}
{\sf \overline{BS}}_4 &=& \sum_{\tau_1=1}^N \frac{\ds 4^{-\tau_1} \big(
        2 \tau_1\big)!}{\big(
        \tau_1!\big)^2 \big(
        1+\tau_1\big)^2}
=
3
-\frac{(1+2 N) (3+2 N)}{(1+N)^2}
\frac{1}{4^{N}} \binom{2 N}{N}
-2 {\sf BS}_3,
\\
{\sf \overline{BS}}_6 &=& 
\sum_{\tau_1=1}^N \frac{\ds 4^{\tau_1} \big(
        \tau_1!\big)^2 
\sum_{\tau_2=1}^{\tau_1} \frac{\ds 4^{-\tau_2} \big(
        2 \tau_2\big)!}{\big(
        \tau_2!\big)^2 \big(
        1+\tau_2\big)^2}}{\big(
        2 \tau_1\big)! \tau_1^2}
= -2 S_2 - 2 S_3 
-\frac{N (4+3 N)}{(1+N)^2}
\nonumber\\ &&
+3 {\sf BS}_4 - 2 {\sf BS}_3 {\sf BS}_4 + 2 {\sf BS}_6,
\\
{\sf \overline{BS}}_9 &=& \sum_{\tau_1=1}^N \frac{\ds
\sum_{\tau_2=1}^{\tau_1} \frac{\ds 4^{\tau_2} \big(
        \tau_2!\big)^2 
\sum_{\tau_3=1}^{\tau_2} \frac{\ds 4^{-\tau_3} \big(
        2 \tau_3\big)!}{\big(
        \tau_3!\big)^2 \big(
        1+\tau_3\big)^2}}{\big(
        2 \tau_2\big)! \tau_2^2}}{\tau_1} =
 2 {\sf BS}_7
+3 {\sf BS}_8
-2 {\sf BS}_3 {\sf BS}_8
-2 {\sf BS}_9
\nonumber\\ && 
+\frac{N (5+4 N)}{(1+N)^2}
+2 {\sf BS}_6 S_1
-\big(
        3
        +2 S_2
        +2 S_3
\big) S_1
-S_2
-2 S_3
-2 S_4
+2 S_{2,1}
\nonumber\\ &&
+2 S_{3,1},
\\
{\sf \overline{BS}}_{11} &=& 
\sum_{i_1=3}^N \frac{2^{2 i_1} \big(
        \big(
                -2+i_1\big)!\big)^2}{\big(
        2 i_1\big)!} = -\frac{44}{3}
+3 {\sf BS}_4
+\frac{(-1+3 N)}{N^2} 2^{1+2 N} \binom{2 N}{N}^{-1},
\\
{\sf \overline{BS}}_{12} &=& 
\sum_{i_1=3}^N \frac{2^{2 i_1} \big(
        \big(
                -2+i_1\big)!\big)^2}{\big(
        2 i_1\big)! i_1} = -\frac{67}{3}
+ 4 {\sf BS}_4
+  {\sf BS}_5
+\frac{(-1+4 N)}{N^2} 2^{1+2 N} \binom{2 N}{N}^{-1},
\\
{\sf \overline{BS}}_{13} &=& 
\sum_{i_1=3}^N \frac{\ds 2^{2 i_1} \big(
        \big(
                -2+i_1\big)!\big)^2 S_1\big({i_1}\big)}{\big(
        2 i_1\big)!}
=
-23
+3 {\sf BS}_5
-3 {\sf BS}_8
+ {\sf BS}_4 \big(
        -2
        +3 S_1
\big)
\nonumber\\ &&
+\frac{2(-1+4 N)}{N^2} 4^N \binom{2 N}{N}^{-1}
+\frac{2 (-1+3 N)}{N^2} 4^N  \binom{2 N}{N}^{-1} S_1.
\end{eqnarray}

The above functions emerge in products in part and one has to calculate the corresponding 
Mellin convolutions, also with harmonic sums. Besides the above binomial sums, the following  34 
harmonic sums up to weight {\sf w = 5} contribute
\begin{eqnarray}
&&
\{
S_1,
S_{-1},
S_2,
S_{-2},
S_3,
S_{-3},
S_4,
S_{-4},
S_{-5},
S_5,
S_{2,1},
S_{-2,1},
S_{2,-1},
S_{-2,-1},
S_{-2,2},
S_{3,1},
S_{-3,1},
S_{4,1},
\nonumber\\ &&
S_{-4,1},
S_{2,3},
S_{2,-3},
S_{-2,3},
S_{-2,-3},
S_{3,1,1},
S_{-3,1,1},
S_{2,1,1},
S_{-2,1,1},
S_{2,2,1},
S_{2,1,-2},
S_{-2,1,-2},
S_{-2,2,1},
\nonumber\\ &&
S_{-2,1,1,1},
S_{2,1,1,1}\big\}
\end{eqnarray}
after algebraic reduction \cite{Blumlein:2003gb}. Here we suppressed
the argument $N$ of the harmonic sums for brevity.

All logarithmic terms to the OMEs and the contributions to the constant term implied by renormalization 
were given in Refs.~\cite{Behring:2014eya,Blumlein:2021xlc} before. 
\section{\boldmath Results for $a_{gg,Q}^{(3)}(N)$ and $\Delta 
a_{gg,Q}^{(3)}(N)$}
\label{sec:3}

\vspace{1mm}
\noindent
We now present the constant contributions of the unrenormalized 
massive OMEs $(\Delta) 
\Ahathat_{gg,Q}^{(3)}$, denoted by 
$a_{gg,Q}^{(3)}$ and $\Delta a_{gg,Q}^{(3)}$. The expression for 
$a_{gg,Q}^{(3)}(N)$, valid for even moments $N \in 
\mathbb{N}, N \geq 2$, reads 
\begin{eqnarray}
\lefteqn{
a_{gg,Q}^{(3)}(N) = \frac{1}{2}\left(1 + (-1)^N\right)} \nonumber\\ &&
\times \Biggl\{
    \textcolor{blue}{C_A} \Biggl[
        \textcolor{blue}{C_F} \textcolor{blue}{T_F} \Biggl(
            \frac{32 S_{-2,2} P_3}{(N-1) N^2 (N+1)^2 (N+2)}
            -\frac{64 S_{-2,1,1} P_{21}}{3 (N-1) N^2 (N+1)^2 (N+2)}
            \nonumber\\
            &&-\frac{16 S_{-4} P_{35}}{3 (N-1) N^2 (N+1)^2 (N+2)}
            +\frac{4 S_4 P_{73}}{3 (N-2) (N-1) N^2 (N+1)^2 (N+2)}
            \nonumber\\
            &&
            +\frac{32 [ {\rm {\sf BS}}_7 - {\rm {\sf BS}}_9 + 7 \zeta_3 {\rm {\sf BS}}_3 ] P_{75}}{3 (N-1) N^2 (N+1)^2 (N+2)}
            -\frac{4 \big(2+N+N^2\big) S_1^3 P_{98}}{27 (N-1)^2 N^3 (N+1)^3 (N+2)^2}
            \nonumber\\
            &&+\frac{16 S_{3,1} P_{116}}{3 (N-2) (N-1) N^2 (N+1)^2 (N+2)}
            -\frac{16 S_{2,1,1} P_{122}}{3 (N-2) (N-1) N^2 (N+1)^2 (N+2)}
            \nonumber\\
            &&
            +\frac{32 [ S_{-1} S_2 - S_{2,-1} + S_{-2,-1} ] P_{135}}{5 (N-3) (N-2) (N-1)^2 N^2 (N+1)^3 (N+2)}
            \nonumber\\
            &&
            -\frac{16 S_{-2,1} P_{157}}{3 (N-3) (N-2) (N-1)^2 N^3 (N+1)^3 (N+2)^2}
            \nonumber\\
            &&
            +\frac{\ds [ {\rm {\sf BS}}_8 - {\rm {\sf BS}}_4 S_1 + 7 \zeta_3] 2^{2-2 N} \binom{2 N}{N} 
P_{185}}{15 (N-3) (N-2) (N-1)^2 N^3 (N+1)^3 (N+2)^2}
            \nonumber\\
            &&
            +\frac{8 S_3 P_{189}}{135 (N-3) (N-2) (N-1)^2 N^3 (N+1)^3 (N+2)^2}
            \nonumber\\
            &&-
            \frac{8 S_{2,1} P_{197}}{15 (N-3) (N-2) (N-1)^2 N^3 (N+1)^3 (N+2)^2}
            \nonumber\\
            &&+\frac{4 S_2 P_{207}}{45 (N-3) (N-2)^2 (N-1)^2 N^4 (N+1)^4 (N+2)^3}
            \nonumber\\
            &&+\frac{P_{216}}{2430 (N-3) (N-2)^2 (N-1)^2 N^6 (N+1)^6 (N+2)^5}
            + {\sf B}_4 \biggl(
                \frac{32 \big(4+N+N^2\big)\big(6+N+N^2\big)}{3 (N-1) N^2 (N+1)^2 (N+2)}
                \nonumber\\
                &&-\frac{64}{3} S_1
            \biggr)
            + \zeta_4 \biggl(
                -\frac{48 \big(4+N+N^2\big)\big(6+N+N^2\big)}{(N-1) N^2 (N+1)^2 (N+2)}
                +96 S_1
            \biggr)
            +\biggl(
                \frac{64 S_{-2,1} P_{22}}{3 (N-1) N^2 (N+1)^2 (N+2)}
                \nonumber\\
                &&+\frac{32 S_3 P_{69}}{9 (N-2) (N-1) N^2 (N+1)^2 (N+2)}
                +\frac{32 S_{2,1} P_{77}}{3 (N-1) N^2 (N+1)^2 (N+2)}
                \nonumber\\
                &&+\frac{4 S_2 P_{159}}{3 (N-2) (N-1)^2 N^3 (N+1)^3 (N+2)^2}
                \nonumber\\
                &&+\frac{8 P_{213}}{405 (N-3) (N-2) (N-1)^2 N^5 (N+1)^5 (N+2)^4}
            \biggr) S_1
            +\biggl(
                -\frac{4 S_2 P_{41}}{3 (N-1) N^2 (N+1)^2 (N+2)}
                \nonumber\\
                &&+\frac{4 P_{178}
                }{27 (N-2) (N-1)^2 N^3 (N+1)^4 (N+2)^3}
            \biggr) S_1^2
            -\frac{2 \big(2+N+N^2\big)^2 S_1^4}{9 (N-1) N^2 (N+1)^2 (N+2)}
            \nonumber\\
            &&+\frac{2 \big(
                2+N+N^2
            \big)
    \big(-6+5 N+5 N^2\big) S_2^2}{(N-1) N^2 (N+1)^2 (N+2)}
            +\biggl(
                -\frac{32 S_1^2 P_{23}}{3 (N-1) N^2 (N+1)^2 (N+2)}
                \nonumber\\
                &&-\frac{32 S_2 P_{50}}{3 (N-2) (N-1) N^2 (N+1)^2 (N+2)}
                -\frac{32 S_{-1} P_{135}}{5 (N-3) (N-2) (N-1)^2 N^2 (N+1)^3 (N+2)}
                \nonumber\\
                &&-\frac{16 S_1 P_{158}}{3 (N-3) (N-2) (N-1)^2 N^3 (N+1)^3 (N+2)^2}
                \nonumber\\
                &&-\frac{16 P_{202}}{15 (N-3) (N-2)^2 (N-1)^2 N^4 (N+1)^4 (N+2)^3}
            \biggr) S_{-2}
            \nonumber\\
            &&-\frac{32 \big(
                2+N+N^2
            \big)
    \big(13+2 N+2 N^2\big) S_{-2}^2}{3 (N-1) N^2 (N+1)^2 (N+2)}
            +\biggl(
                -\frac{32 S_1 P_2}{3 (N-1) N^2 (N+1)^2 (N+2)}
                \nonumber\\
                &&+\frac{8 P_{162}}{3 (N-3) (N-2) (N-1)^2 N^3 (N+1)^3 (N+2)^2}
            \biggr) S_{-3}
            \nonumber\\
                &&
            +\frac{32 \big(-1-8 N+N^2\big)\big(8+10 N+N^2\big) S_{-3,1}}{3 (N-1) N^2 (N+1)^2 (N+2)}
           +\biggl(
                \frac{4 S_1 P_{167}}{3 (N-1)^2 N^3 (N+1)^3 (N+2)^2}
                \nonumber\\
                &&
                +\frac{4 P_{196}}{9 (N-1)^2 N^4 (N+1)^4 (N+2)^3}
                -\frac{4 \big(
                    2+N+N^2\big)^2 S_1^2}{(N-1) N^2 (N+1)^2 (N+2)}
                    \nonumber\\
                    &&-
                \frac{12 \big(
                    2+N+N^2\big)^2 S_2}{(N-1) N^2 (N+1)^2 (N+2)}
                -\frac{24 \big(
                    2+N+N^2\big)^2 S_{-2}}{(N-1) N^2 (N+1)^2 (N+2)}
            \biggr) \zeta_2
            \nonumber\\
                &&
            +\biggl(
                -\frac{8 S_1 P_{128}}{9 (N-2) (N-1) N^2 (N+1)^2 (N+2)}
                \nonumber\\
                &&+\frac{P_{200}}{180 (N-3) (N-2) (N-1)^2 N^3 (N+1)^3 (N+2)^2}
            \biggr) \zeta_3
        \Biggr)
        \nonumber\\
                &&
        +\textcolor{blue}{T_F^2} \Biggl(
            -\frac{4 S_1^2 P_{72}}{135 (N-1) N^2 (N+1)^2 (N+2)}
            +\frac{4 S_2 P_{111}}{135 (N-1) N^2 (N+1)^2 (N+2)}
            \nonumber\\
            &&
            +\frac{\ds [ {\rm {\sf BS}}_8 - {\rm {\sf BS}}_4 S_1 + 7 \zeta_3 ] 2^{2-2 N} \binom{2 N}{N} 
P_{140}}{45 (N-1) N (N+1)^2 (N+2) (2 N -3) (2 N -1)}
            \nonumber\\
            &&+\frac{P_{195}}{3645 (N-1) N^4 (N+1)^4 (N+2) (2 N -3) (2 N -1)}
            +\textcolor{blue}{N_F} \biggl[
                -\frac{4 S_1^2 P_{70}}{27 (N-1) N^2 (N+1)^2 (N+2)}
                \nonumber\\
                &&+\frac{4 S_2 P_{106}}{27 (N-1) N^2 (N+1)^2 (N+2)}
                -\frac{8 S_1 P_{144}}{729 (N-1) N^3 (N+1)^3 (N+2)}
                \nonumber\\
                &&-\frac{2 P_{172}
                }{729 (N-1) N^4 (N+1)^4 (N+2)}
                +\biggl(
                    \frac{4 P_{93}}{27 (N-1) N^2 (N+1)^2 (N+2)}
                    -\frac{160}{27} S_1
                \biggr) \zeta_2
                    \nonumber\\
                    &&                +\biggl(
-\frac{896 \big(
                        1+N+N^2\big)}{27 (N-1) N (N+1) (N+2)}
                    +\frac{448}{27} S_1
                \biggr) \zeta_3
            \biggr]
            -\frac{16 \big(1-7 N+4 N^2+4 N^3\big) [ S_3 - S_{2,1} ]}{15 (N-1) N (N+1)}
            \nonumber\\
            &&
            -\frac{8 P_{174}}{3645 (N-1) N^3 (N+1)^3 (N+2) (2 N -3) (2 N -1)} S_1
            +\biggl(
                \frac{4 P_{108}}{27 (N-1) N^2 (N+1)^2 (N+2)}
                \nonumber\\
            &&
                -\frac{560}{27} S_1
            \biggr) \zeta_2
            +\biggl(
                -\frac{7 P_{49}}{270 (N-1) N (N+1) (N+2)}
                -\frac{1120}{27} S_1
            \biggr) \zeta_3
        \Biggr)
    \Biggr]
    \nonumber\\
    &&
    +\textcolor{blue}{C_F} \textcolor{blue}{T_F^2} \Biggl[
        \frac{16 S_1^2 P_{82}}{27 (N-1) N^3 (N+1)^3 (N+2)}
        -\frac{16 S_2 P_{82}}{9 (N-1) N^3 (N+1)^3 (N+2)}
        \nonumber\\
        &&
        +\frac{\ds [ {\rm {\sf BS}}_8 - {\rm {\sf BS}}_4 S_1 + 7 \zeta_3 ] 2^{4-2 N} \binom{2 N}{N} 
P_{92}}{3 (N-1) N (N+1)^2 (N+2) (2 N -3) (2 N -1)}
        \nonumber\\
        &&-\frac{2 P_{203}}{243 (N-1) N^5 (N+1)^5 (N+2) (2 N -3) (2 N -1)}
        +\textcolor{blue}{N_F} \biggl[
            -\frac{16 S_2 P_{82}}{9 (N-1) N^3 (N+1)^3 (N+2)}
            \nonumber\\
            &&+\frac{16 S_1^2 P_{107}}{27 (N-1) N^3 (N+1)^3 (N+2)}
            -\frac{2 P_{192}}{243 (N-1) N^5 (N+1)^5 (N+2)}
            \nonumber\\
                &&
            -\biggl(
                 \frac{32 P_{141}}{81 (N-1) N^4 (N+1)^4 (N+2)}
                +\frac{16 \big(2+N+N^2\big)^2 S_2}{3 (N-1) N^2 (N+1)^2 (N+2)}
            \biggr) S_1
            \nonumber\\
            &&
            -\frac{112 \big(2+N+N^2\big)^2 S_1^3}{27 (N-1) N^2 (N+1)^2 (N+2)}
            +\frac{160 \big(2+N+N^2\big)^2 S_3}{27 (N-1) N^2 (N+1)^2 (N+2)}
            \nonumber\\
                &&
            -\biggl(
                 \frac{4 P_{136}}{9 (N-1) N^3 (N+1)^3 (N+2)}
                +\frac{16 \big(2+N+N^2\big)^2 S_1}{3 (N-1) N^2 (N+1)^2 (N+2)}
            \biggr) \zeta_2
            \nonumber\\
                &&
            -\frac{448 \big(2+N+N^2\big)^2 \zeta_3}{9 (N-1) N^2 (N+1)^2 (N+2)}
        \biggr]
        -\biggl(
             \frac{32 P_{165}}{81 (N-1) N^4 (N+1)^4 (N+2) (2 N -3) (2 N -1)}
             \nonumber\\
                &&
            +\frac{16 \big(2+N+N^2\big)^2 S_2}{3 (N-1) N^2 (N+1)^2 (N+2)}
        \biggr) S_1
        +\frac{16 \big(2+N+N^2\big)^2 S_1^3}{27 (N-1) N^2 (N+1)^2 (N+2)}
        \nonumber\\
                &&
        -\frac{352 \big(2+N+N^2\big)^2 S_3}{27 (N-1) N^2 (N+1)^2 (N+2)}
        +\frac{64 \big(2+N+N^2\big)^2 S_{2,1}}{3 (N-1) N^2 (N+1)^2 (N+2)}
        \nonumber\\
        &&+\biggl(
            -\frac{8 P_{139}}{9 (N-1) N^3 (N+1)^3 (N+2)}
            +\frac{16 \big(
                2+N+N^2\big)^2 S_1}{3 (N-1) N^2 (N+1)^2 (N+2)}
        \biggr) \zeta_2
        \nonumber\\
                &&
        +\frac{P_{74}}{9 (N-1) N^2 (N+1)^2 (N+2)} \zeta_3
    \Biggr]
    +\textcolor{blue}{C_F^2} \textcolor{blue}{T_F} \Biggl[
        \frac{32 [ {\sf B}_4 + {\rm {\sf BS}}_7 - {\rm {\sf BS}}_9 + 7 \zeta_3 {\rm {\sf BS}}_3  ] 
\big(2+N+N^2\big)}{N^2 (N+1)^2}
        \nonumber\\
        &&-\frac{144 \big(2+N+N^2\big) \zeta_4}{N^2 (N+1)^2}
        +\frac{4 \big(2+N+N^2\big) S_1^3 P_{19}}{9 (N-1) N^3 (N+1)^3 (N+2)}
        \nonumber\\
        &&-\frac{64 S_{-1} S_2 P_{125}}{15 (N-3) (N-2) (N-1)^2 N^2 (N+1)^3}
        +\frac{64 S_{2,-1} P_{125}}{15 (N-3) (N-2) (N-1)^2 N^2 (N+1)^3}
        \nonumber\\
        &&-\frac{64 S_{-2,-1} P_{125}}{15 (N-3) (N-2) (N-1)^2 N^2 (N+1)^3}
        +\frac{32 S_{-2,1}P_{149}}{3 (N-3) (N-2) (N-1)^2 N^3 (N+1)^3 (N+2)}
        \nonumber\\
        &&+
        \frac{16 S_3 P_{152}}{45 (N-3) (N-2) (N-1)^2 N^3 (N+1)^3 (N+2)}
        -\frac{16 S_{2,1} P_{153}}{15 (N-3) (N-2) (N-1)^2 N^3 (N+1)^3}
        \nonumber\\
        &&
        -\frac{\ds [ {\rm {\sf BS}}_8 - {\rm {\sf BS}}_4 S_1 + 7 \zeta_3 ] 2^{5-2 N} \binom{2 N}{N} 
P_{166}}{15 (N-3) (N-2) (N-1)^2 N^3 (N+1)^3 (N+2)}
        \nonumber\\
        &&+\frac{4 S_2 P_{177}}{15 (N-3) (N-2) (N-1)^2 N^4 (N+1)^4 (N+2)}
        \nonumber\\
        &&
        +\frac{2 P_{208}}{45 (N-3) (N-2) (N-1) N^6 (N+1)^6 (N+2)}
        +\biggl(
            \frac{4 S_2 P_{91}}{3 (N-1) N^3 (N+1)^3 (N+2)}
            \nonumber\\
            &&+\frac{8 P_{186}}{15 (N-3) (N-2) (N-1) N^5 (N+1)^5 (N+2)}
            +\frac{16 \big(2+N+N^2\big)\big(-46+13 N+13 N^2\big) S_3}{9 (N-1) N^2 (N+1)^2 (N+2)}
            \nonumber\\
            &&+\frac{32 \big(2+N+N^2\big)\big(-2+3 N+3 N^2\big) S_{2,1}}{3 (N-1) N^2 (N+1)^2 (N+2)}
            +\frac{256 \big(2+N+N^2\big) S_{-2,1}}{(N-1) N^2 (N+1)^2 (N+2)}
        \biggr) S_1
        \nonumber\\
            &&
        +\biggl(
            -\frac{4 P_{133}}{3 (N-1) N^4 (N+1)^4 (N+2)}
            +\frac{4 \big(2+N+N^2\big)\big(-22+5 N+5 N^2\big) S_2}{3 (N-1) N^2 (N+1)^2 (N+2)}
        \biggr) S_1^2
        \nonumber\\
        &&+
        \frac{2 \big(2+N+N^2\big)^2 S_1^4}{9 (N-1) N^2 (N+1)^2 (N+2)}
        -\frac{2 \big(2+N+N^2\big)\big(30+31 N+31 N^2\big) S_2^2}{3 (N-1) N^2 (N+1)^2 (N+2)}
        \nonumber\\
        &&-\frac{4 \big(2+N+N^2\big)\big(54+11 N+11 N^2\big) S_4}{3 (N-1) N^2 (N+1)^2 (N+2)}
        +\biggl(
            \frac{64 S_{-1} P_{125}}{15 (N-3) (N-2) (N-1)^2 N^2 (N+1)^3}
            \nonumber\\
            &&-\frac{32 S_1 P_{149}}{3 (N-3) (N-2) (N-1)^2 N^3 (N+1)^3 (N+2)}
            -\frac{128 \big(2+N+N^2\big) S_1^2}{(N-1) N^2 (N+1)^2 (N+2)}
            \nonumber\\
            &&
            +\frac{32 P_{161}}{15 (N-3) (N-2) (N-1) N^4 (N+1)^4 (N+2)}
            -\frac{256 \big(2+N+N^2\big) S_2}{3 (N-1) N^2 (N+1)^2 (N+2)}
        \biggr) S_{-2}
        \nonumber\\
        &&
        +\frac{64 \big(2+N+N^2\big) S_{-2}^2}{(N-1) N^2 (N+1)^2 (N+2)}
        +\biggl(
            -\frac{16 P_{148}}{3 (N-3) (N-2) (N-1)^2 N^3 (N+1)^3 (N+2)}
            \nonumber\\
            &&
            -\frac{128 \big(2+N+N^2\big) S_1}{3 (N-1) N^2 (N+1)^2 (N+2)}
        \biggr) S_{-3}
        +\frac{128 \big(2+N+N^2\big) S_{-4}}{3 (N-1) N^2 (N+1)^2 (N+2)}
        \nonumber\\
        &&
        +\frac{128 \big(-1+N+N^2\big)\big(2+N+N^2\big) S_{3,1}}{3 (N-1) N^2 (N+1)^2 (N+2)}
        +\frac{256 \big(2+N+N^2\big) S_{-2,2}}{3 (N-1) N^2 (N+1)^2 (N+2)}
        \nonumber\\
        &&+
        \frac{128 \big(2+N+N^2\big) S_{-3,1}}{3 (N-1) N^2 (N+1)^2 (N+2)}
        -\frac{32 \big(2+N+N^2\big)\big(-10+7 N+7 N^2\big) S_{2,1,1}}{3 (N-1) N^2 (N+1)^2 (N+2)}
        \nonumber\\
        &&-\frac{256 \big(2+N+N^2\big) S_{-2,1,1}}{(N-1) N^2 (N+1)^2 (N+2)}
        +\biggl(
            \frac{4 \big(2+N+N^2\big) S_1 P_{20}}{(N-1) N^3 (N+1)^3 (N+2)}
            \nonumber\\
            &&
            -\frac{2 \big(2+N+N^2\big) P_{105}}{(N-1) N^4 (N+1)^4 (N+2)}
            +\frac{4 \big(2+N+N^2\big)^2 S_1^2}{(N-1) N^2 (N+1)^2 (N+2)}
            \nonumber\\
            &&
            -\frac{12 \big(2+N+N^2\big)^2 S_2}{(N-1) N^2 (N+1)^2 (N+2)}
        \biggr) \zeta_2
        +\biggl(
            \frac{P_{180}}{15 (N-3) (N-2) (N-1)^2 N^3 (N+1)^3 (N+2)}
            \nonumber\\
            &&
            -\frac{16 \big(-118+N+N^2\big)\big(2+N+N^2\big) S_1}{3 (N-1) N^2 (N+1)^2 (N+2)}
        \biggr) \zeta_3
    \Biggr]
    +\textcolor{blue}{C_A^2} \textcolor{blue}{T_F} \Biggl[
        \frac{256 S_{-2,2} P_7}{9 (N-1) N^2 (N+1)^2 (N+2)}
        \nonumber\\
        &&
        -\frac{16 S_{-2,3} P_{11}}{3 N (N+1)}
        -\frac{16 S_{-4,1} P_{12}}{3 N (N+1)}
        +\frac{32 S_{-5} P_{16}}{9 N (N+1)}
        -\frac{32 S_{2,1,-2} P_{28}}{9 N (N+1)}
        +\frac{64 S_{-2,-3} P_{29}}{9 N (N+1)}
        -\frac{32 S_{-2,1,-2} P_{30}}{9 N (N+1)}
        \nonumber\\
        &&+\frac{16 S_{2,-3} P_{31}}{9 N (N+1)}
        -\frac{16 S_5 P_{38}}{9 N (N+1)}
        +\frac{16 S_{2,3} P_{39}}{9 N (N+1)}
        +\frac{32 S_{-2,1,1} P_{52}}{9 (N-1) N^2 (N+1)^2 (N+2)}
        \nonumber\\
        &&+\frac{16 S_{-3,1} P_{64}}{9 (N-1) N^2 (N+1)^2 (N+2)}
        +\frac{8 S_2^2 P_{67}}{9 (N-1) N^2 (N+1)^2}
        -\frac{8 [ {\rm {\sf BS}}_7 - {\rm {\sf BS}}_9 + 7 \zeta_3 {\rm {\sf BS}}_3 ] P_{84}}{3 (N-1) N^2 (N+1)^2 (N+2)}
        \nonumber\\
        &&
        -\frac{16 S_{2,1,1} P_{117}}{3 (N-2) (N-1) N^2 (N+1)^2 (N+2)}
        +\frac{4 S_4 P_{120}}{9 (N-2) (N-1) N^2 (N+1)^2 (N+2)}
        \nonumber\\
        &&
        -\frac{8 S_{3,1} P_{123}}{9 (N-2) (N-1) N^2 (N+1)^2 (N+2)}
        -\frac{16 [ S_{-1} S_2 - S_{2,-1} + S_{-2,-1} ] P_{137}}{15 (N-3) (N-2) (N-1)^2 N^2 (N+1)^3 (N+2)}
        \nonumber\\
        &&
        -\frac{16 S_{-2,1} P_{181}}{81 (N-3) (N-2) (N-1)^2 N^3 (N+1)^3 (N+2)}
        \nonumber\\
        &&
        -\frac{\ds [ {\rm {\sf BS}}_8 - {\rm {\sf BS}}_4 S_1 + 7 \zeta_3 ] 2^{1-2 N} \binom{2 N}{N} 
P_{184}}{15 (N-3) (N-2) (N-1)^2 N^3 (N+1)^3 (N+2)^2}
        \nonumber\\
        &&-\frac{4 S_3 P_{198}}{405 (N-3) (N-2) (N-1)^2 N^3 (N+1)^3 (N+2)^2}
        \nonumber\\
        &&-\frac{2 S_{2,1} P_{199}}{45 (N-3) (N-2) (N-1)^2 N^3 (N+1)^3 (N+2)^2}
        \nonumber\\
        &&+\frac{P_{215}}{14580 (N-3) (N-2)^2 (N-1)^2 N^5 (N+1)^5 (N+2)^5}
        + \zeta_4 \biggl(
            \frac{96 P_{13}}{(N-1) N^2 (N+1)^2 (N+2)}
            \nonumber\\
            &&-96 S_1
        \biggr)
        + {\rm B}_4 \biggl(
            -\frac{8 P_{25}}{3 (N-1) N^2 (N+1)^2 (N+2)}
            +\frac{32}{3} S_1
        \biggr)
        +\biggl(
            -\frac{32 S_{-2,1} P_{56}}{9 (N-1) N^2 (N+1)^2 (N+2)}
            \nonumber\\
            &&+\frac{8 S_{2,1} P_{96}}{9 (N-1) N^2 (N+1)^2 (N+2)}
            -\frac{16 S_3 P_{126}}{27 (N-2) (N-1) N^2 (N+1)^2 (N+2)}
            \nonumber\\
            &&+\frac{4 S_2 P_{183}}{81 (N-2) (N-1)^2 N^3 (N+1)^3 (N+2)^2}
            +\frac{16}{9} S_2^2
            +\frac{592}{9} S_4
            -\frac{832}{9} S_{3,1}
            -\frac{128}{9} S_{-2,2}
            \nonumber\\
            &&
            +\frac{2 P_{214}}{3645 (N-3) (N-2) (N-1)^2 N^5 (N+1)^5 (N+2)^4}
            -\frac{128}{9} S_{-3,1}
            +\frac{32}{3} S_{2,1,1}
            +\frac{256}{9} S_{-2,1,1}
        \biggr) S_1
        \nonumber\\
            &&
        +\biggl(
            \frac{4 S_2 P_{54}}{9 (N-1) N^2 (N+1)^2 (N+2)}
            +\frac{P_{190}}{27 (N-2) (N-1)^2 N^4 (N+1)^3 (N+2)^3}
            +\frac{272}{9} S_3
            \nonumber\\
            &&
            -\frac{32}{9} S_{2,1}
            -\frac{128}{9} S_{-2,1}
        \biggr) S_1^2
        +\biggl(
            -\frac{64 (2 N + 1) P_6}{27 (N-1)^2 N^2 (N+1)^2 (N+2)^2}
            +\frac{64}{27} S_2
        \biggr) S_1^3
        \nonumber\\
            &&
        +\biggl(
            -\frac{64 S_{-2,1} P_5}{3 N (N+1)}
            -\frac{16 S_3 P_{43}}{27 N (N+1)}
            +\frac{P_{209}}{405 (N-3) (N-2)^2 (N-1)^2 N^4 (N+1)^4 (N+2)^3}
            \nonumber\\
            &&
            -\frac{80}{9} S_{2,1}
        \biggr) S_2
        +\biggl(
            \frac{64 S_{2,1} P_{18}}{9 N (N+1)}
            +\frac{16 S_1^2 P_{58}}{9 (N-1) N^2 (N+1)^2 (N+2)}
            \nonumber\\
            &&
            +\frac{16 S_2 P_{114}}{9 (N-2) (N-1) N^2 (N+1)^2 (N+2)}
           +\frac{16 S_{-1} P_{137}}{15 (N-3) (N-2) (N-1)^2 N^2 (N+1)^3 (N+2)}
           \nonumber\\
            &&
            +\frac{16 S_1 P_{182}}{81 (N-3) (N-2) (N-1)^2 N^3 (N+1)^3 (N+2)}
            +\frac{128}{27} S_1^3
            -\frac{320}{27} S_3
            +\frac{64}{9} S_{-2,1}
            \nonumber\\
            &&+\frac{16 P_{204}}{405 (N-3) (N-2)^2 (N-1)^2 N^4 (N+1)^4 (N+2)^3}
        \biggr) S_{-2}
        +\biggl(
            \frac{16 P_{100}}{27 (N-1) N^2 (N+1)^2 (N+2)}
            \nonumber\\
            &&
            -\frac{64}{9} S_1
        \biggr) S_{-2}^2
        +\biggl(
            \frac{32 S_2 P_{17}}{9 N (N+1)}
            -\frac{16 S_{-2} P_{32}}{9 N (N+1)}
            -\frac{16 S_1 P_{62}}{9 (N-1) N^2 (N+1)^2 (N+2)}
            +\frac{64}{9} S_1^2
            \nonumber\\
            &&+\frac{8 P_{191}}{81 (N-3) (N-2) (N-1)^2 N^3 (N+1)^3 (N+2)^2}     
        \biggr) S_{-3}
        +\biggl(
            \frac{16 P_{79}}{9 (N-1) N^2 (N+1)^2 (N+2)}
            \nonumber\\
            &&
            +\frac{64}{9} S_1
        \biggr) S_{-4}
        -96 S_{4,1}
        +16 S_{2,2,1}
        +\frac{416}{3} S_{3,1,1}
        +\frac{128}{9} S_{-2,2,1}
        +\frac{128}{9} S_{-3,1,1}
        -\frac{160}{9} S_{2,1,1,1}
        \nonumber\\
            &&
        -\frac{256}{9} S_{-2,1,1,1}
        +\biggl(
            -\frac{4 P_{176}}{27 (N-1)^2 N^3 (N+1)^3 (N+2)^3}
            +\biggl[
                -\frac{16 P_{130}}{27 (N-1)^2 N^2 (N+1)^2 (N+2)^2}
                \nonumber\\
            &&
                +\frac{32}{3} S_2
            \biggr] S_1
            -\frac{64 \big(1+N+N^2\big) S_2}{3 (N-1) N (N+1) (N+2)}
            +\frac{16}{3} S_3
            +\biggl(
                -\frac{64 \big(1+N+N^2\big)}{3 (N-1) N (N+1) (N+2)}
                \nonumber\\
            &&
                +\frac{32}{3} S_1
            \biggr) S_{-2}
            +\frac{16}{3} S_{-3}
            -\frac{32}{3} S_{-2,1}
        \biggr) \zeta_2
        +\biggl(
            -\frac{32 S_2 P_5}{3 N (N+1)}
            -\frac{32 S_{-2} P_{11}}{3 N (N+1)}
            -\frac{80}{3} S_1^2
            \nonumber\\
            &&
            +\frac{4 S_1 P_{129}}{27 (N-2) (N-1) N^2 (N+1)^2 (N+2)}
            \nonumber\\
            &&+\frac{P_{201}}{1080 (N-3) (N-2) (N-1)^2 N^3 (N+1)^3 (N+2)^2}
        \biggr) \zeta_3
    \Biggr]
    +\frac{64}{27} \textcolor{blue}{T_F^3} \zeta_3
\Biggr\},
\label{aggQUNP}
\end{eqnarray}


\noindent
with
\begin{eqnarray}
{\sf B_4} = - 4 \zeta_2 \ln^2(2) + \frac{2}{3} \ln^4(2) - \frac{13}{2} \zeta_4 + 16 
\Li_4\left(\frac{1}{2}\right).
\end{eqnarray}
The polynomials $P_i$ are listed in Appendix~\ref{sec:A}.
Eq.~(\ref{aggQUNP}) possesses a removable pole at $N=2$. By a series 
expansion one obtains 
Eq.~(8.67) of Ref.~\cite{Bierenbaum:2009mv}. Furthermore, also the even moments for $N=4$ to $N=10$ agree with 
the result in Ref.~\cite{Bierenbaum:2009mv}. There are removable poles also at $N = 3, N = 1/2$ 
and $N = 
3/2$, 
see also Ref.~\cite{Ablinger:2014uka}. To see their cancellation, one has to expand the Mellin inversion to 
$x$--space around $x=0$. In the present case one finds the most singular terms $\propto 
\ln(x)/x$ and 
$\propto 1/x$, i.e. it is 
proven that rightmost pole is situated at $N=1$, as expected in the unpolarized gluonic case. Note that in 
general one may not expect to cancel the above poles by just using the $N$--space 
representation.

It is also instructive to see how accurate the asymptotic expansion of $a_{gg,Q}^{(3)}(N)$ 
represents higher moments. We compare the expressions for the $N_F$--independent part only, 
since the $N_F$--dependent 
part solely consists of harmonic sums. One may represent $a_{gg,Q}^{(3)}(N)$ by
\begin{eqnarray}
a_{gg,Q}^{(3)}(N) &=& a_{gg,Q,\delta}^{(3)} +  a_{gg,Q,\rm pl}^{(3)}(N) +
\tilde{a}_{gg,Q}^{(3)}(N).
\label{eq:aggQtil}
\end{eqnarray}
Here $a_{gg,Q,\delta}^{(3)}$ denotes the $N$--independent part,~cf.~(\ref{eq:4.6}),
$a_{gg,Q,\rm pl}^{(3)}(N)$ the part $\propto L$ with
\begin{eqnarray}
L = \ln(N) + \gamma_E, 
\label{eq:LG}
\end{eqnarray}
where $\gamma_E$ denotes the Euler--Mascheroni constant. 
The explicit 
expressions for the asymptotic expansion of 
$(\Delta) \tilde{a}_{gg,Q}^{(3)}(N)$ for the $N_F$--independent part are given in Appendix~\ref{sec:D}.
The asymptotic representation for $\tilde{a}_{gg,Q}^{(3)}(N)$ for positive even integer values converges 
very quickly, as shown in Table~1. Already for $N = 4$ a reasonable approximation is obtained by 
expanding to $O(1/N^{10})$.
\begin{table}[H]
\centering
\begin{tabular}{|r||r|r|}
\hline \hline
$N$  & complete expression & rel. asymp. accuracy \\
\hline \hline
4     & --430.337594532836914 & $ -2.10169$D--3  \\
6     & --261.554324759832203 & $ -3.13097$D--5   \\
8     & --193.698203673549029 & $ -1.29985$D--6  \\
10    & --156.572494406521072 & $ -9.09255$D--8  \\
12    & --132.845604857074259 & $ -7.66447$D--9  \\
22    & --79.9576964391278831 & $  3.81320$D--11   \\
42    & --48.0359673792636099 & $  1.41090$D--13  \\
102   & --24.2090141858051135 & $  3.14898$D--17  \\
\hline
\hline
\end{tabular}
\label{TAB1}
\caption[]{\sf 
Numerical comparison of $\tilde{a}_{gg,Q}^{(3)}(N)$ in QCD              with
its asymptotic
representation for $N_F = 0$ retaining 10 terms of the asymptotic expansion.}
\end{table}

We turn now to $\Delta a_{gg,Q}(N)$ which is given in the Larin scheme  \cite{Larin:1993tq} by
\begin{eqnarray}
\lefteqn{
\Delta a_{gg,Q}^{(3)}(N) = \frac{1}{2}\left(1 - (-1)^N\right)} \nonumber\\ &&
\times \Biggl\{
    \textcolor{blue}{C_A} \Biggl[
        \textcolor{blue}{C_F} \textcolor{blue}{T_F} \Biggl(
            \frac{32 S_{-2,2} P_8}{(N-1) N^2 (N+1)^2 (N+2)}
            +\frac{32 S_{-3,1} P_{10}}{3 (N-1) N^2 (N+1)^2 (N+2)}
            \nonumber\\
            &&
            +\frac{32 [ {\rm {\sf BS}}_7 -  {\rm {\sf BS}}_9 + 7 \zeta_3 {\rm {\sf BS}}_3 ] P_{14}}{3 N^2 (N+1)^2}
            -\frac{64 S_{-2,1,1} P_{27}}{3 (N-1) N^2 (N+1)^2 (N+2)}
            -\frac{4 S_1^3 P_{34}}{27 N^3 (N+1)^3}
            \nonumber\\
            &&-\frac{16 S_{-4} P_{36}}{3 (N-1) N^2 (N+1)^2 (N+2)}
            +\frac{4 S_4 P_{47}}{3 (N-1) N^2 (N+1)^2 (N+2)}
            -\frac{16 S_{2,1,1} P_{66}}{3 (N-1) N^2 (N+1)^2}
            \nonumber\\
            &&
            +\frac{32 [ S_{-1} S_2 - S_{2,-1} + S_{-2,-1} ] P_{81}}{(N-2) (N-1) N^3 (N+1)^2 (N+2)}
            +\frac{16 S_{3,1} P_{83}}{3 (N-1) N^2 (N+1)^2 (N+2)}
            \nonumber\\
            &&-\frac{16 S_{-2,1} P_{115}}{3 (N-2) (N-1) N^3 (N+1)^3 (N+2)}
            +\frac{\ds [ {\rm {\sf BS}}_8 - {\rm {\sf BS}}_4 S_1 + 7 \zeta_3 ] 2^{2-2 N} \binom{2 N}{N} 
P_{132}}{3 (N-2) (N-1) N^3 (N+1)^3}
            \nonumber\\
            &&+\frac{8 S_3 P_{150}}{27 (N-2) (N-1) N^3 (N+1)^3 (N+2)}
            -\frac{8 S_{2,1} P_{160}}{3 (N-2) (N-1) N^3 (N+1)^3 (N+2)}
            \nonumber\\
            &&+\frac{4 S_2 P_{188}}{9 (N-2) (N-1)^2 N^4 (N+1)^4 (N+2)}
            +\frac{P_{211}}{486 (N-2) (N-1)^2 N^6 (N+1)^6 (N+2)^2}
            \nonumber\\
            &&+\biggl(
                \frac{32 (N-3) (N+4)}{3 N^2 (N+1)^2}
                -\frac{64 S_1}{3}
            \biggr) {\rm {\sf B}}_4
            +\biggl(
                -\frac{48 (N-3) (N+4)}{N^2 (N+1)^2}
                +96 S_1
            \biggr) \zeta_4
            \nonumber\\
                &&
            +\biggl(
                \frac{64 S_{-2,1} P_{26}}{3 (N-1) N^2 (N+1)^2 (N+2)}
                +\frac{32 S_3 P_{37}}{9 (N-1) N^2 (N+1)^2 (N+2)}
                \nonumber\\
                &&
                +\frac{32 S_{2,1} P_{76}}{3 (N-1) N^2 (N+1)^2 (N+2)}
                +
                \frac{4 S_2 P_{121}}{3 (N-1) N^3 (N+1)^3 (N+2)}
                \nonumber\\
                &&+\frac{8 P_{205}}{81 (N-2) (N-1)^2 N^5 (N+1)^5 (N+2)^2}
            \biggr) S_1
            +\biggl(
                -\frac{4 S_2 P_{40}}{3 (N-1) N^2 (N+1)^2 (N+2)}
                \nonumber\\
                &&+\frac{4 P_{127}}{27 (N-1) N^3 (N+1)^4 (N+2)}
            \biggr) S_1^2
            -\frac{2 (N-1) (N+2) S_1^4}{9 N^2 (N+1)^2}
            +\frac{2 \big(6+5 N+5 N^2\big) S_2^2}{N^2 (N+1)^2}
            \nonumber\\
            &&+\biggl(
                -\frac{32 S_1^2 P_{24}}{3 (N-1) N^2 (N+1)^2 (N+2)}
                -\frac{32 S_{-1} P_{81}}{(N-2) (N-1) N^3 (N+1)^2 (N+2)}
                \nonumber\\
                &&+\frac{16 S_1 P_{147}}{3 (N-2) (N-1)^2 N^3 (N+1)^3 (N+2)^2}
                -\frac{16 P_{164}}{3 (N-2) (N-1)^2 N^3 (N+1)^4 (N+2)^2}
                \nonumber\\
                &&-\frac{32 \big(2-15 N-12 N^2+N^3\big) S_2}{3 (N-1) N^2 (N+1)^2}
            \biggr) S_{-2}
            -\frac{32 \big(-13+2 N+2 N^2\big) S_{-2}^2}{3 N^2 (N+1)^2}
            \nonumber\\
                &&
            +\biggl(
                -\frac{32 S_1 P_9}{3 (N-1) N^2 (N+1)^2 (N+2)}
                +\frac{8 P_{124}}{3 (N-2) (N-1) N^3 (N+1)^3 (N+2)}
            \biggr) S_{-3}
            \nonumber\\
                &&
            +\biggl(
                \frac{4 S_1 P_{101}}{3 N^3 (N+1)^3}
                +\frac{4 P_{138}}{9 N^4 (N+1)^4}
                -\frac{4 (N-1) (N+2) S_1^2}{N^2 (N+1)^2}
                -\frac{12 (N-1) (N+2) S_2}{N^2 (N+1)^2}
                \nonumber\\
                &&
                -\frac{24 (N-1) (N+2) S_{-2}}{N^2 (N+1)^2}
            \biggr) \zeta_2
            +\biggl(
                -\frac{8 S_1 P_{110}}{9 (N-1) N^2 (N+1)^2 (N+2)}
                \nonumber\\
                &&
                +\frac{P_{170}}{36 (N-2) (N-1) N^3 (N+1)^3 (N+2)}
            \biggr) \zeta_3
        \Biggr)
        +\textcolor{blue}{T_F^2} \Biggl(
            \frac{4 S_2 P_{48}}{135 N^2 (N+1)^2}
            \nonumber\\
            &&
            +\frac{\ds [ {\rm {\sf BS}}_8 - {\rm {\sf BS}}_4 S_1 + 7 \zeta_3 ] 2^{2-2 N} \binom{2 N}{N} 
P_{109}}{45 N (N+1)^2 (2 N -3) (2 N -1)}
           +\frac{P_{175}}{3645 N^4 (N+1)^4 (2 N -3) (2 N -1)}
           \nonumber\\
           &&
            +\textcolor{blue}{N_F} \Biggl(
                \frac{4 S_2 P_{44}}{27 N^2 (N+1)^2}
                -\frac{8 S_1 P_{113}}{729 N^3 (N+1)^3}
                -\frac{2 P_{145}}{729 N^4 (N+1)^4}
                +\frac{4 \big(16-9 N-25 N^2-16 N^3\big) S_1^2}{27 N^2 (N+1)^2}
                \nonumber\\
                &&+\biggl(
                    \frac{4 P_{33}}{27 N^2 (N+1)^2}
                    -\frac{160 S_1}{27}
                \biggr) \zeta_2
                +\biggl(
                    -\frac{896}{27 N (N+1)}
                    +\frac{448 S_1}{27}
                \biggr) \zeta_3
            \Biggr)
            -\frac{64 (N+2) S_3}{15 (N+1)}
            \nonumber\\
                &&
            -\frac{8 S_1 P_{146}}{3645 N^3 (N+1)^3 (2 N -3) (2 N -1)}
            -\frac{4 \big(89-108 N+25 N^2+70 N^3\big) S_1^2}{135 N^2 (N+1)^2}
            +\frac{64 (N+2) S_{2,1}}{15 (N+1)}
            \nonumber\\
            &&
            +\biggl(
                \frac{4 P_{45}}{27 N^2 (N+1)^2}
                -\frac{560 S_1}{27}
            \biggr) \zeta_2
            +\biggl(
                -\frac{7 \big(-3200+2439 N+1287 N^2\big)}{270 N (N+1)}
                -\frac{1120}{27} S_1
            \biggr) \zeta_3
        \Biggr)
    \Biggr]
    \nonumber\\
    &&+\textcolor{blue}{C_F} \textcolor{blue}{T_F^2} \Biggl[
        -\frac{2 P_{194}}{243 N^5 (N+1)^5 (2 N -3) (2 N -1)}
        +\textcolor{blue}{N_F} \Biggl(
            -\frac{2 P_{171}}{243 N^5 (N+1)^5}
            \nonumber\\
                &&
            -\biggl(
                \frac{32 (N-1) (N+2) P_{46}}{81 N^4 (N+1)^4}
                +\frac{16 (N-1) (N+2) S_2}{3 N^2 (N+1)^2}
            \biggr) S_1
            +\frac{160 (N-1) (N+2) S_3}{27 N^2 (N+1)^2}
            \nonumber\\
            &&+\frac{16 (N-1) (N+2) \big(
                6+20 N+29 N^2\big) S_1^2}{27 N^3 (N+1)^3}
            -\frac{112 (N-1) (N+2) S_1^3}{27 N^2 (N+1)^2}
            \nonumber\\
            &&-\frac{16 (N-1) (N+2) \big(-6-8 N+N^2\big) S_2}{9 N^3 (N+1)^3}
            -\biggl(
                 \frac{4 P_{95}}{9 N^3 (N+1)^3}
                +\frac{16 (N-1) (N+2) S_1}{3 N^2 (N+1)^2}
            \biggr) \zeta_2
            \nonumber\\
                &&
            -\frac{448 (N-1) (N+2) \zeta_3}{9 N^2 (N+1)^2}
        \Biggr)
        -\frac{\ds [ {\rm {\sf BS}}_8 - {\rm {\sf BS}}_4 S_1 + 7 \zeta_3 ] (N-1) (N+2) \big(8+9 N-9 
N^2\big) 2^{4-2 N} \binom{2 N}{N}}{3 N (N+1)^2 (2 N -3) (2 N -1)}
        \nonumber\\
        &&+\biggl(
            \frac{32 (N-1) (N+2) P_{103}}{81 N^4 (N+1)^4 (2 N -3) (2 N -1)}
            -\frac{16 (N-1) (N+2) S_2}{3 N^2 (N+1)^2}
        \biggr) S_1
        \nonumber\\
        &&
        +\frac{16 (N-1) (N+2) \big(-6-8 N+N^2\big) S_1^2}{27 N^3 (N+1)^3}
        +\frac{16 (N-1) (N+2) S_1^3}{27 N^2 (N+1)^2}
        \nonumber\\
        &&-\frac{16 (N-1) (N+2) \big(
            -6-8 N+N^2\big) S_2}{9 N^3 (N+1)^3}
        -\frac{352 (N-1) (N+2) S_3}{27 N^2 (N+1)^2}
        +\frac{64 (N-1) (N+2) S_{2,1}}{3 N^2 (N+1)^2}
        \nonumber\\
        &&+\biggl(
            -\frac{8 P_{102}}{9 N^3 (N+1)^3}
            +\frac{16 (N-1) (N+2) S_1}{3 N^2 (N+1)^2}
        \biggr) \zeta_2
        + \frac{P_1}{9 N^2 (N+1)^2}\zeta_3
    \Biggr]
    \nonumber\\
    &&+\textcolor{blue}{C_F^2} \textcolor{blue}{T_F} \Biggl[
        \frac{32 [ {\rm {\sf B}}_4 + {\rm {\sf BS}}_7 - {\rm {\sf BS}}_9 + 7\zeta_3 {\rm {\sf BS}}_3] 
\big(2+N+N^2\big)}{N^2 (N+1)^2}
        -\frac{144 \big(2+N+N^2\big) \zeta_4}{N^2 (N+1)^2}
        +\frac{4 S_1^3 P_{15}}{9 N^3 (N+1)^3}
        \nonumber\\
        &&
        -\frac{64 [ S_{-1} S_2 - S_{2,-1} + S_{-2,-1} ] P_{60}}{3 (N-2) (N-1) N^3 (N+1) (N+2)}
        -\frac{\ds [ {\rm {\sf BS}}_8 - {\rm {\sf BS}}_4 S_1 + 7 \zeta_3 ] 2^{5-2 N} \binom{2 N}{N} 
P_{85}}{3 (N-2) (N-1) N^3 (N+1)^2}
        \nonumber\\
        &&+\frac{32 S_{-2,1} P_{89}}{3 (N-2) (N-1) N^3 (N+1)^3}
        +\frac{16 S_3 P_{119}}{9 (N-2) (N-1) N^3 (N+1)^3 (N+2)}
        \nonumber\\
        &&-\frac{16 S_{2,1} P_{134}}{3 (N-2) (N-1) N^3 (N+1)^3 (N+2)}
        +\frac{4 S_2 P_{151}}{3 (N-2) (N-1) N^4 (N+1)^4 (N+2)}
        \nonumber\\
        &&+\frac{2 P_{210}}{9 (N-2) (N-1)^2 N^6 (N+1)^6 (N+2)^2}
        +\biggl(
            -\frac{4 S_2 P_{51}}{3 N^3 (N+1)^3 (N+2)}
            \nonumber\\
            &&
            +\frac{8 P_{163}}{3 (N-2) N^5 (N+1)^5 (N+2)}
            +\frac{16 \big(46+13 N+13 N^2\big) S_3}{9 N^2 (N+1)^2}
            +\frac{32 \big(2+3 N+3 N^2\big) S_{2,1}}{3 N^2 (N+1)^2}
            \nonumber\\
        &&
            -\frac{256 S_{-2,1}}{N^2 (N+1)^2}
        \biggr) S_1
        +\biggl(
           -\frac{4 P_{90}}{3 N^4 (N+1)^4}
            +\frac{4 \big(
                22+5 N+5 N^2\big) S_2}{3 N^2 (N+1)^2}
        \biggr) S_1^2
        +\frac{2 (N-1) (N+2) S_1^4}{9 N^2 (N+1)^2}
        \nonumber\\
        &&-\frac{2 \big(-30+31 N+31 N^2\big) S_2^2}{3 N^2 (N+1)^2}
        -\frac{4 \big(-54+11 N+11 N^2\big) S_4}{3 N^2 (N+1)^2}
        \nonumber\\
            &&
        +\biggl(
            \frac{64 S_{-1} P_{60}}{3 (N-2) (N-1) N^3 (N+1) (N+2)}
            +\frac{32 P_{65}}{3 (N-2) (N-1)^2 (N+1)^2 (N+2)^2}
            \nonumber\\
            &&-\frac{32 S_1 P_{89}}{3 (N-2) (N-1) N^3 (N+1)^3}
            +\frac{128 S_1^2}{N^2 (N+1)^2}
            +\frac{256 S_2}{3 N^2 (N+1)^2}
        \biggr) S_{-2}
        -\frac{64 S_{-2}^2}{N^2 (N+1)^2}
        \nonumber\\
            &&
        +\biggl(
            -\frac{16 P_{118}}{3 (N-2) (N-1) N^3 (N+1)^3 (N+2)}
            +\frac{128 S_1}{3 N^2 (N+1)^2}
        \biggr) S_{-3}
        -\frac{128 S_{-4}}{3 N^2 (N+1)^2}
        \nonumber\\
        &&+\frac{128 \big(1+N+N^2\big) S_{3,1}}{3 N^2 (N+1)^2}
        -\frac{256 S_{-2,2}}{3 N^2 (N+1)^2}
        -\frac{128 S_{-3,1}}{3 N^2 (N+1)^2}
        -\frac{32 \big(10+7 N+7 N^2\big) S_{2,1,1}}{3 N^2 (N+1)^2}
        \nonumber\\
        &&+\frac{256 S_{-2,1,1}}{N^2 (N+1)^2}
        +\biggl(
            -\frac{2 (N-1) (N+2) P_{42}}{N^4 (N+1)^4}
            +\frac{4 (N-1) (N+2) \big(-4-3 N+3 N^2\big) S_1}{N^3 (N+1)^3}
            \nonumber\\
            &&+\frac{4 (N-1) (N+2) S_1^2}{N^2 (N+1)^2}
            -\frac{12 (N-1) (N+2) S_2}{N^2 (N+1)^2}
        \biggr) \zeta_2
        +\biggl(
             \frac{P_{154}}{3 (N-2) (N-1) N^3 (N+1)^3 (N+2)}
\nonumber\\ &&   
         -\frac{16 \big(118+N+N^2\big) S_1}{3 N^2 (N+1)^2}
        \biggr) \zeta_3
    \Biggr]
    +\textcolor{blue}{C_A^2} \textcolor{blue}{T_F} \Biggl[
        \frac{256 S_{-2,2} P_4}{9 (N-1) N^2 (N+1)^2 (N+2)}
        \nonumber\\
        &&
        -\frac{8 [ {\rm {\sf BS}}_7 - {\rm {\sf BS}}_9 + 7 \zeta_3 {\rm {\sf BS}}_3 ] (N-1) \big(18+21 N+16 N^2+4 N^3\big)}{3 N^2 (N+1)^2}
        \nonumber\\
        &&
        +\frac{32 S_{-2,1,1} P_{53}}{9 (N-1) N^2 (N+1)^2 (N+2)}
        +\frac{16 S_{-3,1} P_{63}}{9 (N-1) N^2 (N+1)^2 (N+2)}
        +\frac{8 S_2^2 P_{68}}{9 N^2 (N+1)^2 (N+2)}
        \nonumber\\
        &&
        -\frac{16 S_{2,1,1} P_{86}}{3 (N-1) N^2 (N+1)^2 (N+2)}
        -\frac{16 [ S_{-1} S_2 - S_{2,-1} + S_{-2,-1} ] P_{87}}{3 (N-2) (N-1) N^3 (N+1)^2 (N+2)}
        \nonumber\\
        &&+\frac{4 S_4 P_{88}}{9 (N-1) N^2 (N+1)^2 (N+2)}
        -\frac{8 S_{3,1} P_{94}}{9 (N-1) N^2 (N+1)^2 (N+2)}
        \nonumber\\
        &&
        -\frac{\ds [ {\rm {\sf BS}}_8 - {\rm {\sf BS}}_4 S_1 + 7 \zeta_3]  2^{1-2 N} \binom{2 N}{N} 
P_{131}}{3 (N-2) (N-1) N^3 (N+1)^3}
        -\frac{16 S_{-2,1} P_{155}}{81 (N-2) (N-1) N^3 (N+1)^3 (N+2)}
        \nonumber\\
        &&
        -\frac{2 S_{2,1} P_{168}}{9 (N-2) (N-1) N^3 (N+1)^3 (N+2)}
        -\frac{4 S_3 P_{169}}{81 (N-2) (N-1) N^3 (N+1)^3 (N+2)}
        \nonumber\\
        &&
        +\frac{P_{212}}{2916 (N-2) (N-1)^2 N^6 (N+1)^6 (N+2)^2}
        +{\rm {\sf B}}_4 \biggl(
            -\frac{8 \big(-18+5 N+5 N^2\big)}{3 N^2 (N+1)^2}
            +\frac{32}{3} S_1
        \biggr)
        \nonumber\\
        &&
        +\zeta_4 \biggl(
            \frac{96 \big(-3+2 N+2 N^2\big)}{N^2 (N+1)^2}
            -96 S_1
        \biggr)
        +\biggl(
            -\frac{32 S_{-2,1} P_{57}}{9 (N-1) N^2 (N+1)^2 (N+2)}
            \nonumber\\
            &&
            +\frac{8 S_{2,1} P_{97}}{9 (N-1) N^2 (N+1)^2 (N+2)}
            -\frac{16 S_3 P_{104}}{27 (N-1) N^2 (N+1)^2 (N+2)}
            +\frac{16}{9} S_2^2
            +\frac{592}{9} S_4
            -\frac{832}{9} S_{3,1}
            \nonumber\\
            &&
            +\frac{4 S_2 P_{143}}{81 (N-1) N^3 (N+1)^3 (N+2)}
            +\frac{2 P_{206}}{729 (N-2) (N-1)^2 N^5 (N+1)^5 (N+2)^2}
            \nonumber\\
            &&-\frac{128}{9} S_{-2,2}
            -\frac{128}{9} S_{-3,1}
            +\frac{32}{3} S_{2,1,1}
            +\frac{256}{9} S_{-2,1,1}
        \biggr) S_1
        +\biggl(
            \frac{4 S_2 P_{55}}{9 (N-1) N^2 (N+1)^2 (N+2)}
            \nonumber\\
            &&+\frac{P_{142}}{27 (N-1) N^4 (N+1)^3 (N+2)}
            +\frac{272}{9} S_3
            -\frac{32}{9} S_{2,1}
            -\frac{128}{9} S_{-2,1}
        \biggr) S_1^2
        +\biggl(
            -\frac{64 (2 N +1)}{27 N^2 (N+1)^2}
            \nonumber\\
            &&
            +\frac{64}{27} S_2
        \biggr) S_1^3
        +\biggl(
            \frac{P_{193}}{81 (N-2) (N-1)^2 N^4 (N+1)^4 (N+2)}
            -\frac{16 \big(-108+175 N+175 N^2\big) S_3}{27 N (N+1)}
            \nonumber\\
            &&
            -\frac{80}{9} S_{2,1}
            -\frac{64 \big(
                -3+N+N^2\big) S_{-2,1}}{3 N (N+1)}
        \biggr) S_2
        -\frac{16 \big(-54+55 N+55 N^2\big) S_5}{9 N (N+1)}
        \nonumber\\
        &&
        +\biggl(
            \frac{16 S_1^2 P_{59}}{9 (N-1) N^2 (N+1)^2 (N+2)}
            +\frac{16 S_2 P_{80}}{9 (N-1) N^2 (N+1)^2 (N+2)}
            \nonumber\\
            &&+\frac{16 S_{-1} P_{87}}{3 (N-2) (N-1) N^3 (N+1)^2 (N+2)}
            +\frac{16 S_1 P_{179}}{81 (N-2) (N-1)^2 N^3 (N+1)^3 (N+2)^2}
            \nonumber\\
            &&+\frac{16 P_{187}}{81 (N-2) (N-1)^2 N^4 (N+1)^4 (N+2)^2}
            +\frac{128}{27} S_1^3
            -\frac{320}{27} S_3
            +\frac{64 \big(
                -9+2 N+2 N^2\big) S_{2,1}}{9 N (N+1)}
            \nonumber\\
            &&+\frac{64}{9} S_{-2,1}
        \biggr) S_{-2}
        +\biggl(
            \frac{16 P_{71}}{27 N^2 (N+1)^2 (N+2)}
            -\frac{64}{9} S_1
        \biggr) S_{-2}^2
        +\biggl(
            -\frac{16 S_1 P_{61}}{9 (N-1) N^2 (N+1)^2 (N+2)}
            \nonumber\\
            &&+\frac{8 P_{156}}{81 (N-2) (N-1) N^3 (N+1)^3 (N+2)}
            +\frac{64}{9} S_1^2
            -\frac{32 \big(
                9+2 N+2 N^2\big) S_2}{9 N (N+1)}
            \nonumber\\
            &&-\frac{16 \big(
                -18+19 N+19 N^2\big) S_{-2}}{9 N (N+1)}
        \biggr) S_{-3}
        +\biggl(
            \frac{16 P_{78}}{9 (N-1) N^2 (N+1)^2 (N+2)}
            +\frac{64}{9} S_1
        \biggr) S_{-4}
        \nonumber\\
        &&-\frac{32 \big(9+5 N+5 N^2\big) S_{-5}}{9 N (N+1)}
        +\frac{16 \big(-54+77 N+77 N^2\big) S_{2,3}}{9 N (N+1)}
        +\frac{16 \big(-18+17 N+17 N^2\big) S_{2,-3}}{9 N (N+1)}
        \nonumber\\
        &&-96 S_{4,1}
        -\frac{16 (N-2) (N+3) S_{-2,3}}{3 N (N+1)}
        +\frac{64 \big(-18+11 N+11 N^2\big) S_{-2,-3}}{9 N (N+1)}
        \nonumber\\
        &&-\frac{16 (N-1) (N+2) S_{-4,1}}{N (N+1)}
        -\frac{32 \big(-18+5 N+5 N^2\big) S_{2,1,-2}}{9 N (N+1)}
        +16 S_{2,2,1}
        +\frac{416}{3} S_{3,1,1}
        \nonumber\\
        &&-\frac{32 \big(-18+13 N+13 N^2\big) S_{-2,1,-2}}{9 N (N+1)}
        +\frac{128}{9} S_{-2,2,1}
        +\frac{128}{9} S_{-3,1,1}
        -\frac{160}{9} S_{2,1,1,1}
        \nonumber\\
        &&-\frac{256}{9} S_{-2,1,1,1}
        +\biggl[
            -\frac{4 P_{99}}{27 N^3 (N+1)^3}
            +\biggl(
                -\frac{16 \big(36+72 N+N^2+2 N^3+N^4\big)}{27 N^2 (N+1)^2}
                \nonumber\\
                &&+\frac{32}{3} S_2
            \biggr) S_1
            -\frac{64 S_2}{3 N (N+1)}
            +\frac{16}{3} S_3
            +\biggl(
                -\frac{64}{3 N (N+1)}
                +\frac{32 S_1}{3}
            \biggr) S_{-2}
            +\frac{16}{3} S_{-3}
            \nonumber\\
            &&-\frac{32}{3} S_{-2,1}
        \biggr] \zeta_2
        +\biggl(
            -\frac{32 \big(-3+N+N^2\big) S_2}{3 N (N+1)}
            +\frac{P_{173}}{216 (N-2) (N-1) N^3 (N+1)^3 (N+2)}
                      \nonumber\\
&&  +\frac{4 S_1 P_{112}}{27 (N-1) N^2 (N+1)^2 (N+2)}
            -\frac{80}{3} S_1^2
-\frac{32 (N-2) (N+3) S_{-2}}{3 N (N+1)}
        \biggr) \zeta_3
    \Biggr]
    +\frac{64}{27} \textcolor{blue}{T_F^3} \zeta_3
\Biggr\}.
\end{eqnarray}


\noindent
One may represent $\Delta a_{gg,Q}^{(3)}(N)$ in an analogous way to Eq.~(\ref{eq:aggQtil}). As in the 
unpolarized case, the expansion 
converges very quickly for the positive odd integer values, cf.~Table~2, with a reasonable description 
down to $N = 3$.
\begin{table}[H]
\centering
\begin{tabular}{|r||r|r|}
\hline \hline
$N$  & complete expression & rel. asymp. accuracy \\
\hline \hline
3     & --429.345090408771279 & $ -9.39608$D--4  \\
5     & --264.713704879676430 & $ -7.20184$D--6  \\
7     & --195.879611926179533 & $ -2.77043$D--7  \\
9     & --158.086523949063478 & $ -2.36920$D--8  \\
11    & --133.968747075663141 & $ -3.28594$D--9  \\
21    & --80.4139205966345783 & $ -5.45018$D--12  \\
41    & --48.2243776140486971 & $ -7.23295$D--15  \\
101   & --24.2616935237270326 & $ -1.03115$D--18  \\
\hline
\hline
\end{tabular}
\label{TAB2}
\caption[]{\sf Numerical comparison of $\Delta \tilde{a}_{gg,Q}^{(3)}(N)$ in QCD 
with 
its asymptotic 
representation for $N_F = 0$ retaining 10 terms of the asymptotic expansion.}
\end{table}
\section{The \boldmath $x$-space representation}
\label{sec:4}

\vspace{1mm}
\noindent
We perform an analytic inverse Mellin transform to $x$--space by using algorithms 
implemented in {\tt 
HarmonicSums}.
In momentum fraction space, the quantities $(\Delta) a_{gg,Q}(x)$ depend besides  
harmonic polylogarithms, $\HA_{\vec{a}}(x)$, on $\GA$--functions up to weight {\sf w = 5} 
at arguments $x$ and $1$ over the alphabet (\ref{eq:alph}), e.g. 
\begin{eqnarray}
\GA\left(\left\{
\sqrt{(1-y) y}, 
\frac{1}{y},
\sqrt{(1-y) y}, 
\frac{1}{y},\frac{1}{1-y}\right\},x\right) = \GA(\{5,1,5,1,2\},x),
\end{eqnarray}
and 17 similar functions both in the unpolarized and polarized case. The appearing 
constants can all be calculated analytically by using {\tt HarmonicSums} and only 
multiple zeta values remain at the end. In addition to this, the following 48 harmonic 
polylogarithms contribute
\begin{eqnarray}
&&\bigl\{
\HA_0, \HA_{-1}, 
\HA_1, 
\HA_{-1,1}, 
\HA_{0,-1},\HA_{0,1},\HA_{0,-1,-1},
\HA_{0,-1,1},\HA_{0,0,-1},
\HA_{0,0,1},\HA_{0,1,-1},\HA_{0,1,1},\HA_{0,-1,-1,-1},
\nonumber\\ && 
\HA_{0,-1,-1,1},\HA_{0,-1,0,1},
\HA_{0,-1,1,-1},\HA_{0,-1,1,1},\HA_{0,0,-1,-1},\HA_{0,0,-1,1},\HA_{0,0,0,-1},
\HA_{0,0,0,1},\HA_{0,0,1,-1},\HA_{0,0,1,1},
\nonumber\\ && 
\HA_{0,1,-1,-1},\HA_{0,1,-1,1},\HA_{0,1,1,-1},
\HA_{0,1,1,1},\HA_{0,-1,-1,0,1},\HA_{0,-1,0,-1,-1},\HA_{0,-1,0,1,1},
\HA_{0,-1,1,0,1},
\nonumber\\ && 
\HA_{0,0,-1,-1,-1},
\HA_{0,0,-1,0,-1},\HA_{0,0,-1,0,1},\HA_{0,0,-1,1,1},\HA_{0,0,0,-1,-1},\HA_{0,0,0,-1,1},
\HA_{0,0,0,0,-1},\HA_{0,0,0,0,1},
\nonumber\\ && 
\HA_{0,0,0,1,-1},\HA_{0,0,0,1,1},
\HA_{0,0,1,0,-1},
\HA_{0,0,1,0,1},\HA_{0,0,1,1,-1},
\HA_{0,0,1,1,1},\HA_{0,1,0,1,-1},
\HA_{0,1,0,1,1},\HA_{0,1,1,1,1}\Bigr\},
\end{eqnarray}
both after algebraic reduction for the $\GA$- and ${\rm H}$-functions \cite{Blumlein:2003gb}, where we 
suppressed
the argument $x$ of the harmonic polylogarithms.

Furthermore, denominator structures of the kind
\begin{eqnarray}
\frac{1}{(1 \pm x)^k},~~~k = 2,3 
\end{eqnarray}
appear, which are also known from other massive calculations 
\cite{Berends:1987ab,Blumlein:2020jrf}. Since the 
corresponding expressions are very lengthy, we present them only in an ancillary file in 
computer--readable form. Here we discuss their principal structure. The expressions for $(\Delta) 
a_{gg,Q}^{(3)}(x)$ have the following form
\begin{eqnarray}
(\Delta) a_{gg,Q}^{(3)}(x) = (\Delta) a_{gg,Q,\delta}^{(3)} \delta(1-x) + [(\Delta) a_{gg,Q,\rm 
plus}^{(3)}(x)]_+
+ (\Delta) a_{gg,Q,\rm reg}^{(3)}(x)],
\end{eqnarray}
where 
\begin{eqnarray}
\int_0^1 dx g(x) [f(x)]_+ := \int_0^1 [g(x) - g(1)] f(x).
\end{eqnarray}
For the $\delta(1-x)$ and $+$-function contributions we obtain 
\begin{eqnarray}
\label{eq:4.6}
(\Delta) a_{gg,Q,\delta}^{(3)} &=& 
\textcolor{blue}{T_F} \Biggl\{
        \textcolor{blue}{C_F} \Biggl[
                \textcolor{blue}{C_A} \Biggl(
                        \frac{16541}{162}
                        -\frac{64 {\sf B}_4}{3}
                        +\frac{128 \zeta_4}{3}
                        +52 \zeta_2
                        -\frac{2617 \zeta_3}{12}
                \Biggr)
                +\textcolor{blue}{T_F} \Biggl(
                        -\frac{1478}{81}
\nonumber\\ &&                   
      +\textcolor{blue}{N_F} \Biggl(
                                -\frac{1942}{81}
                                -\frac{20 \zeta_2}{3}
                        \Biggr)
                        -\frac{88 \zeta_2}{3}
                        -7 \zeta_3
                \Biggr)
        \Biggr]
        +\textcolor{blue}{C_A^2} \Biggl[
                \frac{34315}{324}
                +\frac{32 {\sf B}_4}{3}
                -\frac{3778 \zeta_4}{27}
\nonumber\\ &&              
   +\frac{992}{27} \zeta_2
                +\Biggl(
                        \frac{20435}{216}
                        +24 \zeta_2
                \Biggr) \zeta_3
                -\frac{304}{9} \zeta_5
        \Biggr]
        +\textcolor{blue}{C_A T_F} \Biggl[
                \frac{2587}{135}
                +\textcolor{blue}{N_F} \Biggl(
                        -\frac{178}{9}
                        +\frac{196 \zeta_2}{27}
                \Biggr)
\nonumber\\ &&            
     +\frac{572 \zeta_2}{27}
                -\frac{291 \zeta_3}{10}
        \Biggr]
        +\textcolor{blue}{C_F^2} \Biggl[
                \frac{274}{9}
                +\frac{95 \zeta_3}{3}
        \Biggr]
        +\frac{64}{27} \textcolor{blue}{T_F^2} \zeta_3
\Biggr\},
\\
(\Delta) a_{gg,Q,\rm plus}^{(3)} &=& 
\frac{\textcolor{blue}{T_F}}{1-x} \Biggl\{
        \textcolor{blue}{C_A T_F}  \Biggl[
                \frac{35168}{729}
                +\textcolor{blue}{N_F} \Biggl(
                        \frac{55552}{729}
                        +\frac{160 \zeta_2}{27}
                        -\frac{448 \zeta_3}{27}
                \Biggr)
                +\frac{560}{27} \zeta_2
                +\frac{1120}{27} \zeta_3
        \Biggr]
\nonumber\\ && 
        +\textcolor{blue}{C_A^2} \Biggl[
                -\frac{32564}{729}
                -\frac{32 {\sf B}_4}{3}
                +104 \zeta_4
                -\frac{3248 \zeta_2}{81}
                -\frac{1796 \zeta_3}{27}
        \Biggr]
        +\textcolor{blue}{C_A C_F}  \Biggl[
                -\frac{6152}{27}
                +\frac{64 {\sf B}_4}{3}
\nonumber\\ &&               
  -96 \zeta_4
                -40 \zeta_2
                +\frac{1208 \zeta_3}{9}
        \Biggr]
\Biggr\}.
\end{eqnarray}
They are the same in both cases.
The regular parts $(\Delta) a_{gg,Q,\rm reg}^{(3)}(x)$ are given in an ancillary file 
to the present paper.

We further expand $(\Delta) a_{gg,Q}(x)$  in the small $x$ and large $x$ regions, where
much simpler structures ruled by logarithms are obtained. 
The principal $x$--space structure to higher powers in $x$ is
\begin{eqnarray}
(\Delta) a_{gg,Q}^{x \rightarrow 0}(x) &=& 
c_1 \frac{\ln(x)}{x} + c_2 \frac{1}{x} +
\sum_{k=0}^\infty \Biggl[c_{3,k} + c_{4,k} \ln(x) 
+ c_{5,k} \ln^6(x) + c_{7,k} \ln^3(x)
+ c_{8,k} \ln^4(x) 
\nonumber\\ &&
+ c_{9,k} \ln^5(x)\Biggr] x^k
\\
(\Delta) a_{gg,Q}^{x \rightarrow 1}(x) &=& d_1 \delta(1-x) + \frac{d_2}{1-x}  + 
\sum_{k=0}^\infty
\Biggl[d_{3,k} + d_{4,k} \ln(1-x) + d_{5,k} \ln^2(1-x) 
\nonumber\\ &&
+ d_{6,k} \ln^3(1-x)+ d_{7,k} \ln^4(1-x)\Biggr] x^k,
\end{eqnarray}
where some of the coefficients can be zero.
The leading behavior for the two expansions around $x=0$ and $x=1$ in the unpolarized case
are given by
\begin{eqnarray}
\lefteqn{ a_{gg,Q}^{x \rightarrow 0}(x) \propto} \nonumber\\ && 
\frac{1}{x}
\Biggl\{
\ln(x) 
\Biggl[
        \textcolor{blue}{C_A^2 T_F}
        \biggl(
                -\frac{11488}{81}
                +\frac{224 \zeta_2}{27}
                +\frac{256 \zeta_3}{3}
        \biggr) 
        + \textcolor{blue}{C_A C_F T_F}
        \biggl(
                -\frac{15040}{243}
                -\frac{1408 \zeta_2}{27}
\nonumber\\ &&      
           -\frac{1088 \zeta_3}{9}
        \biggr)
\Biggr]
+\textcolor{blue}{C_A T_F^2} 
\Biggl[
        \frac{112016}{729}
        +\frac{1288}{27} \zeta_2
        +\frac{1120}{27} \zeta_3
        +\biggl(
                \frac{108256}{729}
                +\frac{368 \zeta_2}{27}
                -\frac{448 \zeta_3}{27}
        \biggr) 
\nonumber\\ && \times
\textcolor{blue}{N_F}
\Biggr]
+ \textcolor{blue}{C_F} 
\Biggl[
        \textcolor{blue}{T_F^2} 
        \Biggl(
                -\frac{107488}{729}
                -\frac{656}{27} \zeta_2
                +\frac{3904}{27} \zeta_3
                +\biggl(
                        \frac{116800}{729}
                        +\frac{224 \zeta_2}{27}
                        -\frac{1792 \zeta_3}{27}
                \biggr) \textcolor{blue}{N_F}
        \Biggr)
\nonumber\\ && 
        + \textcolor{blue}{C_A T_F}
        \Biggl(
                -\frac{5538448}{3645}
                +\frac{1664 {\sf B}_4}{3}
                -\frac{43024 \zeta_4}{9}
                +\frac{12208}{27} \zeta_2
                +\frac{211504}{45} \zeta_3
        \Biggr) 
\Biggr]
\nonumber\\ && 
+ \textcolor{blue}{C_A^2 T_F}
\biggl(
        -\frac{4849484}{3645}
        -\frac{352 {\sf B}_4}{3}
        +\frac{11056 \zeta_4}{9}
        -\frac{1088}{81} \zeta_2
        -\frac{84764}{135} \zeta_3
\biggr) 
\nonumber\\ && 
+ \textcolor{blue}{C_F^2 T_F}
\biggl(
        \frac{10048}{5}
        -640 {\sf B}_4
        +\frac{51104 \zeta_4}{9}
        -\frac{10096}{9} \zeta_2
        -\frac{280016}{45} \zeta_3
\biggr) 
\Biggr\}
\nonumber\\ &&
+
\Biggl[
        -\frac{4}{3}  \textcolor{blue}{C_F C_A T_F} 
        +\frac{2}{15} \textcolor{blue}{C_F^2 T_F}
\Biggr] \ln^5(x)
+ \Biggl[
-\frac{40}{27} \textcolor{blue}{C_A^2 T_F} 
+\frac{4}{9} \textcolor{blue}{C_F^2 T_F}
+\textcolor{blue}{C_F} \Biggl(
        -\frac{296}{27} \textcolor{blue}{C_A T_F} 
\nonumber\\ && 
        +\Biggl(
                \frac{28}{27}
+\frac{56}{27} \textcolor{blue}{N_F}\Biggr) \textcolor{blue}{T_F^2}
\Biggr)
\Biggr] \ln^4(x)
+ \Biggl[
\frac{112}{81} \textcolor{blue}{C_A} (1+2 \textcolor{blue}{N_F}) \textcolor{blue}{T_F^2}
+\textcolor{blue}{C_F} \Biggl(
        \Biggl(
                \frac{1016}{81}+\frac{496}{81} \textcolor{blue}{N_F}\Biggr) \textcolor{blue}{T_F^2}
\nonumber\\ &&            
        +\textcolor{blue}{C_A T_F} \Biggl(
     -\frac{10372}{81}
                -\frac{328 \zeta_2}{9}
        \Biggr)
\Biggr)
+\textcolor{blue}{C_F^2 T_F} \Biggl[
        -\frac{2}{3}
        +\frac{4 \zeta_2}{9}
\Biggr]
+\textcolor{blue}{C_A^2 T_F} \Biggl[
        -\frac{1672}{81}
        +8 \zeta_2
\Biggr]
\Biggr] \ln^3(x)
\nonumber\\ && 
+ \Biggl[
\frac{8}{81} \textcolor{blue}{C_A} (155+118 \textcolor{blue}{N_F}) \textcolor{blue}{T_F^2}
+\textcolor{blue}{C_F} \Biggl[
        \textcolor{blue}{T_F^2} \Biggl(
                -\frac{32}{81}
                +\textcolor{blue}{N_F} \Biggl(
                        \frac{3872}{81}
                        -\frac{16 \zeta_2}{9}
                \Biggr)
                +\frac{232 \zeta_2}{9}
        \Biggr)
\nonumber\\ && 
        +\textcolor{blue}{C_A T_F}  \Biggl(
                -\frac{70304}{81}
                -\frac{680 \zeta_2}{9}
                +\frac{80 \zeta_3}{3}
        \Biggr)
\Biggr]
+\textcolor{blue}{C_A^2 T_F} \Biggl[
        \frac{4684}{81}
        +\frac{20 \zeta_2}{3}
\Biggr]
+\textcolor{blue}{C_F^2 T_F} \Biggl[
        56
\nonumber\\ &&         
+\frac{8 \zeta_2}{3}
        -40 \zeta_3
\Biggr]
\Biggr] \ln^2(x)
+ \Biggl[
\textcolor{blue}{C_F} \Biggl[
        \textcolor{blue}{T_F^2} \Biggl(
                \frac{140992}{243}
                +\textcolor{blue}{N_F} \Biggl(
                        \frac{182528}{243}
                        -\frac{400 \zeta_2}{27}
                        -\frac{640 \zeta_3}{9}
                \Biggr)
\nonumber\\ &&                
 -\frac{728}{27} \zeta_2
                -\frac{224}{9} \zeta_3
        \Biggr)
        +\textcolor{blue}{C_A T_F} \Biggl(
                -\frac{514952}{243}
                +\frac{152 \zeta_4}{3}
                -\frac{21140 \zeta_2}{27}
                -\frac{2576 \zeta_3}{9}
        \Biggr)
\Biggr]
\nonumber\\ && 
+\textcolor{blue}{C_A T_F^2}  \Biggl[
        \frac{184}{27}
        +\textcolor{blue}{N_F} \Biggl(
                \frac{656}{27}
                -\frac{32 \zeta_2}{27}
        \Biggr)
        +\frac{464 \zeta_2}{27}
\Biggr]
+\textcolor{blue}{C_A^2 T_F} \Biggl[
        -\frac{42476}{81}
        -92 \zeta_4
        +\frac{4504 \zeta_2}{27}
\nonumber\\ &&  
       +\frac{64 \zeta_3}{3}
\Biggr]
+\textcolor{blue}{C_F^2 T_F} \Biggl[
        -\frac{1036}{3}
        -\frac{976 \zeta_4}{3}
        -\frac{58 \zeta_2}{3}
        +\frac{416 \zeta_3}{3}
\Biggr]
\Biggr] \ln(x),
\end{eqnarray}
and 
\begin{eqnarray}
a_{gg,Q}^{(3),x \rightarrow 1}(x) &\propto& 
a_{gg,Q,\delta}^{(3)} \delta(1-x)
+ a_{gg,Q,\rm plus}^{(3)}(x)
+ \Biggl[
        -\frac{32}{27} \textcolor{blue}{C_A T_F^2} (17+12 \textcolor{blue}{N_F}) 
        +\textcolor{blue}{C_A C_F T_F} \Biggl(
                56
                -\frac{32 \zeta_2}{3}
        \Biggr)
\nonumber\\ && 
        +\textcolor{blue}{C_A^2 T_F} \Biggl(
                \frac{9238}{81}
                -\frac{104 \zeta_2}{9}
                +16 \zeta_3
        \Biggr)
\Biggr] \ln(1-x)
+\Biggl[
        -\frac{8}{27} \textcolor{blue}{C_A T_F^2} (7+8 \textcolor{blue}{N_F}) 
\nonumber\\ &&       
 +\textcolor{blue}{C_A^2 T_F} \Biggl(
                \frac{314}{27}
                -\frac{4 \zeta_2}{3}
        \Biggr)
\Biggr] \ln^2(1-x)
+\frac{32}{27} \textcolor{blue}{C_A^2 T_F} \ln^3(1-x).
\end{eqnarray}
In the polarized case one has
\begin{eqnarray}
\Delta a_{gg,Q}^{x \rightarrow 0}(x) &\propto& 
\Biggl[
        -\frac{4}{15} \textcolor{blue}{C_A^2} \textcolor{blue}{T_F}
        +\frac{12}{5} \textcolor{blue}{C_A} \textcolor{blue}{C_F} \textcolor{blue}{T_F}
        +\frac{2}{15} \textcolor{blue}{C_F^2} \textcolor{blue}{T_F}
\Biggr] \ln^5(x)
+\Biggl[
        \frac{34}{27} \textcolor{blue}{C_A^2} \textcolor{blue}{T_F}
        +\frac{14}{9} \textcolor{blue}{C_F^2} \textcolor{blue}{T_F}
        \nonumber \\ &&
        +\textcolor{blue}{C_F} \Biggl(
                \frac{760}{27} \textcolor{blue}{C_A} \textcolor{blue}{T_F}
                + \textcolor{blue}{T_F^2}
                \biggl(
                        \frac{28}{27}
                        +\frac{56 \textcolor{blue}{N_F}}{27}
                \biggr)
        \Biggr)
\Biggr] \ln^4(x)
+\Biggl[
        \frac{112}{81} \textcolor{blue}{C_A} (1+2 \textcolor{blue}{N_F}) \textcolor{blue}{T_F^2}
        \nonumber \\ &&
        +\textcolor{blue}{C_F} 
        \Biggl(
                \textcolor{blue}{T_F^2} 
                \biggl(
                        \frac{968}{81}
                        +\frac{1552}{81} \textcolor{blue}{N_F}
                \biggr) 
                +\textcolor{blue}{C_A} \textcolor{blue}{T_F} 
                \biggl(
                        \frac{13484}{81}
                        -\frac{184 \zeta_2}{9}
                \biggr)
        \Biggr)
        +\textcolor{blue}{C_F^2} \textcolor{blue}{T_F} 
        \Biggl(
                -\frac{70}{9}
                +\frac{4 \zeta_2}{9}
        \Biggr)
        \nonumber \\ &&
        +\textcolor{blue}{C_A^2} \textcolor{blue}{T_F} 
        \Biggl(
                \frac{2848}{81}
                +8 \zeta_2
        \Biggr)
\Biggr] \ln^3(x)
+\Biggl[
        \frac{16}{81} \textcolor{blue}{C_A} (85+146 \textcolor{blue}{N_F}) \textcolor{blue}{T_F^2}
        +\textcolor{blue}{C_F} 
        \Biggl(
                \textcolor{blue}{T_F^2} 
                \biggl(
                        \frac{2680}{81}
                        \nonumber \\ &&
                        +\textcolor{blue}{N_F} 
                        \biggl(
                                \frac{6704}{81}
                                -\frac{16 \zeta_2}{9}
                        \biggr)
                        +\frac{232 \zeta_2}{9}
                \biggr)
                +\textcolor{blue}{C_A} \textcolor{blue}{T_F} 
                \biggl(
                        \frac{44476}{81}
                        -\frac{1184 \zeta_2}{9}
                        -\frac{544 \zeta_3}{3}
                \biggr)
        \Biggr)
        \nonumber \\ &&
        +\textcolor{blue}{C_F^2} \textcolor{blue}{T_F} 
        \Biggl(
                -\frac{358}{3}
                -\frac{412 \zeta_2}{3}
                +\frac{8 \zeta_3}{3}
        \Biggr)
        +\textcolor{blue}{C_A^2} \textcolor{blue}{T_F} 
        \Biggl(
                \frac{2588}{27}
                +\frac{244 \zeta_2}{9}
                +112 \zeta_3
        \Biggr)
\Biggr] \ln^2(x)
\nonumber \\ &&
+ \Biggr[
        \textcolor{blue}{C_F} 
        \Biggl(
                \textcolor{blue}{T_F^2} 
                \biggl(
                        \frac{53920}{243}
                        +\textcolor{blue}{N_F} \biggl(
                                \frac{121280}{243}
                                -\frac{304 \zeta_2}{27}
                                -\frac{640 \zeta_3}{9}
                        \biggr)
                        +\frac{2776}{27} \zeta_2
                        -\frac{224}{9} \zeta_3
                \biggr)
                \nonumber \\ &&
                +\textcolor{blue}{C_A} \textcolor{blue}{T_F} 
                \biggl(
                        -\frac{638672}{243}
                        +\frac{4400 \zeta_4}{3}
                        -\frac{16484 \zeta_2}{27}
                        +\frac{5968 \zeta_3}{9}
                \biggr)
        \biggr)
        +\textcolor{blue}{C_A} \textcolor{blue}{T_F^2} 
        \Biggl(
                \frac{4880}{81}
                \nonumber \\ &&
                +\textcolor{blue}{N_F} 
                \biggl(
                        \frac{9760}{81}
                        -\frac{32 \zeta_2}{27}
                \biggr)
                +\frac{464 \zeta_2}{27}
        \Biggr)
        +\textcolor{blue}{C_F^2} \textcolor{blue}{T_F} 
        \Biggr(
                -\frac{2152}{3}
                -\frac{1744 \zeta_4}{3}
                -\frac{530 \zeta_2}{3}
                -\frac{3136 \zeta_3}{3}
        \Biggl)
        \nonumber \\ &&
        +\textcolor{blue}{C_A^2} \textcolor{blue}{T_F} 
        \Biggl(
                \frac{964}{9}
                -\frac{452 \zeta_4}{3}
                +\frac{2240 \zeta_2}{27}
                +\frac{1904 \zeta_3}{9}
        \Biggr)
\Biggr] \ln(x),
\end{eqnarray}
and 
\begin{eqnarray}
\Delta a_{gg,Q}^{x \rightarrow 1}(x) =
a_{gg,Q}^{x \rightarrow 1}(x)
\end{eqnarray}
for all terms up to $\propto \ln(1-x)$.
There are no predictions for these limits in the literature. As will be shown in Section~\ref{sec:5}, 
both the 
formally most leading small $x$
and large $x$ contributions are insufficient approximations to $(\Delta) a_{gg,Q}(x)$, as expected from other 
results
\cite{Ablinger:2014nga,SMX}. This also applies to the leading large $x$ result due to its limited reach.
\section{Numerical Results}
\label{sec:5}

\vspace{1mm}
\noindent
For the numerical representations one needs to calculate the $\GA$--functions in a new way, while for 
the harmonic polylogarithms different numerical codes exist, 
cf.~e.g.~Refs.~\cite{Gehrmann:2001pz,Vollinga:2004sn,Ablinger:2018sat}. One way to 
proceed in the present case would be to 
rationalize the letters of the $\GA$--functions by a formalism available in the package {\tt HarmonicSums} leading 
to the cyclotomic letters \cite{Ablinger:2011te}
\begin{eqnarray} 
\Biggl\{
\frac{1}{1+t^2}, \frac{t}{1+t^2} \Biggr\}. 
\end{eqnarray}
The corresponding functions can also be decomposed into letters belonging to generalized harmonic polylogarithms 
\cite{Moch:2001zr,Ablinger:2013cf},
\begin{eqnarray} 
\Biggl\{\frac{1}{t}, \frac{1}{1-t}, \frac{1}{1+t},
\frac{1}{1 + i t}, \frac{1}{1 - i t} \Biggr\}. 
\end{eqnarray}
Iterated integrals over the latter letters can be calculated numerically using the H\"older convolution 
\cite{Borwein:1999js}\footnote{This is a particular convolution in which the H\"older condition \cite{NASM} and  
H\"older mean \cite{HOELDER} are used.}
implemented in Ref.~\cite{Vollinga:2004sn}.

Since the above representation is still somewhat slow, we have alternatively expanded the final expressions for
$a_{gg,Q}^{(3)}(x)$ and
$\Delta a_{gg,Q}^{(3)}(x)$ in series around $x=0$ and $x=1$ analytically to 50 terms. Both 
representations are matched
in the middle of the $x$ range $]0,1]$ at an accuracy of $5 \cdot 10^{-15}$. The corresponding routines are found in an
attachment to the present paper. Since our expansions are fully analytic, they can be extended to an even higher 
accuracy if needed.

\begin{figure}[H]   
\centering
\includegraphics[width=0.7\textwidth]{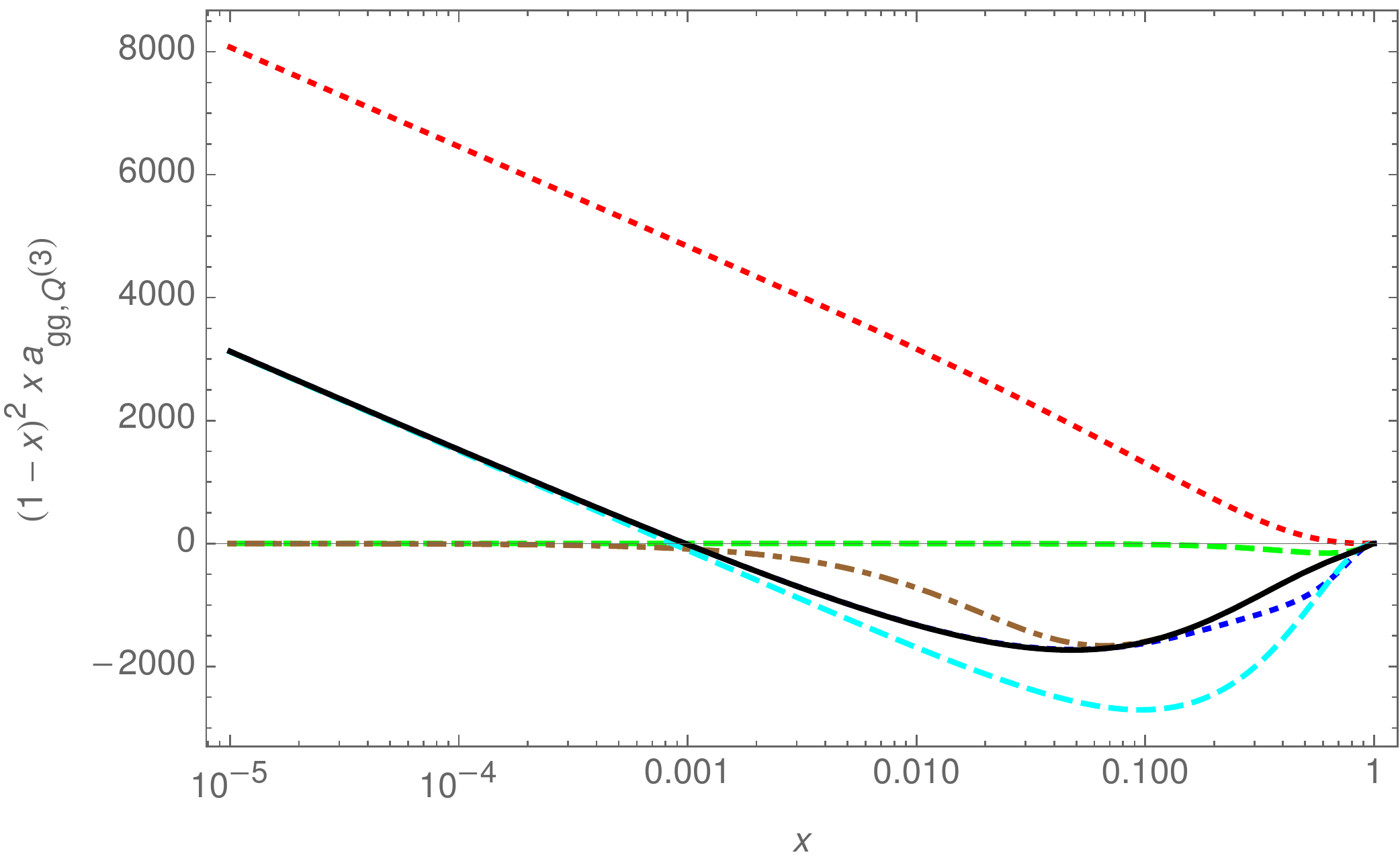}
\caption{\sf 
The non--$N_F$ terms of $a_{gg,Q}^{(3)}(N)$ (rescaled) as a function of $x$. 
Full line (black): complete result;
upper dotted line (red): term $\propto \ln(x)/x$;
lower dashed line (cyan): small $x$ terms $\propto 1/x$;
lower dotted line (blue): small $x$ terms including all $\ln(x)$ terms
up to the constant term; 
upper dashed line (green): large $x$ contribution up to the constant 
term;
dash-dotted line (brown): complete large $x$ contribution.}
\label{fig:1} 
\end{figure}
\begin{figure}[H]   
\centering
\includegraphics[width=0.7\textwidth]{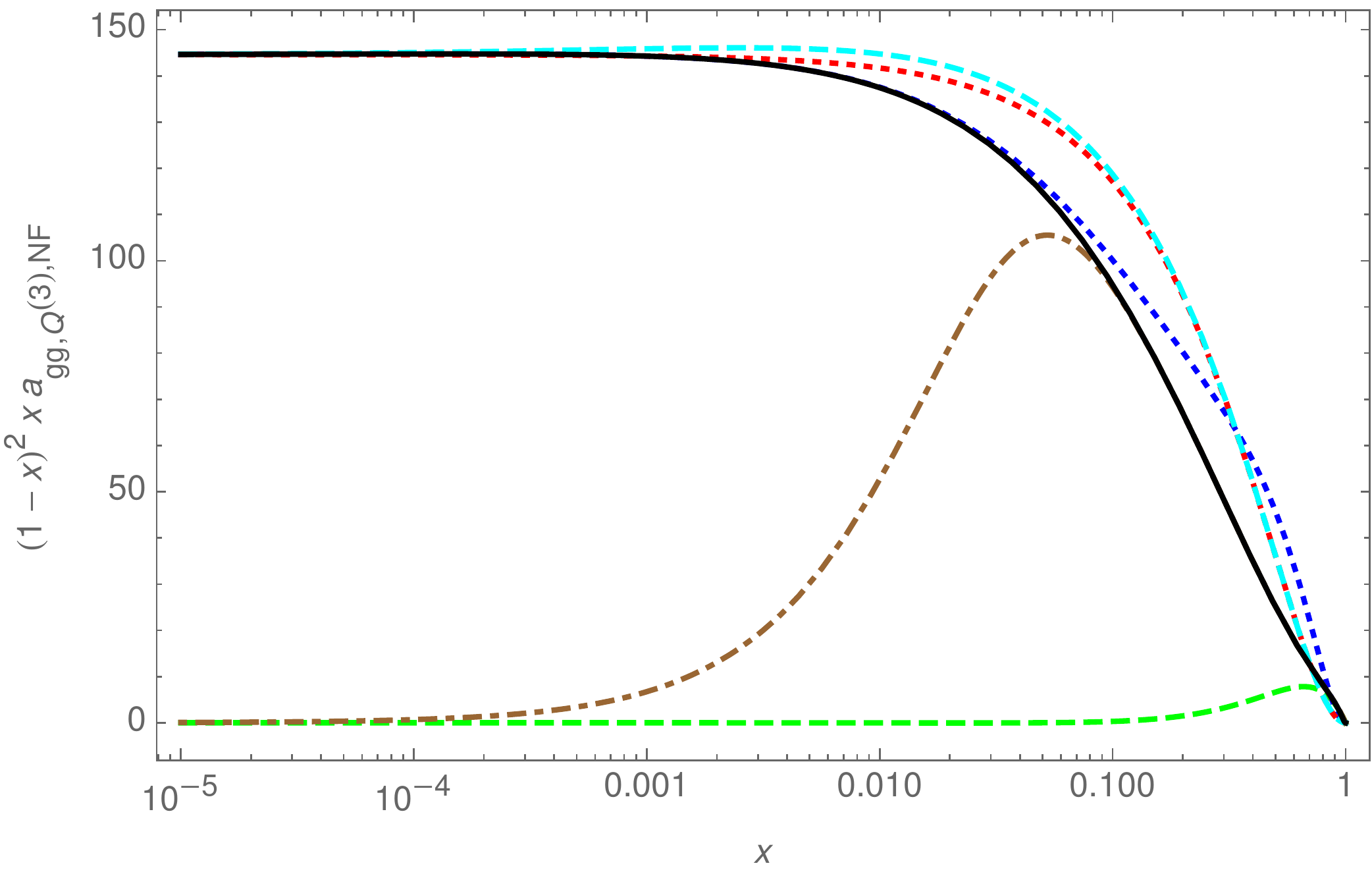}
\caption{\sf 
The $N_F$ terms of $a_{gg,Q}^{(3)}(N)$ (rescaled) as a function of $x$. 
Full line (black): complete result; 
upper dotted line (red): term $\propto \ln(x)/x$;
upper dashed line (cyan): small $x$ terms $\propto 1/x$;
lower dotted line (blue): small $x$ terms including all $\ln(x)$ terms
up the constant term;
lower dashed line (green): large $x$ contribution
up to the constant term;
dash-dotted line (brown): full large $x$ contribution.}
\label{fig:2} 
\end{figure}

We split the discussion of $(\Delta) a_{gg,Q}^{(3)}(x)$ into the terms free of $N_F$ and the linear $N_F$ term, since
the parameter $N_F$ is arbitrary. In the unpolarized case
the corresponding results are given in Figures~\ref{fig:1} and \ref{fig:2}, where 
we have used the rescaling factor $x (1-x)^2$ for better visibility.

The formally leading small $x$ term $\propto \ln(x)/x$ of the $N_F$--independent part deviates from the 
complete result everywhere, cf.~Figure~\ref{fig:1}.\footnote{This has also been observed before for other 
OMEs in Refs.~\cite{Blumlein:1998mg,Ablinger:2014nga}.} Adding the $1/x$ allows to describe the region 
$x < 0.001$. Adding also the remaining terms of the expansion around $x=0$ up to the constant term one 
covers the region $x < 0.1$. The large $x$ singular and logarithmic contributions $\propto \ln^l(1-x)$ 
up to the constant term stop to agree with the complete result at $x \sim 0.7$. 
The matching between the small $x$ and large $x$ expansions, taking into account 50 expansion
terms in both cases, can be performed with a
relative accuracy of $O(10^{-14})$ and finally allows to describe the whole region $x \in ]0,1[$.

Both the $N_F$--dependent terms in the unpolarized and polarized cases are smaller than the corresponding 
$N_F$--independent parts. The unpolarized $N_F$--dependent part of $a_{gg,Q}^{(3)}(x)$ behaves like 
$1/x$ in the small $x$ limit, as shown in Figure~\ref{fig:2}. The expansions around $x=0$ and $x=1$ to 
$O(x^{50})$ can be matched at $x \sim 0.5$ at a relative accuracy of $O(10^{-14})$. Just keeping 
the divergent and $\ln^l(1-x)$ up to the constant term in the large $x$ region yields
agreement with the complete result for $x \gsim 0.7$. The leading small $x$ term, which in view of the 
complete expression for $a_{gg,Q}^{(3)}(x)$ is subleading, starts to deviate for values $x > 0.002$ 
from the complete expression. Accounting in addition for the next term $\propto \ln^4(x)$ the range is 
even only $x < 0.0002$. Including all small $x$ logarithmic terms up to the constant term cover the 
region $x < 0.02$.

In the polarized case, we rescale $\Delta a_{gg,Q}^{(3)}(N)$ by the factor $\sqrt{x} (1-x)^2$ for 
better visibility, which is different from the one in the unpolarized case, cf.~Figures~3,4.
\begin{figure}[H] 
\centering 
\includegraphics[width=0.7\textwidth]{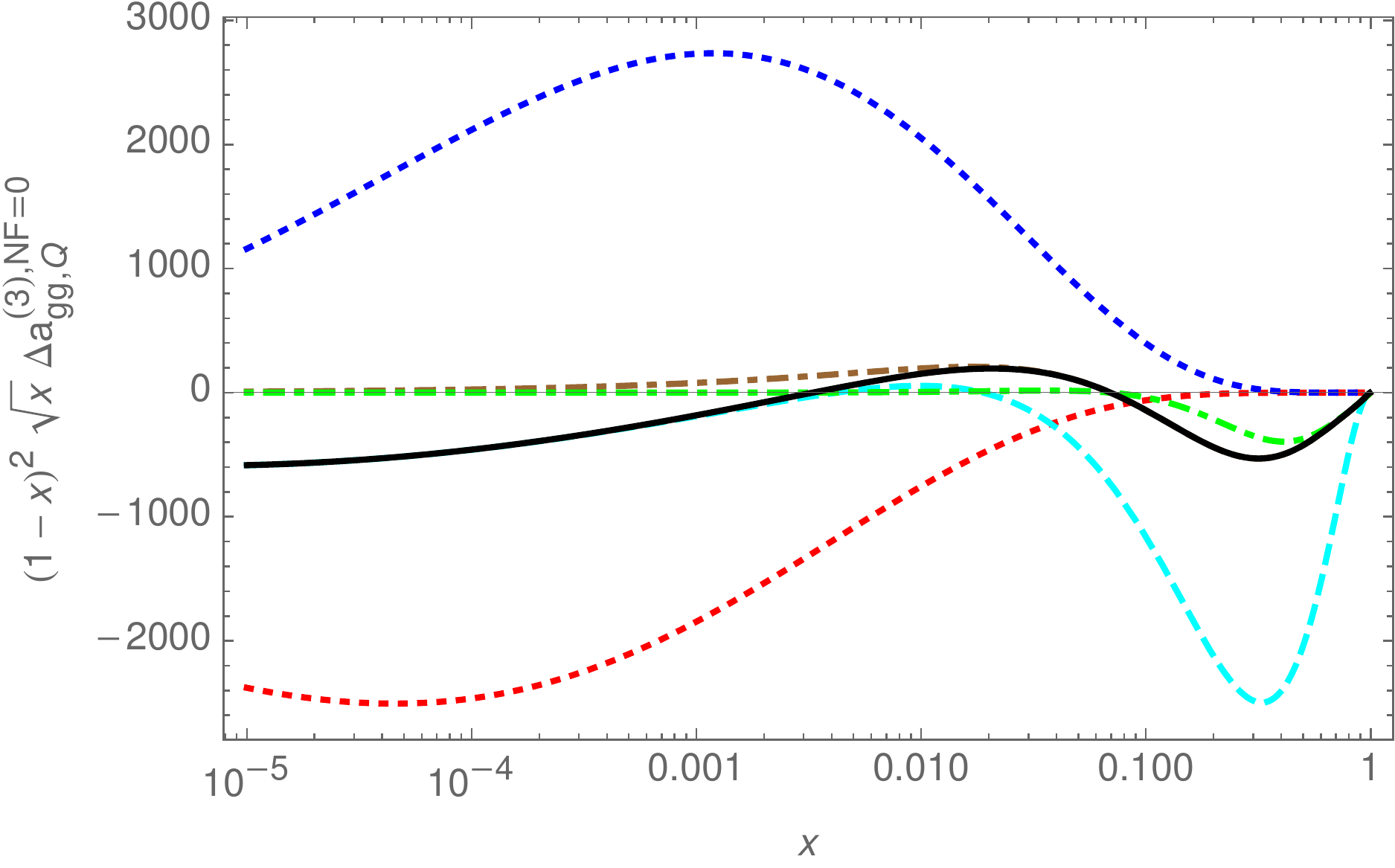} 
\caption{\sf The 
non--$N_F$ terms of $\Delta a_{gg,Q}^{(3)}(N)$ (rescaled) as a function of $x$. 
Full line (black): complete result; 
lower dotted line (red): term $\ln^5(x)$; 
upper dotted line (blue): small $x$ terms $\propto \ln^5(x)$ and $\ln^4(x)$;
upper dashed line (cyan): small $x$ terms including all $\ln(x)$ terms up to the constant term; 
lower dash-dotted line (green): large $x$ contribution up to the constant term; 
dash-dotted line (brown): full large $x$ contribution.} 
\label{fig:3} 
\end{figure}
In the $N_F$--independent case, the leading small $x$ 
result $\propto \ln^5(x)$ deviates from the complete result significantly, which is also the case adding one subleading 
term. Taking all small $x$ logarithms into account 
one covers the region $x < 0.002$. $x \sim 0.5$ is 
the ideal matching point of the 50--term  small $x$ and large $x$ expansions at a relative accuracy of $ \lsim 
10^{-14}$. The 
large $x$ expansion of the leading terms $\propto 1/(1-x)$ and $\ln^k(1-x)$ down to the constant term range to $0.7 < x$, 
while the 50--term 
expansion covers $0.02 < x$ and does then deviate from the complete result.
\begin{figure}[H]   
\centering
\includegraphics[width=0.7\textwidth]{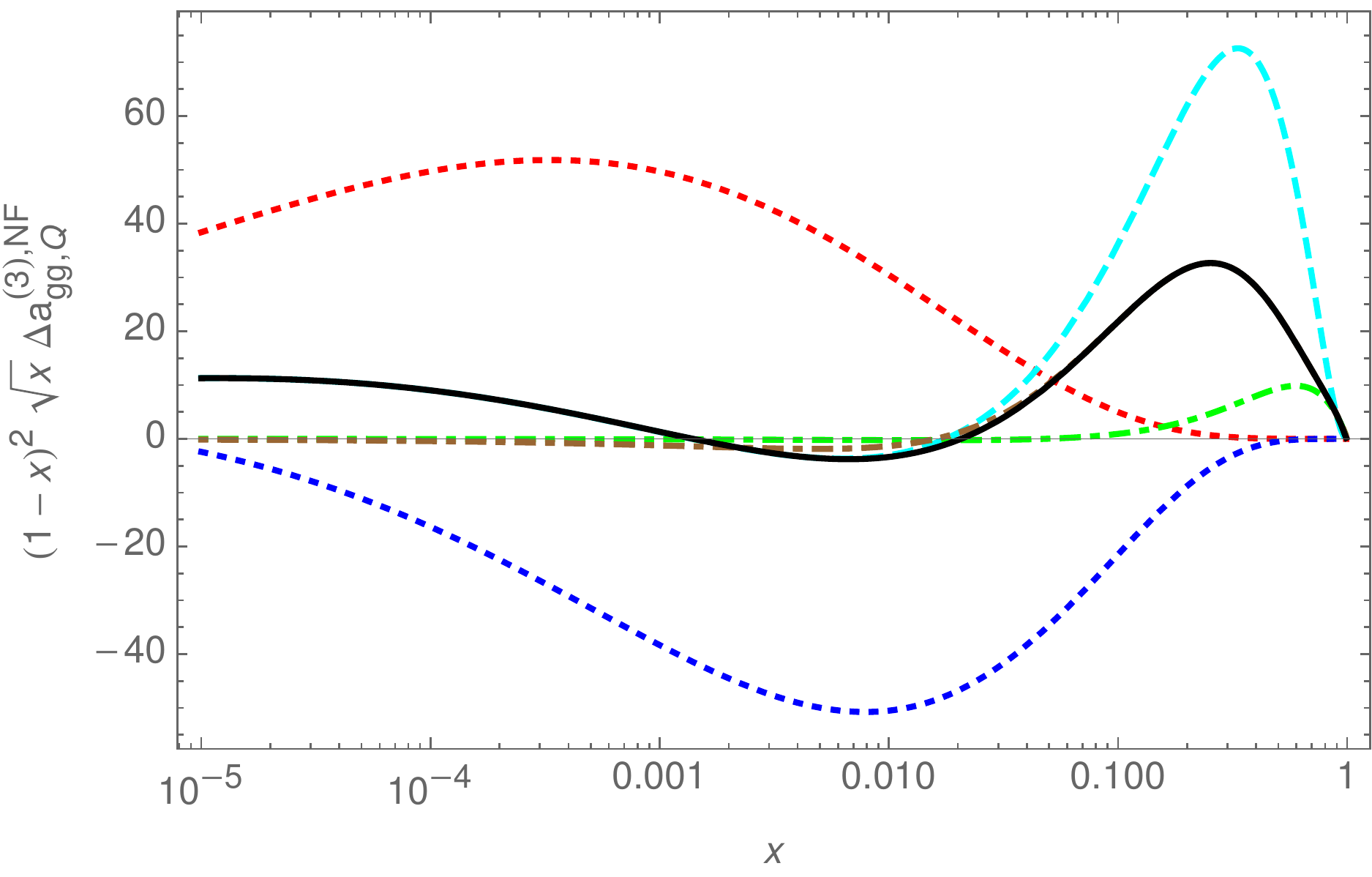}
\caption{\sf The $\Delta a_{gg,Q}^{(3)}(N)$  term  $\propto N_F$ (rescaled) as a 
function of $x$.
Full line (black): complete result;
upper dotted line (red): term $\propto \ln^4(x)$;
lower dotted line (blue): small $x$ terms $\propto \ln^4(x)$ and $\ln^3(x)$;
upper dashed line (cyan): small $x$ terms including all $\ln(x)$ terms up to the constant term;
lower dashed line (green): large $x$ contribution up to the constant term;
dash-dotted line (brown): full large $x$ contribution.}
\label{fig:4} 
\end{figure}

A similar behavior of the different curves in the $N_F$--dependent part in the polarized case to 
the $N_F$--independent part is observed in Figure~\ref{fig:4}. Again, the leading small $x$ term 
$\propto \ln^4(x)$ does not describe the complete expression anywhere, 
as well as taking into account the next logarithmic order,
while the complete set of the small 
$x$ logarithms $\ln^k(x)$ covers the range $x  < 0.008$. The 50--term large $x$ expansion describes the 
complete result for $x \gsim 0.3$, while the leading large $x$ singular contributions agree with it only 
for $x > 0.8$. 

For the numerical representation used we present the regular parts expanded in terms of power series 
to $O(x^{50})$ around $x=0$ and $x=1$ in a {\tt Mathematica} .m file attached to this paper. 
Since the corresponding 
expansions are performed analytically, it can be extended to work at an even higher precision, if 
needed.
\section{Conclusions}
\label{sec:6}

\vspace*{1mm}
\noindent
We have calculated the single heavy quark mass OMEs $A_{gg,Q}^{(3)}$ and $\Delta A_{gg,Q}^{(3)}$ in the 
unpolarized and polarized cases, and the previously known logarithmic contributions are now supplemented 
by the functions $a_{gg,Q}^{(3)}$ and $\Delta a_{gg,Q}^{(3)}$. In Mellin $N$--space, they contain besides 
nested harmonic sums also nested finite binomial sums. In momentum fraction $x$--space, theses 
quantities are represented by iterated integrals, the $\GA$--functions, over an alphabet containing also 
square root--valued letters. We provided recursions and asymptotic expansions in 
Mellin $N$--space for 
these quantities, allowing also the analytic continuation of $a_{gg,Q}^{(3)}$ and 
$\Delta 
a_{gg,Q}^{(3)}$ from the even (odd) moments to $N \in \mathbb{C}$. In $x$--space it is convenient to
use highly accurate Taylor series expansions. Large expressions both in $N$-- and $x$--space are 
provided in computer--readable ancillary files to the present paper. We find that leading small and 
large $x$ expansions of $a_{gg,Q}^{(3)}$ and $\Delta  a_{gg,Q}^{(3)}$ are of limited use, since they 
cover rather small regions in $x$ only.

The present results complete the transition matrix elements in the single-- and double--mass
variable flavor number scheme for the gluon distributions in the unpolarized and polarized cases at 
three--loop order.

\appendix
\section{\boldmath The contributing polynomials in $N$-space}
\label{sec:A}

\vspace*{1mm}
\noindent
The polynomials $P_i$ occurring in the expressions of $(\Delta) a_{gg,Q}^{(3)}$ in 
Section~\ref{sec:3} 
are given by
\begin{eqnarray}
    P_1 &=& -63 N^4-126 N^3-431 N^2-368 N+736 
    , \\ 
   P_2 &=& N^4+2 N^3-61 N^2-62 N-8 
    , \\ 
   P_3 &=& N^4+2 N^3-23 N^2-24 N-4 
    , \\ 
   P_4 &=& N^4+2 N^3-16 N^2-17 N+3 
    , \\ 
   P_5 &=& N^4+2 N^3-N^2-2 N+3 
    , \\ 
   P_6 &=& N^4+2 N^3+3 N^2+2 N-2 
    , \\ 
   P_7 &=& N^4+2 N^3+17 N^2+16 N+3 
    , \\ 
   P_8 &=& N^4+2 N^3+25 N^2+24 N-4 
    , \\ 
   P_9 &=& N^4+2 N^3+63 N^2+62 N-8 
    , \\ 
   P_{10} &=& N^4+2 N^3+75 N^2+74 N-8 
    , \\ 
   P_{11} &=& 2 N^4+4 N^3-3 N^2-5 N+6 
    , \\ 
   P_{12} &=& 2 N^4+4 N^3-N^2-3 N+6 
    , \\ 
   P_{13} &=& 2 N^4+4 N^3+7 N^2+5 N+6 
    , \\ 
   P_{14} &=& 2 N^4+6 N^3+N^2-3 N-12 
    , \\ 
   P_{15} &=& 3 N^4-12 N^3-37 N^2-10 N+8 
    , \\ 
   P_{16} &=& 3 N^4+6 N^3-11 N^2-14 N+9 
    , \\ 
   P_{17} &=& 3 N^4+6 N^3-8 N^2-11 N+9 
    , \\ 
   P_{18} &=& 3 N^4+6 N^3-4 N^2-7 N+9 
    , \\ 
   P_{19} &=& 5 N^4-8 N^3-23 N^2-22 N-8 
    , \\ 
   P_{20} &=& 5 N^4+4 N^3+N^2-10 N-8 
    , \\ 
   P_{21} &=& 5 N^4+10 N^3-65 N^2-70 N-16 
    , \\ 
   P_{22} &=& 5 N^4+10 N^3-29 N^2-34 N-16 
    , \\ 
   P_{23} &=& 5 N^4+10 N^3-N^2-6 N-16 
    , \\ 
   P_{24} &=& 5 N^4+10 N^3+11 N^2+6 N-16 
    , \\ 
   P_{25} &=& 5 N^4+10 N^3+25 N^2+20 N+36 
    , \\ 
   P_{26} &=& 5 N^4+10 N^3+39 N^2+34 N-16 
    , \\ 
   P_{27} &=& 5 N^4+10 N^3+75 N^2+70 N-16 
    , \\ 
   P_{28} &=& 6 N^4+12 N^3-7 N^2-13 N+18 
    , \\ 
   P_{29} &=& 6 N^4+12 N^3-N^2-7 N+18 
    , \\ 
   P_{30} &=& 6 N^4+12 N^3+N^2-5 N+18 
    , \\ 
   P_{31} &=& 6 N^4+12 N^3+5 N^2-N+18 
    , \\ 
   P_{32} &=& 6 N^4+12 N^3+7 N^2+N+18 
    , \\ 
   P_{33} &=& 9 N^4+18 N^3+113 N^2+104 N-24 
    , \\ 
   P_{34} &=& 11 N^4+4 N^3-59 N^2-88 N-12 
    , \\ 
   P_{35} &=& 11 N^4+22 N^3-5 N^2-16 N+68 
    , \\ 
   P_{36} &=& 11 N^4+22 N^3+27 N^2+16 N+68 
    , \\ 
   P_{37} &=& 13 N^4+62 N^3+63 N^2+14 N+88 
    , \\ 
   P_{38} &=& 18 N^4+36 N^3+19 N^2+N+54 
    , \\ 
   P_{39} &=& 18 N^4+36 N^3+41 N^2+23 N+54 
    , \\ 
   P_{40} &=& 29 N^4+58 N^3+9 N^2-20 N+20 
    , \\ 
   P_{41} &=& 29 N^4+58 N^3+49 N^2+20 N+20 
    , \\ 
   P_{42} &=& 35 N^4+64 N^3+28 N^2-13 N-6 
    , \\ 
   P_{43} &=& 36 N^4+72 N^3+103 N^2+67 N+108 
    , \\ 
   P_{44} &=& 40 N^4+72 N^3+5 N^2-27 N+48 
    , \\ 
   P_{45} &=& 99 N^4+198 N^3+463 N^2+364 N-84 
    , \\ 
   P_{46} &=& 130 N^4+269 N^3+142 N^2-24 N-18 
    , \\ 
   P_{47} &=& 131 N^4+454 N^3+471 N^2+148 N+236 
    , \\ 
   P_{48} &=& 220 N^4+330 N^3-25 N^2-198 N+249 
    , \\ 
   P_{49} &=& 1287 N^4+3726 N^3-3047 N^2-7214 N-2624 
    , \\ 
   P_{50} &=& N^5-12 N^4-11 N^3-54 N^2-52 N-8 
    , \\ 
   P_{51} &=& N^5+46 N^4+305 N^3+484 N^2+156 N-16 
    , \\ 
   P_{52} &=& 3 N^5-7 N^4-25 N^3-269 N^2-254 N-72 
    , \\ 
   P_{53} &=& 3 N^5-7 N^4-25 N^3+259 N^2+274 N-72 
    , \\ 
   P_{54} &=& 3 N^5-5 N^4-21 N^3-79 N^2-66 N-24 
    , \\ 
   P_{55} &=& 3 N^5-5 N^4-21 N^3+89 N^2+102 N-24 
    , \\ 
   P_{56} &=& 3 N^5-4 N^4-19 N^3-146 N^2-134 N-60 
    , \\ 
   P_{57} &=& 3 N^5-4 N^4-19 N^3+142 N^2+154 N-60 
    , \\ 
   P_{58} &=& 3 N^5-N^4-13 N^3-47 N^2-38 N-48 
    , \\ 
   P_{59} &=& 3 N^5-N^4-13 N^3+49 N^2+58 N-48 
    , \\ 
   P_{60} &=& 3 N^5+7 N^4+8 N^3+56 N^2-32 N+48 
    , \\ 
   P_{61} &=& 3 N^5+16 N^4+21 N^3-190 N^2-198 N-12 
    , \\ 
   P_{62} &=& 3 N^5+16 N^4+21 N^3+194 N^2+186 N-12 
    , \\ 
   P_{63} &=& 3 N^5+25 N^4+39 N^3-253 N^2-270 N+24 
    , \\ 
   P_{64} &=& 3 N^5+25 N^4+39 N^3+275 N^2+258 N+24 
    , \\ 
   P_{65} &=& 4 N^5+15 N^4+102 N^3+223 N^2-148 N-340 
    , \\ 
   P_{66} &=& 8 N^5+16 N^4-25 N^3-40 N^2-69 N+62 
    , \\ 
   P_{67} &=& 9 N^5+18 N^4-19 N^3-30 N^2-20 N+18 
    , \\ 
   P_{68} &=& 9 N^5+45 N^4+62 N^3-6 N^2-50 N-36 
    , \\ 
   P_{69} &=& 13 N^5+36 N^4+55 N^3+60 N^2+116 N-176 
    , \\ 
   P_{70} &=& 16 N^5+41 N^4+2 N^3+47 N^2+70 N+32 
    , \\ 
   P_{71} &=& 29 N^5+107 N^4+160 N^3+142 N^2-30 N-180 
    , \\ 
   P_{72} &=& 70 N^5+95 N^4-223 N^3-751 N^2-629 N-142 
    , \\ 
   P_{73} &=& 131 N^5+192 N^4+35 N^3+270 N^2+532 N-472 
    , \\ 
   P_{74} &=& -63 N^6-189 N^5-431 N^4-547 N^3-1714 N^2-1472 N-1472 
    , \\ 
   P_{75} &=& 2 N^6+8 N^5+3 N^4-14 N^3-5 N^2+6 N+24 
    , \\ 
   P_{76} &=& 2 N^6+8 N^5+9 N^4-2 N^3-17 N^2-12 N+36 
    , \\ 
   P_{77} &=& 2 N^6+8 N^5+9 N^4-2 N^3+7 N^2+12 N+36 
    , \\ 
   P_{78} &=& 3 N^6+3 N^5-11 N^4-19 N^3+86 N^2+94 N+60 
    , \\ 
   P_{79} &=& 3 N^6+3 N^5-5 N^4+17 N^3-64 N^2-86 N+60 
    , \\ 
   P_{80} &=& 3 N^6+9 N^5-9 N^4-73 N^3-84 N^2-26 N-36 
    , \\ 
   P_{81} &=& 3 N^6+9 N^5+10 N^4+40 N^3-12 N^2+8 N+32 
    , \\ 
   P_{82} &=& 4 N^6+3 N^5-50 N^4-129 N^3-100 N^2-56 N-24 
    , \\ 
   P_{83} &=& 4 N^6+16 N^5-53 N^4-218 N^3-217 N^2-64 N-44 
    , \\ 
   P_{84} &=& 4 N^6+16 N^5+9 N^4-22 N^3-7 N^2+12 N+36 
    , \\ 
   P_{85} &=& 5 N^6-2 N^5-22 N^4+138 N^3+5 N^2-40 N+12 
    , \\ 
   P_{86} &=& 5 N^6+20 N^5+25 N^4+8 N^3+24 N^2+26 N-36 
    , \\ 
   P_{87} &=& 6 N^6+17 N^5+15 N^4+56 N^3-60 N^2+8 N+48 
    , \\ 
   P_{88} &=& 6 N^6+45 N^5-419 N^4-1287 N^3-583 N^2+246 N-168 
    , \\ 
   P_{89} &=& 7 N^6+11 N^5+55 N^4+69 N^3-54 N^2+40 N+64 
    , \\ 
   P_{90} &=& 8 N^6+27 N^5-64 N^4-309 N^3-334 N^2-104 N+8 
    , \\ 
   P_{91} &=& 9 N^6-15 N^5-89 N^4-177 N^3+36 N^2+28 N-16 
    , \\ 
   P_{92} &=& 9 N^6+9 N^5-53 N^4+47 N^3+44 N^2-104 N-80 
    , \\ 
   P_{93} &=& 9 N^6+27 N^5+161 N^4+277 N^3+358 N^2+224 N+48 
    , \\ 
   P_{94} &=& 9 N^6+60 N^5-259 N^4-900 N^3-632 N^2-42 N+36 
    , \\ 
   P_{95} &=& 15 N^6+45 N^5+13 N^4-13 N^3+80 N^2+4 N-24 
    , \\ 
   P_{96} &=& 15 N^6+60 N^5+43 N^4-76 N^3-112 N^2-38 N-132 
    , \\ 
   P_{97} &=& 15 N^6+60 N^5+43 N^4-76 N^3-64 N^2+10 N-132 
    , \\ 
   P_{98} &=& 17 N^6+33 N^5-27 N^4+59 N^3+130 N^2-44 N-24 
    , \\ 
   P_{99} &=& 27 N^6+81 N^5+148 N^4+161 N^3+253 N^2-390 N-144 
    , \\ 
   P_{100} &=& 29 N^6+78 N^5+71 N^4+90 N^3+206 N^2+138 N+180 
    , \\ 
   P_{101} &=& 30 N^6+90 N^5+79 N^4+8 N^3+23 N^2+70 N+12 
    , \\ 
   P_{102} &=& 33 N^6+99 N^5+88 N^4+47 N^3+68 N^2-29 N-42 
    , \\ 
   P_{103} &=& 37 N^6+18 N^5-15 N^4-323 N^3-174 N^2-36 N-54 
    , \\ 
   P_{104} &=& 38 N^6+96 N^5+233 N^4+426 N^3+35 N^2-216 N+108 
    , \\ 
   P_{105} &=& 38 N^6+108 N^5+151 N^4+106 N^3+21 N^2-28 N-12 
    , \\ 
   P_{106} &=& 40 N^6+112 N^5-3 N^4-166 N^3-301 N^2-210 N-96 
    , \\ 
   P_{107} &=& 44 N^6+123 N^5+386 N^4+543 N^3+520 N^2+248 N+24 
    , \\ 
   P_{108} &=& 99 N^6+297 N^5+631 N^4+767 N^3+1118 N^2+784 N+168 
    , \\ 
   P_{109} &=& 100 N^6+439 N^5+8 N^4-1286 N^3-858 N^2+1519 N-498 
    , \\ 
   P_{110} &=& 199 N^6+645 N^5+505 N^4-81 N^3-1052 N^2-912 N+2712 
    , \\ 
   P_{111} &=& 220 N^6+550 N^5-135 N^4-883 N^3-1621 N^2-1329 N-462 
    , \\ 
   P_{112} &=& 421 N^6+831 N^5+637 N^4+1473 N^3-626 N^2-1872 N+5184 
    , \\ 
   P_{113} &=& 6944 N^6+19536 N^5+17781 N^4+5108 N^3+1791 N^2+576 N-432 
    , \\ 
   P_{114} &=& 3 N^7+3 N^6-21 N^5-31 N^4-64 N^3-122 N^2-104 N+72 
    , \\ 
   P_{115} &=& 3 N^7+41 N^6+335 N^5+471 N^4-302 N^3-740 N^2+552 N+288 
    , \\ 
   P_{116} &=& 4 N^7+8 N^6-85 N^5-112 N^4+59 N^3-46 N^2-172 N+88 
    , \\ 
   P_{117} &=& 5 N^7+10 N^6-15 N^5-42 N^4-4 N^3+86 N^2+32 N+72 
    , \\ 
   P_{118} &=& 5 N^7+23 N^6+63 N^5+189 N^4+148 N^3-28 N^2+144 N+128 
    , \\ 
   P_{119} &=& 6 N^7+3 N^6-17 N^5+843 N^4+1463 N^3+218 N^2-348 N+136 
    , \\ 
   P_{120} &=& 6 N^7+33 N^6-533 N^5-545 N^4+767 N^3+740 N^2-180 N+336 
    , \\ 
   P_{121} &=& 8 N^7-21 N^6-41 N^5+387 N^4+373 N^3-526 N^2-428 N+56 
    , \\ 
   P_{122} &=& 8 N^7+16 N^6-57 N^5-104 N^4+87 N^3-26 N^2-28 N-248 
    , \\ 
   P_{123} &=& 9 N^7+42 N^6-403 N^5-478 N^4+760 N^3+310 N^2-456 N-72 
    , \\ 
   P_{124} &=& 9 N^7+43 N^6+277 N^5+477 N^4-106 N^3-844 N^2+408 N+288 
    , \\ 
   P_{125} &=& 15 N^7-25 N^6-192 N^5+442 N^4+107 N^3+2391 N^2+1030 N+1032 
    , \\ 
   P_{126} &=& 38 N^7+20 N^6+77 N^5+104 N^4-385 N^3-466 N^2+36 N-216 
    , \\ 
   P_{127} &=& 109 N^7+877 N^6+2129 N^5+2215 N^4+908 N^3-620 N^2-878 N-276 
    , \\ 
   P_{128} &=& 199 N^7+247 N^6-785 N^5-1091 N^4+1510 N^3-56 N^2+888 N-5424 
    , \\ 
   P_{129} &=& 421 N^7-11 N^6-881 N^5+775 N^4+172 N^3-3356 N^2+2880 N-10368 
    , \\ 
   P_{130} &=& N^8+4 N^7+2 N^6+64 N^5+173 N^4+292 N^3+256 N^2-72 N-72 
    , \\ 
   P_{131} &=& 4 N^8+25 N^7+32 N^6-152 N^5+678 N^4+359 N^3-882 N^2-112 N+144 
    , \\ 
   P_{132} &=& 4 N^8+45 N^7+44 N^6-248 N^5+1142 N^4+931 N^3-1022 N^2-224 N+192 
    , \\ 
   P_{133} &=& 5 N^8+41 N^7+41 N^6+25 N^5-14 N^4-54 N^3-84 N^2-72 N-16 
    , \\ 
   P_{134} &=& 8 N^8+21 N^7-33 N^6+5 N^5+601 N^4+698 N^3-156 N^2-88 N+96 
    , \\ 
   P_{135} &=& 15 N^8-252 N^6+228 N^5+631 N^4+1780 N^3+3822 N^2+1032 N+744 
    , \\ 
   P_{136} &=& 15 N^8+60 N^7+4 N^6-162 N^5-311 N^4-186 N^3-220 N^2-80 N+48 
    , \\ 
   P_{137} &=& 30 N^8-5 N^7-514 N^6+626 N^5+902 N^4+2735 N^3+5654 N^2+4 N+168 
    , \\ 
   P_{138} &=& 33 N^8+132 N^7+46 N^6-225 N^5-296 N^4-285 N^3-185 N^2-456 N-108 
    , \\ 
   P_{139} &=& 33 N^8+132 N^7+106 N^6-108 N^5-74 N^4+282 N^3+245 N^2+148 N+84 
    , \\ 
   P_{140} &=& 100 N^8+539 N^7+283 N^6-2094 N^5+452 N^4+219 N^3-1495 N^2+712 N
\nonumber\\ &&
+996 
    , \\ 
   P_{141} &=& 205 N^8+856 N^7+3169 N^6+6484 N^5+7310 N^4+4722 N^3+1534 N^2
   \nonumber \\ &&
   +48 N-72 
    , \\ 
   P_{142} &=& 266 N^8+717 N^7-697 N^6-4325 N^5-4481 N^4+560 N^3+1120 N^2
   \nonumber \\ &&
   +1512 N+864 
    , \\ 
   P_{143} &=& 1720 N^8+6898 N^7+6007 N^6-4079 N^5-5207 N^4-335 N^3-252 N^2
   \nonumber \\ &&
   -648 N-1512 
    , \\ 
   P_{144} &=& 6944 N^8+26480 N^7+23321 N^6-15103 N^5-39319 N^4-27001 N^3
   \nonumber \\ &&
   -11178 N^2-2016 N+864 
    , \\ 
   P_{145} &=& 7209 N^8+28836 N^7-39838 N^6-199272 N^5-187779 N^4-51844 N^3
   \nonumber \\ &&
   -3888 N^2+3168 N+6048 
    , \\ 
   P_{146} &=& 96020 N^8+84383 N^7-200790 N^6-241078 N^5-14299 N^4+60396 N^3
   \nonumber \\ &&
   +35730 N^2-12960 N+6480 
    , \\ 
   P_{147} &=& 3 N^9+32 N^8+302 N^7+584 N^6-377 N^5-1224 N^4+1176 N^3+2144 N^2
   \nonumber \\ &&
   -1104 N-768 
    , \\ 
   P_{148} &=& 5 N^9+3 N^8-66 N^7-82 N^6+469 N^5+1099 N^4+2392 N^3+1092 N^2
   \nonumber \\ &&
   +656 N+192 
    , \\ 
   P_{149} &=& 7 N^9-3 N^8-78 N^7-46 N^6+439 N^5+1285 N^4+2112 N^3+1068 N^2
   \nonumber \\ &&
   +592 N+384 
    , \\ 
   P_{150} &=& 27 N^9+36 N^8-1166 N^7-1760 N^6+4331 N^5+88 N^4-10864 N^3
   \nonumber \\ &&
   +2740 N^2+10192 N-1104 
    , \\ 
   P_{151} &=& 30 N^9+109 N^8+121 N^7+939 N^6+2417 N^5+1188 N^4-932 N^3
   \nonumber \\ &&
   -32 N^2-64 N-128 
    , \\ 
   P_{152} &=& 35 N^9-120 N^8-689 N^7+2546 N^6+3317 N^5+7020 N^4+36669 N^3
   \nonumber \\ &&
   +27874 N^2+13468 N-3720 
    , \\ 
   P_{153} &=& 40 N^9-125 N^8-245 N^7+542 N^6+1938 N^5-2977 N^4+9079 N^3
   \nonumber \\ &&
   +9040 N^2+1188 N+720 
    , \\ 
   P_{154} &=& 95 N^9+446 N^8+344 N^7-1122 N^6+5165 N^5+29844 N^4+27308 N^3
   \nonumber \\ &&
   -8512 N^2-896 N+7232 
    , \\ 
   P_{155} &=& 448 N^9+788 N^8-1990 N^7-4453 N^6-2185 N^5+551 N^4+3637 N^3
   \nonumber \\ &&
   +8244 N^2-2988 N-3024 
    , \\ 
   P_{156} &=& 448 N^9+860 N^8-1558 N^7-4381 N^6-5443 N^5-601 N^4+8767 N^3
   \nonumber \\ &&
   +9036 N^2-3348 N-2160 
    , \\ 
   P_{157} &=& N^{10}+37 N^9-10 N^8-634 N^7+81 N^6+5157 N^5+12472 N^4+9408 N^3
   \nonumber \\ &&
   +896 N^2+4272 N+2880 
    , \\ 
   P_{158} &=& 3 N^{10}-29 N^9-62 N^8+538 N^7+251 N^6-4533 N^5-13200 N^4-11384 N^3
   \nonumber \\ &&
   -432 N^2-3408 N-2304 
    , \\ 
   P_{159} &=& 8 N^{10}-51 N^9-96 N^8+508 N^7+458 N^6-1601 N^5-2194 N^4+152 N^3
   \nonumber \\ &&
   -464 N^2-976 N+224 
    , \\ 
   P_{160} &=& 8 N^{10}+27 N^9+4 N^8-176 N^7-322 N^6-75 N^5-690 N^4-672 N^3+1000 N^2
   \nonumber \\ &&
   +368 N-384 
    , \\ 
   P_{161} &=& 15 N^{10}+10 N^9-238 N^8+88 N^7+2647 N^6+9610 N^5+17712 N^4
   \nonumber \\ &&
   +13108 N^3+5128 N^2-1872 N-1440 
    , \\ 
   P_{162} &=& 15 N^{10}+19 N^9-238 N^8-358 N^7+1087 N^6+4483 N^5+10400 N^4
   \nonumber \\ &&
   +9536 N^3+2176 N^2+5328 N+2880 
    , \\ 
   P_{163} &=& 16 N^{10}+68 N^9-13 N^8-656 N^7-1581 N^6-974 N^5+1800 N^4+3412 N^3
   \nonumber \\ &&
   +2008 N^2+336 N-32 
    , \\ 
   P_{164} &=& 16 N^{10}+89 N^9+294 N^8+554 N^7+48 N^6-835 N^5+330 N^4+1776 N^3
   \nonumber \\ &&
   +688 N^2+336 N-32 
    , \\ 
   P_{165} &=& 23 N^{10}+136 N^9-221 N^8+388 N^7+1470 N^6+2206 N^5+2192 N^4+2564 N^3
   \nonumber \\ &&
   +2082 N^2+1008 N+216 
    , \\ 
   P_{166} &=& 25 N^{10}-35 N^9-295 N^8+185 N^7+1615 N^6+897 N^5+5981 N^4+13197 N^3
   \nonumber \\ &&
   +5802 N^2+1068 N+360 
    , \\ 
   P_{167} &=& 30 N^{10}+150 N^9+163 N^8-248 N^7-562 N^6-296 N^5+33 N^4-30 N^3-48 N^2
   \nonumber \\ &&
   +184 N+48 
    , \\ 
   P_{168} &=& 102 N^{10}+309 N^9-238 N^8-1822 N^7+106 N^6+4733 N^5+1294 N^4-3580 N^3
   \nonumber \\ &&
   -3976 N^2-864 N+2496 
    , \\ 
   P_{169} &=& 270 N^{10}-221 N^9-2926 N^8-1180 N^7+5054 N^6+4823 N^5-2578 N^4-2942 N^3
   \nonumber \\ &&
   +15732 N^2+15288 N-3024 
    , \\ 
   P_{170} &=& 1536 N^{10}-2955 N^9-16182 N^8+14924 N^7+34190 N^6-155541 N^5-442072 N^4
   \nonumber \\ &&
   -107436 N^3+390656 N^2-26624 N-227328 
    , \\ 
   P_{171} &=& 2913 N^{10}+14565 N^9+4234 N^8-60374 N^7-54875 N^6+68545 N^5+112000 N^4
   \nonumber \\ &&
   +39280 N^3-10752 N^2-16272 N-6048 
    , \\ 
   P_{172} &=& 7209 N^{10}+36045 N^9-52924 N^8-417598 N^7-794647 N^6-770095 N^5
   \nonumber \\ &&
-388726 N^4
   -63040 N^3-576 N^2-25344 N-12096 
    , \\ 
   P_{173} &=& 12672 N^{10}+42963 N^9+6 N^8-264652 N^7-183166 N^6+673325 N^5+703736 N^4
   \nonumber \\ &&
   -277684 N^3-833920 N^2+133632 N+525312 
    , \\ 
   P_{174} &=& 96020 N^{10}+180403 N^9-293651 N^8-563492 N^7+196513 N^6+478087 N^5
   \nonumber \\ &&
   -194200 N^4-207066 N^3-7470 N^2-38880 N-12960 
    , \\ 
   P_{175} &=& 149796 N^{10}+331992 N^9+2242307 N^8+877052 N^7-6336162 N^6-4554532 N^5
   \nonumber \\ &&
   +1462595 N^4+1113864 N^3+133200 N^2+246240 N-90720 
    , \\ 
   P_{176} &=& 27 N^{11}+189 N^{10}+631 N^9+1356 N^8+2155 N^7+2207 N^6+211 N^5-4984 N^4
   \nonumber \\ &&
   -8400 N^3-5824 N^2-2544 N-576 
    , \\ 
   P_{177} &=& 75 N^{11}-35 N^{10}+624 N^9-7558 N^8+12763 N^7+46561 N^6+91954 N^5
   \nonumber \\ &&
   +198152 N^4+119160 N^3+5280 N^2-4256 N-1920 
    , \\ 
   P_{178} &=& 76 N^{11}+875 N^{10}+3212 N^9+4756 N^8+1408 N^7-5169 N^6-12976 N^5
   \nonumber \\ &&
   -12806 N^4+112 N^3+3392 N^2-1984 N-1632 
    , \\ 
   P_{179} &=& 448 N^{11}+1236 N^{10}-2116 N^9-7857 N^8-1560 N^7+9270 N^6+6398 N^5
   \nonumber \\ &&
   -237 N^4-12098 N^3-18612 N^2+6984 N+7776 
    , \\ 
   P_{180} &=& 475 N^{11}+330 N^{10}-6255 N^9-4360 N^8-6703 N^7+109282 N^6+63439 N^5
   \nonumber \\ &&
   +360220 N^4+1376628 N^3+1002368 N^2+175616 N+154560 
    , \\ 
   P_{181} &=& 502 N^{11}-1112 N^{10}-4284 N^9+6519 N^8+14409 N^7-12978 N^6-17866 N^5
   \nonumber \\ &&
   -12913 N^4-38013 N^3+7524 N^2-6588 N-12960 
    , \\ 
   P_{182} &=& 502 N^{11}-1112 N^{10}-4248 N^9+6609 N^8+13113 N^7-11466 N^6-14842 N^5
   \nonumber \\ &&
   -12427 N^4-51441 N^3+8028 N^2-2700 N-7776 
    , \\ 
   P_{183} &=& 1936 N^{11}+5826 N^{10}-8779 N^9-34974 N^8+5532 N^7+59112 N^6-4333 N^5
   \nonumber \\ &&
   -41196 N^4+21988 N^3+34344 N^2+6336 N-4320 
    , \\ 
   P_{184} &=& 20 N^{12}+125 N^{11}-20 N^{10}-1835 N^9-3590 N^8+10755 N^7+23456 N^6
   \nonumber \\ &&
   +33335 N^5+108782 N^4+67828 N^3-34184 N^2-5952 N+8640 
    , \\ 
   P_{185} &=& 20 N^{12}+225 N^{11}+40 N^{10}-3295 N^9-5210 N^8+18695 N^7+39964 N^6+64435 N^5
   \nonumber \\ &&
   +209418 N^4+196612 N^3+16504 N^2+4032 N+11520 
    , \\ 
   P_{186} &=& 30 N^{12}+25 N^{11}+120 N^{10}-1204 N^9-2904 N^8-8041 N^7-11950 N^6-3528 N^5
   \nonumber \\ &&
   +6536 N^4+4620 N^3-520 N^2-1520 N-480 
    , \\ 
   P_{187} &=& 54 N^{12}-983 N^{11}-2940 N^{10}+8147 N^9+25836 N^8-7266 N^7-51705 N^6
   \nonumber \\ &&
   -1540 N^5+55887 N^4+26122 N^3+300 N^2+792 N-864 
    , \\ 
   P_{188} &=& 84 N^{12}+192 N^{11}-661 N^{10}-2146 N^9+452 N^8+3894 N^7+303 N^6-3234 N^5
   \nonumber \\ &&
   -5930 N^4-4370 N^3+2560 N^2+3864 N-1920 
    , \\ 
   P_{189} &=& 135 N^{12}-90 N^{11}-7255 N^{10}+1000 N^9+67581 N^8-2322 N^7-181501 N^6
   \nonumber \\ &&
   -48456 N^5-137736 N^4-612388 N^3-150184 N^2+174096 N+113760 
    , \\ 
   P_{190} &=& 266 N^{12}+983 N^{11}-1576 N^{10}-9928 N^9-6696 N^8+7669 N^7-954 N^6-5380 N^5
   \nonumber \\ &&
   +16080 N^4+10832 N^3-2656 N^2+8640 N+6912 
    , \\ 
   P_{191} &=& 394 N^{12}-36 N^{11}-4636 N^{10}-2733 N^9+18141 N^8+12348 N^7-27010 N^6
   \nonumber \\ &&
   -32985 N^5-61373 N^4-57162 N^3-18828 N^2-49032 N-20736 
    , \\ 
   P_{192} &=& 2913 N^{12}+17478 N^{11}+6253 N^{10}-121030 N^9-399973 N^8-664606 N^7
   \nonumber \\ &&
   -829641 N^6-867778 N^5-563504 N^4-110240 N^3+67728 N^2
   \nonumber\\ && 
+45504 N+12096 
    , \\ 
   P_{193} &=& 3300 N^{12}+5838 N^{11}-26849 N^{10}-35867 N^9+83580 N^8+69996 N^7-113901 N^6
   \nonumber \\ &&
   +16689 N^5+167318 N^4-21088 N^3-35688 N^2+5904 N+5184 
    , \\ 
   P_{194} &=& 8868 N^{12}+26604 N^{11}-12901 N^{10}-61985 N^9+24562 N^8-16310 N^7-140777 N^6
   \nonumber \\ &&
   -67997 N^5+55040 N^4+63048 N^3+55008 N^2+4968 N-9072 
    , \\ 
   P_{195} &=& 149796 N^{12}+481788 N^{11}+4037555 N^{10}+6431215 N^9-710852 N^8-14957774 N^7
   \nonumber \\ &&
   -21164117 N^6-11167685 N^5+2360450 N^4+2452488 N^3-1225440 N^2
   \nonumber \\ &&
   -518400 N+181440 
    , \\ 
   P_{196} &=& 33 N^{13}+264 N^{12}+574 N^{11}-470 N^{10}-2978 N^9-912 N^8+8524 N^7+14408 N^6
   \nonumber \\ &&
   +9543 N^5+4750 N^4+4440 N^3+3344 N^2+2544 N+864 
    , \\ 
   P_{197} &=& 40 N^{13}+55 N^{12}-450 N^{11}-1455 N^{10}+1040 N^9+10269 N^8+1382 N^7-28529 N^6
   \nonumber \\ &&
   -34324 N^5-70484 N^4-96232 N^3-16416 N^2+4704 N-11520 
    , \\ 
   P_{198} &=& 270 N^{13}-4885 N^{12}+990 N^{11}+51475 N^{10}-25020 N^9-165453 N^8+159306 N^7
   \nonumber \\ &&
   +313033 N^6-458562 N^5-819322 N^4-269976 N^3+512832 N^2
\nonumber\\ &&
+597312 N+142560 
    , \\ 
   P_{199} &=& 510 N^{13}+765 N^{12}-6830 N^{11}-15855 N^{10}+33390 N^9+79359 N^8-86978 N^7
   \nonumber \\ &&
   -140429 N^6+273876 N^5+415216 N^4+51568 N^3-232496 N^2-95616 N
\nonumber\\ &&
+46080 
    , \\ 
   P_{200} &=& 7680 N^{13}-30135 N^{12}-89760 N^{11}+339115 N^{10}+393290 N^9-770421 N^8
   \nonumber \\ &&
   -1106268 N^7-1744699 N^6-9260454 N^5-32119364 N^4-35024552 N^3
   \nonumber \\ &&
   -4473856 N^2-2520576 N-5575680 
    , \\ 
   P_{201} &=& 51840 N^{13}+93855 N^{12}-531360 N^{11}-2157395 N^{10}+1567910 N^9
   \nonumber \\ &&
   +11256237 N^8-3136644 N^7-17459037 N^6+30905238 N^5+68574308 N^4
   \nonumber \\ &&
   +30890344 N^3-30448128 N^2-5833728 N+14100480 
    , \\ 
   P_{202} &=& 95 N^{14}+125 N^{13}-1134 N^{12}-2050 N^{11}+6499 N^{10}+26893 N^9+38472 N^8
   \nonumber \\ &&
-55832 N^7
   -205660 N^6-56080 N^5+145792 N^4+53472 N^3-6144 N^2
\nonumber\\ &&
-88448 N
-42240 
    , \\ 
   P_{203} &=& 8868 N^{14}+35472 N^{13}-9409 N^{12}-152862 N^{11}+61883 N^{10}+593774 N^9
   \nonumber \\ &&
-379547 N^8
   -1672874 N^7-807075 N^6+89818 N^5-325576 N^4-407328 N^3
   \nonumber \\ &&
-167688 N^2
   -21600 N+18144 
    , \\ 
   P_{204} &=& 540 N^{15}-6940 N^{14}-6255 N^{13}+92984 N^{12}+99855 N^{11}-389419 N^{10}-647943 N^9
   \nonumber \\ &&
   +663238 N^8+1833777 N^7-126095 N^6-1116630 N^5+69928 N^4-480432 N^3
   \nonumber \\ &&
   -718416 N^2-1212192 N-570240 
    , \\ 
   P_{205} &=& 2091 N^{15}+9807 N^{14}+3454 N^{13}-50522 N^{12}-77079 N^{11}+55082 N^{10}+164624 N^9
   \nonumber \\ &&
   -48090 N^8-195178 N^7+109703 N^6+248568 N^5+38380 N^4-49640 N^3
   \nonumber \\ &&
-5280 N^2
   -26208 N-13824 
    , \\ 
   P_{206} &=& 8020 N^{15}+11831 N^{14}-92283 N^{13}-365351 N^{12}-433033 N^{11}+1065225 N^{10}
   \nonumber \\ &&
   +3874583 N^9+2278519 N^8-4959567 N^7-7109812 N^6-895688 N^5+2185524 N^4
   \nonumber \\ &&
   +697824 N^3+281664 N^2+388800 N+77760 
    , \\ 
   P_{207} &=& 420 N^{16}+540 N^{15}-8300 N^{14}-15615 N^{13}+49927 N^{12}+148830 N^{11}-80392 N^{10}
   \nonumber \\ &&
   -672719 N^9-625021 N^8+156216 N^7+823430 N^6+3125340 N^5+4621504 N^4
   \nonumber \\ &&
   +2625824 N^3+429792 N^2+87744 N+46080 
    , \\ 
   P_{208} &=& 685 N^{16}+1370 N^{15}-9010 N^{14}-19290 N^{13}+26146 N^{12}+91966 N^{11}-14748 N^{10}
   \nonumber \\ &&
   -149230 N^9+45035 N^8+174316 N^7-271314 N^6-505068 N^5-281130 N^4
   \nonumber \\ &&
   -52080 N^3+16080 N^2+17280 N+4320 
    , \\ 
   P_{209} &=& 12180 N^{16}+8370 N^{15}-256195 N^{14}-157950 N^{13}+1778081 N^{12}+1177830 N^{11}
   \nonumber \\ &&
   -4307281 N^{10} -3049362 N^9+3710647 N^8+11089008 N^7+11202520 N^6
   \nonumber \\ &&
   -23576760 N^5-52089008 N^4-32240448 N^3-12305664 N^2
   \nonumber \\ &&
   -7993728 N-1866240 
    , \\ 
   P_{210} &=& 137 N^{17}+822 N^{16}+424 N^{15}-5764 N^{14}-8196 N^{13}+16720 N^{12}+41536 N^{11}
   \nonumber \\ &&
   +15066 N^{10}   -25651 N^9-42278 N^8-54216 N^7-37786 N^6+78 N^5
   \nonumber \\ &&
   -396 N^4-8496 N^3-4416 N^2+1248 N+576 
    , \\ 
   P_{211} &=& 39255 N^{17}+240282 N^{16}+41728 N^{15}-2061700 N^{14}-2654870 N^{13}+5644496 N^{12}
   \nonumber \\ &&
   +10599172 N^{11}-8903908 N^{10}-24858601 N^9+2221990 N^8+27458420 N^7
   \nonumber \\ &&
   +6952696 N^6-13270864 N^5-14011424 N^4-5396352 N^3
   \nonumber \\ &&
   -1497600 N^2+2412288 N+1119744 
    , \\ 
   P_{212} &=& 242739 N^{17}+1505034 N^{16}+1803872 N^{15}-6299252 N^{14}-18083694 N^{13}
   \nonumber \\ &&
-3855808 N^{12}
   +41147860 N^{11}+41085852 N^{10}-48824861 N^9-83231194 N^8
   \nonumber \\ &&
+12645540 N^7
   +64164904 N^6+31839104 N^5+22953024 N^4+15952896 N^3
   \nonumber \\ &&
-2543616 N^2
   -4064256 N-746496 
    , \\ 
   P_{213} &=& 10455 N^{18}+59490 N^{17}-15790 N^{16}-741440 N^{15}-1390120 N^{14}+1428397 N^{13}
   \nonumber \\ &&
   +7150867 N^{12}+5281630 N^{11}-7138741 N^{10}-8816137 N^9+10256689 N^8
   \nonumber \\ &&
   +16683860 N^7-3484864 N^6-12146200 N^5-502576 N^4+5909760 N^3
   \nonumber \\ &&
   +4302720 N^2+2851200 N+829440 
    , \\ 
   P_{214} &=& 40100 N^{18}+99255 N^{17}-727380 N^{16}-3314430 N^{15}-1772040 N^{14}+14821596 N^{13}
   \nonumber \\ &&
   +48888776 N^{12}+98029290 N^{11}+59157432 N^{10}-330544971 N^9-879181188 N^8
   \nonumber \\ &&
   -779917140 N^7-105697492 N^6+173719200 N^5+551952 N^4-97381440 N^3
   \nonumber \\ &&
   -68610240 N^2-34525440 N-8398080 
    , \\ 
   P_{215} &=& 1213695 N^{20}+7525170 N^{19}-6722900 N^{18}-132732760 N^{17}-180657906 N^{16}
   \nonumber \\ &&
   +706987500 N^{15}+1986194496 N^{14}-505023504 N^{13}-7245869189 N^{12}
   \nonumber \\ &&
   -7460329438 N^{11}+6529524348 N^{10}+22209128904 N^9+18794760144 N^8
   \nonumber \\ &&
   -4187656992 N^7-23855002304 N^6-26274133120 N^5-18561813504 N^4
   \nonumber \\ &&
   -9634314240 N^3-2690703360 N^2+136028160 N+199065600 
    , \\ 
   P_{216} &=& 196275 N^{22}+1397685 N^{21}-1454770 N^{20}-30923820 N^{19}-41291522 N^{18}
   \nonumber \\ &&
   +225466098 N^{17}+630395612 N^{16}-373372336 N^{15}-3171331361 N^{14}
   \nonumber \\ &&
   -2650077679 N^{13}+4908510270 N^{12}+10281951044 N^{11}+3751227016 N^{10}
   \nonumber \\ &&
   -6343664096 N^9-8882356992 N^8-5272720448 N^7+530329472 N^6
   \nonumber \\ &&
+4243436032 N^5
   +2879400960 N^4+137687040 N^3-705024000 N^2-525864960 N
\nonumber\\ &&
-149299200 .     
\end{eqnarray}

\section{Special constants}
\label{sec:B}

\vspace*{1mm}
\noindent
In the following, we present some examples for $\GA$--functions over the alphabet 
$\mathfrak{A}$
at $x=1$, which appear in the present calculation. These constants can be mapped to cyclotomic numbers. They 
finally reduce to multiple zeta values.
\begin{eqnarray}
\GA\left(\left\{\frac{\sqrt{1-x}}{x}\right\},1\right)&=& -2+2 \ln (2),
\\ 
\GA\left(
        \left\{\sqrt{1-x} \sqrt{x}\right\},1\right)&=& \frac{\pi }{8},
\\
\GA\big(
        \big\{
        \frac{\sqrt{1-x}}{x},\frac{\sqrt{1-x}}{x}\big\},1\big)&=& 2 (1- \ln (2))^2,
\\
\GA\big(
        \big\{\sqrt{1-x} \sqrt{x},\sqrt{1-x} \sqrt{x}\big\},1\big)&=& \frac{\pi ^2}{128},
\\ 
G\left(
        \left\{\frac{1}{x},\frac{\sqrt{1-x}}{x},\frac{\sqrt{1-x}}{x}\right\},1\right)&=& 
6 - 8 \ln(2) + 8 \ln^2(2) - \frac{8}{3} \ln^3(2) - 2 \zeta_2 
\nonumber\\ &&
+ 2 \ln(2) \zeta_2 - \frac{3}{2} \zeta_3,
\\
G\left(
        \left\{\frac{1}{x},\frac{\sqrt{1-x}}{x},\frac{1}{x},\frac{1}{x}\right\},1\right)&=& -48
+48 \ln (2)
-24 \ln^2 (2)
+8 \ln^3 (2)
-2 \ln^4 (2)
\nonumber\\ &&
+12 \zeta_2
-12 \ln (2) \zeta_2
+6 \ln^2 (2) \zeta_2
+\frac{27}{10} \zeta_2^2
+12 \zeta_3
\nonumber\\ &&
-12 \ln (2) \zeta_3,
\\
G\left(
        \left\{\frac{1}{x},\frac{1}{x},\frac{1}{1-x},\frac{\sqrt{1-x}}{x},\frac{1}{1-x}\right\},1\right)&=& 128
-64 \Li_5\left(\frac{1}{2}\right)
-96 \ln (2)
-64 \Li_4\left(\frac{1}{2}\right) \ln (2)
\nonumber\\ &&
+32 \ln^2 (2)
-\frac{32}{15} \ln^5 (2)
-40 \zeta_2
+24 \ln (2) \zeta_2
\nonumber\\ &&
+\frac{32}{3} \ln^3 (2) \zeta_2
-32 \zeta_3
-28 \ln^2 (2) \zeta_3
+\frac{55}{2} \zeta_2 \zeta_3
\nonumber\\ &&
+\frac{93}{8} \zeta_5.
\end{eqnarray}
\section{Mellin inversion of finite binomial sums}
\label{sec:C}

\vspace*{1mm}
\noindent
The following Mellin inversions are obtained for the nested finite binomial sums occurring in the
present paper. We define
\begin{eqnarray} 
\Mvec[f(x)](N) = \int_0^1 dx~x^N f(x)~~\text{and}~~
\Mvec[[g(x)]_+](N) = \int_0^1 dx~(x^N -1 ) g(x).
\end{eqnarray} 
Let
\begin{eqnarray} 
w = \sqrt{1-x}~~~~\text{and}~~~~
r = \sqrt{x(1-x)}.
\end{eqnarray} 
One obtains
\begin{eqnarray} 
\label{eq:GA1}
\Mvec^{-1}[{\sf BS}_0(N)](x) &=& \frac{1}{2 x^{l+3/2}}, 
\\
\Mvec^{-1}[{\sf BS}_1(N)](x) &=& \delta(1-x) - \frac{1}{2} 
\left[\frac{1}{(1-x)^{3/2}}\right]_+ ,
\\ 
\Mvec^{-1}[{\sf BS}_2(N)](x) &=& 
\frac{1}{\pi} \frac{1}{\sqrt{x} \sqrt{1-x}} ,
\\
\Mvec^{-1}[{\sf BS}_{3}(N)](x) &=&
-\Biggl[\frac{\big(
        \pi  (1-x)^{3/2}
        -2 \sqrt{x}
        +8 x^{3/2}
        -10 x^{5/2}
        +4 x^{7/2}
\big)}{\pi 
        (1-x)^{5/2}}
-\frac{8 \GA(\{5\},x)}{\pi  (1-x)}\Biggr]_+ ,
\nonumber\\ &&
\\
\Mvec^{-1}[{\sf BS}_{4}(N)](x) &=& \Biggl[
-\frac{2 \ln(2)}{1-x}
+2 \big(
        -1
        +\sqrt{1-x}
        +x
\big) \frac{1}{(1-x)^{3/2}}
+\frac{\GA(\{4\},x)}{1-x}
\Biggr]_+ ,
\\
\Mvec^{-1}[{\sf BS}_{5}(N)](x) &=& 
\Biggl[\frac{-2 \ln^2(2)
-2 (1 - \ln(2)) \HA_0(x)
+\zeta_2
}{1-x}
+\frac{2 \GA(\{4\},x)}{1-x}
-\frac{\GA(\{1,4\},x)}{1-x}\Biggr]_+ ,
\nonumber\\
\\
\Mvec^{-1}[{\sf BS}_{6}(N)](x) &=&
\Biggl[\GA(\{5\},x)
\Biggl[-\frac{4 \pi }{1-x}
        +16 (1-2 x) \frac{\sqrt{x}}{\sqrt{1-x}}
\Biggr]
-\pi  (1-2 x) \frac{\sqrt{x}}{\sqrt{1-x}}
\nonumber\\ && 
+2 x (1-2 x)^2 
+\frac{64 \GA(\{5,5\},x)}{1-x}\Biggr]_+ ,
\\
\Mvec^{-1}[{\sf BS}_{7}(N)](x) &=& 
\Biggl[\frac{x \big(
        18-33 x+40 x^2-18 x^3\big)}{3 (1-x)}
+\GA(\{5\},x)
\Biggl[
        8 \frac{
                \sqrt{x}-3 x^{3/2}+2 x^{5/2}}{(1-x)^{3/2}} 
\nonumber\\ &&        
	-\frac{8 \pi \ln(2)}{1-x}
 -\frac{28 \zeta_3}{\pi  (1-x)}
\Biggr]
-2 \pi \ln(2)   \frac{\sqrt{x}-3 x^{3/2}+2 x^{5/2}
}{(1-x)^{3/2}} 
\nonumber\\ && 
-2 x (1-2 x)^2 \HA_0(x)
-\frac{7 \big(
        \sqrt{x}-3 x^{3/2}+2 x^{5/2}\big) \zeta_3}{\pi (1-x)^{3/2}}
\nonumber\\ && +16 (1-x) (-1+2 x) \frac{\sqrt{x}}{(1-x)^{3/2}} 
\GA(\{5,1\},x)
+\frac{32 \GA(\{5,5\},x)}{1-x}
\nonumber\\ &&
-\frac{64 \GA(\{5,5,1\},x)}{1-x}\Biggr]_+ ,
\\
\Mvec^{-1}[{\sf BS}_{8}(N)](x) &=& 
\Biggl[
-\frac{4 \big(
        1-\sqrt{1
        -x
        }\big)}{1-x}
+\Biggl(
        \frac{2 (1-\ln(2))}{1-x}
        +\frac{\HA_0(x)}{\sqrt{1-x}}
\Biggr) \HA_1(x)
-\frac{\HA_{0,1}(x)}{\sqrt{1-x}}
\nonumber\\ &&
+\frac{\HA_1(x) \GA(\{6,1\},x)}{2 (1-x)}
-\frac{\GA(\{6,1,2\},x)}{2 (1-x)}
\Biggr]_+,
\nonumber\\ 
\\
\Mvec^{-1}[{\sf BS}_{9}(N)](x) &=& 
\Biggl[
\GA(\{5\},x) \Biggl(
        -\frac{8 \ln(2) \pi }{1-x}
        +8 (1-2 x) \frac{\sqrt{x}}{\sqrt{1-x}}
        +\frac{28 \zeta_3}{\pi  (1-x)}
\Biggr)
\nonumber\\ && 
-\frac{1}{3 \pi  (1-x)} \Biggl[
        \pi  x^2 (21
        +2 x (-16+9 x)
        )
        +36 \ln(2) (1-2 x) \sqrt{(1-x) x} \zeta_2
\nonumber\\ &&   
      -21 (1-2 x) \sqrt{(1-x) x} \zeta_3
\Biggr]
+2 (1-2 x)^2 x \HA_1(x)
+\frac{32 \GA(\{5,5\},x)}{1-x}
\nonumber\\ && 
+\frac{64 \GA(\{5,5,2\},x)}{1-x}
+16 (1-2 x) \frac{\sqrt{x}}{\sqrt{1-x}} \GA(\{5,2\},x)
\Biggr]_+ ,
\nonumber\\
\\
\Mvec^{-1}[{\sf BS}_{10}(N)](x) &=& 
\Biggl[-\frac{1}{1-x} \big[
        -4
        -4 \ln(2) \big(
                -1+\sqrt{1
                -x
                }\big)
        +4 \sqrt{1-x}
        +\zeta_2
\big]
\nonumber\\ && 
+2 (-1+\ln(2)) \big(
        -1
        +\sqrt{1-x}
        +x
\big) \frac{\HA_0(x)}{(1-x)^{3/2}}
-2 \frac{\HA_1(x)}{\sqrt{1-x}}
\nonumber\\ &&
+\frac{\HA_{0,1}(x)}{\sqrt{1-x}}
-\frac{(-2+\ln(2)) \GA\big(\{6,1\},x)}{1-x}
+\frac{\GA(\{6,1,2\},x)}{2 (1-x)}
\nonumber\\ &&
-\frac{\GA\big(\{1,6,1\},x)}{2 (1-x)}\Biggr]_+~.
\label{eq:GA2}
\end{eqnarray}
At lower weight, the $\GA$--functions over the alphabet (\ref{eq:alph}) can be expressed in terms
of known functions, i.e. by elementary functions and polylogarithms with involved
arguments. This is, however, not possible in general at higher weight. For the simplest cases 
one finds.
\begin{eqnarray} 
\GA(\{4\},x)   &=& 
2 (-1 +\ln(2) +w)
+w \HA_0(x)
-(1+w) \HA_{-1}(w) 
-(1-w) \HA_1(w),
\\
\GA(\{1,4\},x)   &=& 
-4
+4 \ln(2)
-2 \ln^2(2)
+4 w
+\frac{1}{2} w \HA_0(x)^2
-2 \ln(2) \HA_1(w)
+\frac{1}{2} (1-w) \HA_1^2(w)
\nonumber\\ &&
+\Biggl(
        2 (-2+\ln(2))
        +(1+w) \HA_1(w)
\Biggr) \HA_{-1}(w)
-\frac{1}{2} (1+w) \HA_{-1}^2(w)
\nonumber\\ && 
-2 \HA_{-1,1}(w)
+\zeta_2,
\\
\GA(\{2,4\},x) &=& 
4
-w \big(
        4
        -\zeta_2
\big)
+\big(
        4 w
        -2 (1-w) \HA_1(w)
        -2 (1+w) \HA_{-1}(w)
\big) \HA_0(w)
\nonumber\\ &&  
 +\big(
        2 (-1
        +\ln(2)
        +w
        )
        +w \HA_0(x)
\big) \HA_1(x)
-(1-w) \HA_1(w) \HA_1(x)
\nonumber\\ && 
-(1+w) \HA_{-1}(w) \HA_1(x)
+2 (1-w) \HA_{0,1}(w)
-w \HA_{0,1}(x)
+2 (1+w) \HA_{0,-1}(w)
\nonumber\\ &&
-3 \zeta_2,
\\
\GA(\{4,4\},x) &=& 
2 \big(
        2
        +\ln^2(2)
        -2 \ln(2) (1-w)
        -2 w
        -x
\big)
-x \HA_0(x)
+\frac{1}{2} \HA_0(x)^2
+(2
\nonumber\\ && 
-2 \ln(2)
-2 w
-x
) \HA_1(w)
+\big(
        2
        -2 \ln(2)
        -2 w
        +x
        +2 \HA_1(w)
\big) \HA_{-1}(w),
\\
\GA(\{5\},x) &=& 
\frac{1}{4} {\rm arcsin}\big(
        \sqrt{x}\big)
-\frac{1}{4} w (1-2 x) \sqrt{x},
\\
\GA(\{5,1\},x) &=& 
\frac{C}{4}
+\frac{1}{32} (1-6 \ln(2)) \pi 
-\frac{i \pi ^2}{48}
-\frac{1}{4} i {\rm arctan}^2\left(
        \frac{1-2 r}{1-2 x}\right)
\nonumber\\ && 
+\ln(2)
 \Biggl[
        \frac{1}{4} {\rm arctan}\left(
                \frac{1-2 r}{1-2 x}\right)
        +\frac{(-1+2 x) \big(
                -r
                -4 (-1+r) x
                +4 (-1+r) x^2
        \big)}{4 (1-2 r)^2}
\Biggr]
\nonumber\\ &&   +\frac{1}{8 (1-2 r)^2} 
\Biggl[
        \big(
                -r
                -4 (-1+r) x
                +4 (-1+r) x^2
        \big)
\Biggl[-3
                +2 x
\nonumber\\ &&           
                +2 (-1+2 x) \ln\left(
                        \frac{2-4 r}{(1-2 x)^2}\right)
 -4 (-1+2 x) \ln\left(
                        -\frac{2 (r
                        -x
                        )}{-1+2 x}\right)
        \Biggr]\Biggr]
\nonumber\\ &&  
+\frac{1}{8} {\rm arctan}\left(
        \frac{1-2 r}{1-2 x}
\right)
\Biggl[-1
        +4 \ln\left(
                \frac{(1+i) (-1+2 x)}{-1
                +(1-i) r
                +(1+i) x
                }\right)
\nonumber\\ && 
        +2 \ln\left(
                \frac{2-4 r}{(1-2 x)^2}\right)\Biggr]
-\frac{1}{2} 
        {\rm arctan}\left(
                \frac{r
                -x
                }{-1
                +r
                +x
                }\right) \ln\left(
                -\frac{2 (r
                -x
                )}{-1+2 x}\right)
\nonumber\\ && 
+\frac{i}{4}  \Li_2\left[
        -\frac{(1+i) (r
        -x
        )}{-1+2 x}\right]
-\frac{i}{4}  \Li_2\left[
        -\frac{(1-i) (r
        -x
        )}{-1+2 x}\right]
\nonumber\\ && 
- \frac{i}{4} 
        \Li_2\left[
                \frac{-1
                +(1+i) r
                +(1-i) x
                }{-i
                +(1+i) r
                +(-1+i) x
                }\right]
\nonumber\\
\\
\GA(\{5,2\},x) &=& 
\frac{1}{8} \Biggl[
        \frac{i \pi ^2}{3}
        +2 i {\rm arccos}^2\big(
                \sqrt{x}\big)
        + {\rm arccos}\big(
                \sqrt{x}
        \big)
\Biggl(1
                -4 \ln\big(
                        2
                        -2 x
                        +2 i \sqrt{x(1-x)}
                \big)
\nonumber\\ &&             
   +2 \ln(1-x)
        \Biggr)
        +(1
        +2 x
        +2 (1-2 x) \ln(1-x)
        ) \sqrt{x(1-x)}
\nonumber\\ &&
        -\frac{1}{2} \pi  (1
-4 \ln(2))
        -2 i \Li_2\left(
                \frac{1}{\big(
                        i \sqrt{1-x}
                        +\sqrt{x}
                \big)^2}\right)
\Biggr],
\\
\GA(\{5,5\},x) &=& 
\frac{1}{32} \Biggl[
        (1-2 x)^2 (1-x) x
        + {\rm arcsin}^2\big(
                \sqrt{x}\big)
        -2 (1-2 x) {\rm arcsin}\big(
                \sqrt{x}\big) 
\nonumber\\ &&
\times
\sqrt{(1-x) x}
\Biggr],
\\
\GA(\{6,1\},x) &=&
4 (-1
+\ln(2)
+w
)
-2 (1-w) \HA_1(w)
-2 (1+w) \HA_{-1}(w),
\\
\GA(\{6,1,2\},x) &=& 
 -8
-2 w \big(
        -4
        +\zeta_2
\big)
+\big(
        -8 w
        +4 (1- w) \HA_1(w)
        +4 (1+w) \HA_{-1}(w)
\big) 
\nonumber\\ &&
\times \HA_0(w)
-4 (1-w) \HA_{0,1}(w)
-4 (1+w) \HA_{0,-1}(w)
+6 \zeta_2,
\\
\GA(\{1,6,1\},x) &=& 
 4 \big(
        -4
        +4 \ln(2)
        -2 \ln^2(2)
        +4 w
        +\zeta_2
\big)
-4 (1
+\ln(2)
-w
) \HA_1(w)
\nonumber\\ &&
+\big(
        4 (-3
        +\ln(2)
        -w
        )
        +4 \HA_1(w)
\big) \HA_{-1}(w)
-8 \HA_{-1,1}(w),
\end{eqnarray}
where $C$ denotes Catalan's constant \cite{CAT}
\begin{eqnarray}
C =  \sum_{k=0}^\infty \frac{(-1)^k}{(2 k + 1)^2}.
\end{eqnarray}
Furthermore, one has
\begin{eqnarray}
{\rm arctan}(x) &=& \frac{i}{2} \ln\left[\frac{1- i x}{1 + i x}\right]
\\
{\rm arcsin}(\sqrt{x}) &=& -i \ln(w + i \sqrt{x})
\\
{\rm arccos}(\sqrt{x}) &=& \frac{\pi}{2} + i \ln(w + i \sqrt{x})
\\
{\rm arctanh}(w) &=& \frac{1}{2}\left[\HA_1(w) + \HA_{-1}(w)\right]
\\
\Li_2\left(\frac{1}{u}\right) &=& 2 \zeta_2 - \frac{1}{2} 
\HA^2_0(u) - 
\HA_{0,1}(u) - 
i\pi \HA_0(u)
\label{ea:C31}
\end{eqnarray}
and in case, one has to consider the analytic continuation of the above quantities.

\section{\boldmath Asymptotic expansion of $\tilde{a}_{gg,Q}^{(3)}(N)$ and $\Delta \tilde{a}_{gg,Q}^{(3)}(N)$}
\label{sec:D}

\vspace*{1mm}
\noindent
Here we present the asymptotic expansion of the finite binomial sums used and for 
$\tilde{a}_{gg,Q}^{(3)}(N)$ and $\Delta \tilde{a}_{gg,Q}^{(3)}(N)$ 
to $O(1/N^{10})$ for QCD, by specifying the color factors to $SU(3)$. One obtains
\begin{eqnarray} 
{\sf BS}_0(N) &\propto& \frac{1}{2N} \sum_{k=0}^\infty \left(\frac{2l+1}{2N}\right)^k,
\\
{\sf BS}_1(N) &\propto& 
\sqrt{\pi} \sqrt{N}
\Biggl[
1
+\frac{1}{8 N}
+\frac{1}{128 N^2}
-\frac{5}{1024 N^3}
-\frac{21}{32768 N^4}
+\frac{399}{262144 N^5}
+\frac{869}{4194304 N^6}
\nonumber\\ &&
-\frac{39325}{33554432 N^7}
-\frac{334477}{2147483648 N^8}
+\frac{28717403}{17179869184 N^9}
+\frac{59697183}{274877906944 N^{10}}
\Biggr], 
\\
{\sf BS}_2(N) &\propto& 
\frac{1}{\sqrt{\pi} \sqrt{N}} \Biggl[1
-\frac{1}{8 N} 
+\frac{1}{128 N^2}
+\frac{5}{1024 N^3}
-\frac{21}{32768 N^4}
-\frac{399}{262144 N^5}
+\frac{869}{4194304 N^6}
\nonumber\\ &&
+\frac{39325}{33554432 N^7}
-\frac{334477}{2147483648 N^8}
-\frac{28717403}{17179869184 N^9}
+\frac{59697183}{274877906944 N^{10}}\Biggr],
\\
{\sf BS}_3(N)
&\propto& 
2 \ln(2)
+ \frac{1}{\sqrt{N} \sqrt{\pi }}
\Biggl[
        -2
+\frac{7}{12 N} 
-\frac{61}{320 N^2}
+\frac{307}{10752 N^3}
+\frac{911}{49152 N^4}
-\frac{12559}{1441792 N^5}
\nonumber\\ && 
-\frac{1404237}{136314880 N^6}
+\frac{386621}{50331648 N^7}
+\frac{223373117}{18253611008 N^8}
-\frac{42971056687}{3427383902208 N^9}
\nonumber\\ &&
-\frac{50702821769}{2061584302080 N^{10}}
\Biggr],
\\
{\sf BS}_4(N) &\propto& 
3 \zeta_2
+
\frac{\sqrt{\pi }}{\sqrt{N}}
\Biggl[
-2
+\frac{5}{12 N}
-\frac{21}{320 N^2}
-\frac{223}{10752 N^3}
+\frac{671}{49152 N^4}
+\frac{11635}{1441792 N^5}
\nonumber\\ && 
-\frac{1196757}{136314880 N^6}
-\frac{376193}{50331648 N^7}
+\frac{201980317}{18253611008 N^8}
+\frac{42437231395}{3427383902208 N^9}
\nonumber\\ &&
-\frac{47256733409}{2061584302080 N^{10}}
\Biggr], 
\\
{\sf BS}_5(N) &\propto&
-\frac{7}{2} \zeta_3
+ 6 \ln(2) \zeta_2
+
\frac{\sqrt{\pi }}{\sqrt{N}}\Biggl[
-\frac{2}{3 N}
+\frac{9}{20 N^2}
-\frac{199}{1344 N^3}
-\frac{145}{4608 N^4}
+\frac{8909}{180224 N^5}
\nonumber\\ &&
+\frac{427409}{25559040 N^6}
-\frac{1375417}{31457280 N^7}
-\frac{17099251}{855638016 N^8}
+\frac{30183885889}{428422987776 N^9}
\nonumber\\ &&
+\frac{110144579981}{2705829396480 N^{10}}
\Biggr],
\\
{\sf BS}_6(N) &\propto& 
\frac{7}{2} \zeta_3
+\frac{2}{N}
-\frac{4}{3 N^2}
+\frac{32}{45 N^3}
-\frac{8}{35 N^4}
-\frac{8}{225 N^5}
+\frac{248}{3465 N^6}
+\frac{6856}{315315 N^7}
\nonumber\\ &&
-\frac{3176}{45045 N^8}
-\frac{11304}{425425 N^9}
+\frac{615256}{4849845 N^{10}}
+\frac{8597752}{160044885 N^{11}}
+\frac{\zeta_2}{\sqrt{N} \sqrt{\pi }}
\Biggl[
-6
+\frac{7}{4 N}
\nonumber\\ && 
-\frac{183}{320 N^2}
+\frac{307}{3584 N^3}
+\frac{911}{16384 N^4}
-\frac{37677}{1441792 N^5}
-\frac{4212711}{136314880 N^6}
\nonumber\\ &&
+\frac{386621}{16777216 N^7}
+\frac{670119351}{18253611008 N^8}
-\frac{42971056687}{1142461300736 N^9}
-\frac{50702821769}{687194767360 N^{10}}
\Biggr], 
\nonumber\\
\\
{\sf BS}_7(N) &\propto& 
- 16 \Li_4\left(\frac{1}{2}\right)
- \frac{2}{3} \ln^4(2) 
+ 4 \ln^2(2) \zeta_2
- 7 \ln(2) \zeta_3
+ \frac{53}{10} \zeta_2^2
+\frac{1}{3 N^2} 
-\frac{23}{45 N^3} 
+\frac{17}{35 N^4} 
\nonumber\\ && 
-\frac{7}{25 N^5} 
+\frac{169}{10395 N^6} 
+\frac{12737}{105105 N^7} 
-\frac{719}{27027 N^8} 
-\frac{1755653}{11486475 N^9} 
+\frac{4601449}{72747675 N^{10}} 
\nonumber\\ && 
+\frac{1}{\sqrt{N} \sqrt{\pi }}
\Biggl\{
\ln(2) \zeta_2 
\Biggl[
-12
+\frac{7}{2 N}
-\frac{183}{160 N^2}
+\frac{307}{1792 N^3}
+\frac{911}{8192 N^4}
-\frac{37677}{720896 N^5}
\nonumber\\ && 
-\frac{4212711}{68157440 N^6}
+\frac{386621}{8388608 N^7}
+\frac{670119351}{9126805504 N^8}
-\frac{42971056687}{571230650368 N^9}
\nonumber\\ && 
-\frac{50702821769}{343597383680 N^{10}}
\Biggr]
+\zeta_3 \Biggl[
7
-\frac{49}{24 N}
+\frac{427}{640 N^2}
-\frac{307}{3072 N^3}
-\frac{6377}{98304 N^4}
\nonumber\\ &&
+\frac{87913}{2883584 N^5}
+\frac{9829659}{272629760 N^6}
-\frac{2706347}{100663296 N^7}
-\frac{1563611819}{36507222016 N^8}
\nonumber\\ &&
+\frac{42971056687}{979252543488 N^9}
+\frac{354919752383}{4123168604160 N^{10}}
\Biggr]\Biggr\}, 
\\
{\sf BS}_8(N) &\propto& 
-7 \zeta_3
+\Biggl[
        +3 (\ln(N) + \gamma_E)
        +\frac{3}{2 N}
        -\frac{1}{4 N^2}
        +\frac{1}{40 N^4}
        -\frac{1}{84 N^6}
        +\frac{1}{80 N^8}
        -\frac{1}{44 N^{10}}
\Biggr] \zeta_2
\nonumber\\ &&
+ \sqrt{\frac{\pi}{N}} \Biggr[
4 
-\frac{23}{18 N} 
+\frac{1163}{2400 N^2}
-\frac{64177}{564480 N^3}
-\frac{237829}{7741440 N^4}
+\frac{5982083}{166526976 N^5}
\nonumber\\ &&
+\frac{5577806159}{438593126400 N^6}
-\frac{12013850977}{377864847360 N^7}
-\frac{1042694885077}{90766080737280 N^8}
\nonumber\\ &&
+\frac{6663445693908281}{127863697547722752 N^9}
+        \frac{23651830282693133}{1363413316298342400 N^{10}}
\Biggr],
\\
{\sf BS}_9(N) &\propto&
7 \ln(2) \zeta_3
+\frac{6}{N}
-\frac{20}{9 N^2}
+\frac{203}{450 N^3}
+\frac{2752}{11025 N^4}
-\frac{8647}{31500 N^5}
+\frac{587381}{24012450 N^6}
\nonumber\\ && 
+\frac{1143813683}{9468909450 N^7}
-\frac{12378799}{2029052025 N^8}
-\frac{3519473211853}{23455841409000 N^9}
\nonumber\\ && 
-\frac{2992510324519}{282251958288300 N^{10}}
+\frac{1509837197342693}{4657157311756950 N^{11}}
+(\ln(N) + \gamma_E)  \Biggl[
\frac{2}{N}
-\frac{4}{3 N^2}
\nonumber\\ && 
+\frac{32}{45 N^3}
-\frac{8}{35 N^4}
-\frac{8}{225 N^5}
+\frac{248}{3465 N^6}
+\frac{6856}{315315 N^7}
-\frac{3176}{45045 N^8}
-\frac{11304}{425425 N^9}
\nonumber\\ && 
+\frac{615256}{4849845 N^{10}}
+        \frac{8597752}{160044885 N^{11}}
\Biggr]
+\frac{1}{\sqrt{\pi N}}
\Biggl\{
        \ln(2) \zeta_2 \Biggl[
-12
+\frac{7}{2 N}
-\frac{183}{160 N^2} 
\nonumber\\ && 
+\frac{307}{1792 N^3}
+\frac{911}{8192 N^4}
-\frac{37677}{720896 N^5}
-\frac{4212711}{68157440 N^6}
+\frac{386621}{8388608 N^7}
\nonumber\\ &&
+\frac{670119351}{9126805504 N^8}
-\frac{42971056687}{571230650368 N^9}
                -\frac{50702821769}{343597383680 N^{10}}
\Biggr] 
        + \zeta_3 \Biggl[
-7 
+\frac{49}{24 N}
\nonumber\\ && 
-\frac{427}{640 N^2} 
+\frac{307}{3072 N^3}
+\frac{6377}{98304 N^4}
-\frac{87913}{2883584 N^5}
-\frac{9829659}{272629760 N^6}
+\frac{2706347}{100663296 N^7}
\nonumber\\ &&
+\frac{1563611819}{36507222016 N^8}
-\frac{42971056687}{979252543488 N^9}
-\frac{354919752383}{4123168604160 N^{10}}
\Biggr]
\Biggr\}, 
\\
{\sf BS}_{10}(N) &\propto&
6 \ln(2) \zeta_2
+\frac{7}{2} \zeta_3
+\sqrt{\frac{\pi}{N}}
\Biggl\{
        -4
        -\frac{7}{18 N}
        +\frac{817}{2400 N^2}
        -\frac{3835}{37632 N^3}
        -\frac{24677}{1105920 N^4}
\nonumber\\ && 
        +\frac{1822519}{71368704 N^5}
        +\frac{797783941}{55820943360 N^6}
        -\frac{149172521}{7927234560 N^7}
        -\frac{45292690253}{2327335403520 N^8}
\nonumber\\ &&    
     +\frac{7730705027207}{293041323638784 N^9}
        +\frac{51644847163033}{1190564934451200 N^{10}}
        -\frac{286475788323757}{18096587003658240 N^{11}}
\nonumber\\ && 
     +(\ln(N) + \gamma_E) \Biggl[
                -2
+\frac{5}{12 N}
-\frac{21}{320 N^2}
-\frac{223}{10752 N^3}
+\frac{671}{49152 N^4}
+\frac{11635}{1441792 N^5}
\nonumber\\ &&
-\frac{1196757}{136314880 N^6}
-\frac{376193}{50331648 N^7}
+\frac{201980317}{18253611008 N^8}
+\frac{42437231395}{3427383902208 N^9}
\nonumber\\ &&
-\frac{47256733409}{2061584302080 N^{10}}
\Biggr]
\Biggr\}.
\end{eqnarray}

The asymptotic expansions of $\tilde{a}_{gg,Q, N_F = 0}^{(3)}(N)$ and 
$\Delta \tilde{a}_{gg,Q, N_F = 0}^{(3)}(N)$ are given by
\begin{eqnarray}
\lefteqn{\tilde{a}_{gg,Q, N_F = 0}^{(3)} \propto \frac{1}{2}(1+(-1)^N)}
\nonumber\\ &&
\times \Biggl\{\frac{1}{N} \Biggl[
        -\frac{16 L^3}{3}
        +L^2 \Biggl(
                \frac{457}{9}
                -6 \zeta_2
        \Biggr)
        +L \Biggl(
                -\frac{5491}{9}
                +\frac{172 \zeta_2}{3}
                -72 \zeta_3
        \Biggr)
        +\frac{1545977}{2430}
        +128 \liFhalf
\nonumber\\ &&        
 +\frac{16 \ln^4(2)}{3}
        -\frac{2501 \zF}{3}
        +\frac{569}{3} \zeta_2
        -32 \ln^2(2) \zeta_2
        +\frac{10261}{45} \zeta_3
\Biggr]
+\frac{1}{N^2} \Biggl[
        -\frac{20 L^4}{81}
        -\frac{16 L^3}{81}
\nonumber\\ && 
        +L^2 \Biggl(
                -\frac{11323}{162}
                -\frac{284 \zeta_2}{27}
        \Biggr)
        +L \Biggl(
                \frac{346327}{486}
                -\frac{9604 \zeta_2}{27}
                -\frac{6280 \zeta_3}{81}
        \Biggr)
        -\frac{40929283}{14580}
        -\frac{2368}{9} \liFhalf
\nonumber\\ &&      
  -\frac{296 \ln^4(2)}{27}
        +\frac{12044 \zF}{9}
        -\frac{23179}{54} \zeta_2
        -\frac{40}{3} \ln(2) \zeta_2
        +\frac{592}{9} \ln^2(2) \zeta_2
        -\frac{596741}{810} \zeta_3
\Biggr]
\nonumber\\ && 
+\frac{1}{N^3} \Biggl[
         \frac{20 L^4}{81}
        -\frac{584 L^3}{27}
        +L^2 \Biggl(
                \frac{110639}{810}
                +\frac{68 \zeta_2}{27}
        \Biggr)
        +L \Biggl(
                -\frac{348873919}{85050}
                +\frac{22421 \zeta_2}{27}
                -\frac{1496 \zeta_3}{81}
        \Biggr)
\nonumber\\ && 
        +\frac{70746883829}{17860500}
        +256 \liFhalf
        +\frac{32 \ln^4(2)}{3}
        -\frac{4021 \zF}{3}
        +\frac{36481}{135} \zeta_2
        -\frac{172}{9} \ln(2) \zeta_2
\nonumber\\ && 
        -64 \ln^{2}(2)\zeta_2
        +\frac{41777}{27} \zeta_3
\Biggr]
+\frac{1}{N^4} \Biggl[
        -\frac{140 L^4}{81}
        +\frac{5012 L^3}{243}
        +L^2 \Biggl(
                -\frac{57859}{810}
                -100 \zeta_2
        \Biggr)
\nonumber\\ && 
        +L \Biggl(
                \frac{84447577}{18900}
                -\frac{91496 \zeta_2}{81}
                -\frac{69208 \zeta_3}{81}
        \Biggr)
        -\frac{102450206417}{8930250}
        -\frac{34528}{45} \liFhalf
\nonumber\\ &&       
  -\frac{4316 \ln^4(2)}{135}
        +\frac{901591 \zF}{270}
        -\frac{983209}{405} \zeta_2
        +\frac{728}{9} \ln(2) \zeta_2
        +\frac{8632}{45} \ln^2(2) \zeta_2
\nonumber\\ && 
        -\frac{645605}{972} \zeta_3
\Biggr]
+\frac{1}{N^5} \Biggl[
         \frac{260 L^4}{81}
        -\frac{237004 L^3}{1215}
        +L^2 \Biggl(
                \frac{1889789}{4050}
                +\frac{4900 \zeta_2}{27}
        \Biggr)
\nonumber\\ &&    
     +L \Biggl(
                -\frac{52226399557}{2338875}
                +\frac{1231751 \zeta_2}{810}
                +\frac{124360 \zeta_3}{81}
        \Biggr)
+        \frac{342439924906957}{8644482000}
\nonumber\\ && 
        +1280 \liFhalf
        +\frac{160 \ln^4(2)}{3}
        -\frac{145781 \zF}{27}
        +
        \frac{9055723 \zeta_2}{5400}
        -\frac{2788}{9} \ln(2) \zeta_2
      -320 \ln^2(2) \zeta_2
\nonumber\\ &&   
        -\frac{2976962 \zeta_3}{1215}
\Biggr]
+\frac{1}{N^6} \Biggl[
        -\frac{700 L^4}{81}
        +\frac{35950 L^3}{81}
        +L^2 \Biggl(
                -\frac{1942289}{1215}
                -\frac{13340 \zeta_2}{27}
        \Biggr)
\nonumber\\ && 
        +L \Biggl(
                \frac{354034895677}{6949800}
                -\frac{438671 \zeta_2}{405}
                -\frac{341240 \zeta_3}{81}
        \Biggr)
        -\frac{3512899304529779}{35777570400}
\nonumber\\ &&   
      -\frac{532288}{189} \liFhalf
        -
        \frac{66536 \ln^4(2)}{567}
        +\frac{739532 \zF}{63}
        -\frac{469123591 \zeta_2}{34020}
        +\frac{6128}{9} \ln(2) \zeta_2
\nonumber\\ && 
        +\frac{133072}{189} \ln^2(2) \zeta_2
        +\frac{75254303 \zeta_3}{5670}
\Biggr]
+\frac{1}{N^7} \Biggl[
         \frac{1460 L^4}{81}
        -\frac{772130 L^3}{567}
        +L^2 \Biggl(
                \frac{326977031}{59535}
\nonumber\\ &&           
      +\frac{9052 \zeta_2}{9}
        \Biggr)
        +L \Biggl(
                -\frac{366567325520251}{2383781400}
                -\frac{735578 \zeta_2}{405}
                +\frac{688744 \zeta_3}{81}
        \Biggr)
\nonumber\\ && 
+        \frac{48350302011570733597}{143169910884000}
        +5376 \liFhalf
        +224 \ln^4(2)
        -\frac{606637 \zF}{27}
\nonumber\\ &&   
      +\frac{4619767781 \zeta_2}{476280}
        -\frac{22108}{9} \ln(2) \zeta_2
        -1344 \ln^2(2) \zeta_2
        -\frac{37472116}{945} \zeta_3
\Biggr]
\nonumber\\ && 
+\frac{1}{N^8} \Biggl[
        -\frac{1060 L^4}{27}
        +\frac{16654490 L^3}{5103}
        +L^2 \Biggl(
                -\frac{294540853}{17010}
                -\frac{58828 \zeta_2}{27}
        \Biggr)
\nonumber\\ &&        
 +L \Biggl(
                \frac{20298162797504269}{52102650600}
                +\frac{13779523 \zeta_2}{1701}
                -\frac{165592 \zeta_3}{9}
        \Biggr)
        -\frac{495344}{45} \liFhalf
\nonumber\\ && 
        -\frac{314238928512294227009}{354652322104080}
        -\frac{61918 \ln^4(2)}{135}
        +\frac{24921641 \zF}{540}
        -\frac{107294041 \zeta_2}{1701}
\nonumber\\ &&    
     +\frac{14408}{9} \ln(2) \zeta_2
        +\frac{123836}{45} \ln^2(2) \zeta_2
        +\frac{22360477051 \zeta_3}{204120}
\Biggr]
+\frac{1}{N^9} \Biggl[
         \frac{6500 L^4}{81}
\nonumber\\ &&  
      -\frac{206201494 L^3}{25515}
        +L^2 \Biggl(
                \frac{4303709917}{85050}
                +\frac{119108 \zeta_2}{27}
        \Biggr)
        +L \Biggl(
                -\frac{2535160339716913903}{2474875903500}
\nonumber\\ &&          
       -
                \frac{739828937 \zeta_2}{34020}
                +\frac{3007240 \zeta_3}{81}
        \Biggr)
+        \frac{24382761321719406012354719}{9144963448540920000}
        +21760 \liFhalf
\nonumber\\ &&  
  +\frac{2720 \ln^4(2)}{3}
        -\frac{824183 \zF}{9}
        +\frac{74753396657 \zeta_2}{1360800}
        -\frac{139924}{9} \ln(2) \zeta_2
        -5440 \ln^2(2) \zeta_2
\nonumber\\ &&  
       -\frac{2513889701 \zeta_3}{10206}
\Biggr]
+\frac{1}{N^{10}} \Biggl[
        -\frac{13340 L^4}{81}
        + \frac{13615043 L^3}{729}
        +L^2 \Biggl(
-\frac{4157236511}{29160}
                -\frac{81236 \zeta_2}{9}
        \Biggr)
\nonumber\\ &&    
     +L \Biggl(
                \frac{2035166415603071}{814438800}
                +\frac{487601393 \zeta_2}{5670}
                -\frac{6147640 \zeta_3}{81}
        \Biggr)
        -\frac{4333888}{99} \liFhalf
\nonumber\\ && 
        -\frac{2408250929100519977653159}{347591370894768000}
        -\frac{541736 \ln^4(2)}{297}
        +\frac{54840506 \zF}{297}
        -\frac{8559461498321 \zeta_2}{22453200}
\nonumber\\ &&   
     -\frac{63136}{9} \ln(2) \zeta_2
        +\frac{1083472}{99} \ln^2(2) \zeta_2
        +\frac{186728120056 \zeta_3}{280665}
\Biggr]\Biggr\},
\end{eqnarray}
where $L$ is defined in Eq.~(\ref{eq:LG}) 
and
\begin{eqnarray}
\lefteqn{\Delta \tilde{a}_{gg,Q, N_F = 0}^{(3)} \propto \frac{1}{2}(1 - (-1)^N)}
\nonumber\\ &&
\times \Biggl\{\frac{1}{N} \Biggl[
        -\frac{16 L^3}{3}
        +L^2 \left(
                \frac{457}{9}
                -6 \zeta_2
        \right)
        +L \left(
                -\frac{5491}{9}
                +\frac{172 \zeta_2}{3}
                -72 \zeta_3
        \right)
+        \frac{1545977}{2430}
        +128 \Li_4\left(\frac{1}{2}\right)
\nonumber\\ && 
        +\frac{16 \ln^4(2)}{3}
        +\frac{569}{3} \zeta_2
        -32 \ln^2(2) \zeta_2
        -\frac{5002}{15} \zeta_2^2
        +\frac{10261}{45} \zeta_3
\Biggr]
+\frac{1}{N^2} \Biggl[
        -\frac{20 L^4}{81}
        +\frac{64 L^3}{81}
\nonumber\\ && 
        +L^2 \left(
                -\frac{10411}{162}
                -\frac{284 \zeta_2}{27}
        \right)
        +L \left(
                \frac{348367}{486}
                -332 \zeta_2
                -\frac{6280 \zeta_3}{81}
        \right)
        -
        \frac{31616491}{14580}
        -\frac{2368}{9} \Li_4\left(\frac{1}{2}\right)
\nonumber\\ && 
        -\frac{296 \ln^4(2)}{27}
        -\frac{12611}{54} \zeta_2
        -\frac{40}{3} \ln(2) \zeta_2
        +\frac{592}{9} \ln^2(2) \zeta_2
        +\frac{24088}{45} \zeta_2^2
        -\frac{686773}{810} \zeta_3
\Biggr]
+\frac{1}{N^3} \Biggl[
         \frac{20 L^4}{81}
\nonumber\\ &&  
      -\frac{1832 L^3}{81}
        +L^2 \left(
                \frac{111599}{810}
                +\frac{68 \zeta_2}{27}
        \right)
        +L \left(
                -\frac{328566919}{85050}
                +\frac{21781 \zeta_2}{27}
                -\frac{1496 \zeta_3}{81}
        \right)
\nonumber\\ &&
+        \frac{62031195029}{17860500}
        +\frac{32 \ln^4(2)}{3}
        +256 \Li_4\left(\frac{1}{2}\right)
        +\frac{4781}{135} \zeta_2
        -
        \frac{172}{9} \ln(2) \zeta_2
        -64 \ln^2(2) \zeta_2
\nonumber\\ && 
        -\frac{8042}{15} \zeta_2^2
        +\frac{675559}{405} \zeta_3
\Biggr]
+\frac{1}{N^4} \Biggl[
         \frac{20 L^4}{81}
        +\frac{9860 L^3}{243}
        +L^2 \left(
                -\frac{50743}{162}
                +\frac{332 \zeta_2}{9}
        \right)
        +L \Biggl(
                \frac{43674185}{6804}
\nonumber\\ &&  
                -\frac{83456 \zeta_2}{81}
                +\frac{31528 \zeta_3}{81}
        \Biggr)
       -\frac{1094756651}{198450}
        -\frac{11488}{45} \Li_4\left(\frac{1}{2}\right)
        -\frac{1436 \ln^4(2)}{135}
        -\frac{2261}{5} \zeta_2
\nonumber\\ && 
        -\frac{488}{9} \ln(2) \zeta_2
        +\frac{2872}{45} \ln^2(2) \zeta_2
       +\frac{345311}{675} \zeta_2^2
        -\frac{12599017 \zeta_3}{4860}
\Biggr]
+\frac{1}{N^5} \Biggl[
        -\frac{20 L^4}{27}
        -\frac{51964 L^3}{1215}
\nonumber\\ && 
        +L^2 \left(
                \frac{3020069}{4050}
                -\frac{2492 \zeta_2}{27}
        \right)
        +L \left(
                -\frac{3048134171}{334125}
                +\frac{1160471 \zeta_2}{810}
                -952 \zeta_3
        \right)
\nonumber\\ && 
+        \frac{16455625566683}{1234926000}
        +256 \Li_4\left(\frac{1}{2}\right)
        +\frac{32 \ln^4(2)}{3}
        +\frac{1333649 \zeta_2}{16200}
        -\frac{940}{9} \ln(2) \zeta_2
        -64 \ln^2(2) \zeta_2
\nonumber\\ &&   
        -\frac{13810}{27} \zeta_2^2
        +\frac{1101248}{243} \zeta_3
\Biggr]
+\frac{1}{N^6} \Biggl[
         \frac{100 L^4}{81}
        +\frac{8858 L^3}{243}
      +L^2 \left(
                -\frac{353827}{243}
                +\frac{5140 \zeta_2}{27}
        \right)
\nonumber\\ &&
        +L \left(
                \frac{532290942707}{48648600}
                -\frac{684791 \zeta_2}{405}
                +\frac{162440 \zeta_3}{81}
        \right)
        -\frac{48448}{189} \Li_4\left(\frac{1}{2}\right)
\nonumber\\ && 
        -\frac{224935849458230479}{8765504748000}
        -\frac{6056 \ln^4(2)}{567}
        -\frac{34195783 \zeta_2}{34020}
        +\frac{1048}{9} \ln(2) \zeta_2
        +\frac{12112}{189} \ln^2(2) \zeta_2
\nonumber\\ &&  
      +\frac{543232}{945} \zeta_2^2
        -\frac{124918583 \zeta_3}{17010}
\Biggr]
+\frac{1}{N^7} \Biggl[
        -\frac{140 L^4}{81}
        -\frac{36406 L^3}{1701}
        +L^2 \left(
                \frac{147944849}{59535}
                -\frac{3268 \zeta_2}{9}
        \right)
\nonumber\\ && 
        +L \left(
                -\frac{27141581812291}{2383781400}
                +\frac{712342 \zeta_2}{405}
                -\frac{318616 \zeta_3}{81}
        \right)
+        \frac{1233346073528967337}{28633982176800}
        +256 \Li_4\left(\frac{1}{2}\right)
\nonumber\\ && 
       +\frac{32 \ln^4(2)}{3}
        +\frac{680314781 \zeta_2}{476280}
        -\frac{6796}{9} \ln(2) \zeta_2
        -64 \ln^2(2) \zeta_2
        -\frac{100714}{135} \zeta_2^2
        +\frac{21870544 \zeta_3}{1701}
\Biggr]
\nonumber\\ &&  
+\frac{1}{N^8} \Biggl[
        +\frac{20 L^4}{9}
        -\frac{13126 L^3}{5103}
        +L^2 \left(
                -\frac{67038673}{17010}
                +\frac{18788 \zeta_2}{27}
        \right)
        +L \Biggl(
                \frac{36270418853929}{4007896200}
\nonumber\\ &&            
                -\frac{15210661 \zeta_2}{8505}
     +\frac{208376 \zeta_3}{27}
        \Biggr)
        -\frac{13106933647387977503}{227341232118000}
        -\frac{11504}{45} \Li_4\left(\frac{1}{2}\right)
        -\frac{1438 \ln^4(2)}{135}
\nonumber\\ &&     
        -\frac{1097947 \zeta_2}{1215}
    +\frac{9784}{9} \ln(2) \zeta_2
        +\frac{2876}{45} \ln^2(2) \zeta_2
        +\frac{1557881 \zeta_2^2}{1350}
        -
        \frac{4263728837 \zeta_3}{204120}
\Biggr]
+\frac{1}{N^9} \Biggl[
        -\frac{220 L^4}{81}
\nonumber\\ && 
        +\frac{900506 L^3}{25515}
        +L^2 \left(
                \frac{57444913}{9450}
                -\frac{36124 \zeta_2}{27}
        \right)
        +L \Biggl(
                -\frac{358561874254051}{190375069500}
                +\frac{75800311 \zeta_2}{34020}
\nonumber\\ &&                
 -\frac{1223672 \zeta_3}{81}
        \Biggr)
        +\frac{37969630085411335836523}{703458726810840000}
        +256 \Li_4\left(\frac{1}{2}\right)
        +\frac{32 \ln^4(2)}{3}
        -\frac{3288591983 \zeta_2}{1360800}
\nonumber\\ &&   
      -\frac{32812}{9} \ln(2) \zeta_2
        -64 \ln^2(2) \zeta_2
        -\frac{90782}{45} \zeta_2^2
        +\frac{1953040931 \zeta_3}{51030}
\Biggr]
+
\frac{1}{N^{10}} \Biggl[
         \frac{260 L^4}{81}
        -\frac{389675 L^3}{5103}
\nonumber\\ &&    
     +L^2 \Biggl(
-\frac{1908055889}{204120}                
                +\frac{7828 \zeta_2}{3}
        \Biggr)
        +L \Biggl(
                -\frac{39361468046155979}{1979900722800}
                -\frac{1383047 \zeta_2}{5670}
                +\frac{2414920 \zeta_3}{81}
        \Biggr)
\nonumber\\ &&
        -\frac{2328298215957116515103723}{76817692967743728000}
        -\frac{25408}{99} \Li_4\left(\frac{1}{2}\right)
        -\frac{3176 \ln^4(2)}{297}
        +\frac{47691873199 \zeta_2}{22453200}
\nonumber\\ &&    
    +\frac{47320}{9} \ln(2) \zeta_2
        +\frac{6352}{99} \ln^2(2) \zeta_2
        +\frac{5656652 \zeta_2^2}{1485}
        -\frac{2480325346 \zeta_3}{40095}
\Biggr]\Biggr\}.
\end{eqnarray}
Furthermore, one finds that
\begin{eqnarray}
\frac{\tilde{a}_{gg,Q, N_F = 0}^{(3)}}{[1+(-1)^N]} 
- \frac{\Delta\tilde{a}_{gg,Q, N_F = 0}^{(3)}}{[1-(-1)^N]} \propto \frac{1}{N^2}. 
\end{eqnarray}
This explains the close agreement of the numbers in column 2 of Tables~1 and 2.

\vspace{5mm}
\noindent
{\bf Acknowledgment.}~
We would like to thank D.~Broadhurst, I. Bierenbaum, S.~Klein, P.~Marquard, C.G.~Raab, M.~Round, M.~Saragnese 
and F.~Wi\ss{}brock for discussions, and M.~Steinhauser for providing the code {\tt MATAD 3.0}. 
This work has received funding from the European Research Council (ERC) under the European
Union's Horizon 2020 research and innovation programme grant agreement 101019620 (ERC Advanced Grant TOPUP). It 
has been supported in part by the Austrian Science Fund (FWF) grants SFB F50 (F5009--N15) and P33530 and by the 
Research Center ``Elementary Forces and Mathematical Foundations (EMG)'' of J. Gutenberg University Mainz and DFG.

\end{document}